\documentclass[useAMS,usenatbib]{mn2e}
\usepackage{color} 
\usepackage[pdftex]{graphicx}

\usepackage{multirow}
\usepackage{amssymb}

\usepackage{amsmath}

\newcommand{\kmsk}{\mbox{$\>{\rm km\, s^{-1}\, kpc^{-1}}$}}
\newcommand{\kkms}{\mbox{$\>{\rm kpc\, km\, s^{-1}}$}}

\newcommand{\kpc}{\mbox{$\>{\rm kpc}$}} 
\newcommand{\pc}{\mbox{$\>{\rm pc}$}} 
\newcommand{\Gyr}{\mbox{$\>{\rm Gyr}$}}
\newcommand{\Myr}{\mbox{$\>{\rm Myr}$}}
\newcommand{\yr}{\mbox{$\>{\rm yr}$}}
\newcommand{\Msun}{\>{\rm M_{\odot}}}

\newcommand{\omp}{\mbox{$\Omega_{\rm p}$}} 
\newcommand{\omg}[1]{\mbox{$\Omega_{#1}$}}
\newcommand\degrees{^\circ}
\newcommand{\avg}[1]{\mbox{$\left<{#1}\right>$}}
\newcommand{\avgt}[1]{\mbox{$\left<{#1}\right>_t$}}
\newcommand{\sig}[1]{\mbox{$\sigma_{#1}$}}
\newcommand{\act}[1]{\mbox{$J_{#1}$}}

\newcommand{\feh}{\mbox{$\rm [Fe/H]$}}
\newcommand{\al}{\mbox{$\rm \alpha$}}
\newcommand{\ofe}{\mbox{$\rm [O/Fe]$}}
\newcommand{\alfe}{\mbox{$\rm [\alpha/Fe]$}}

\newcommand{\age}{\mbox{$\tau$}}
\newcommand{\avag}{\mbox{$\left<\tau\right>$}}
\newcommand{\avfe}{\mbox{$\left<{\feh}\right>$}}

\def\gaia{{\it Gaia}}

\def\eg{{\it e.g.}}

\def\ie{{\it i.e.}}

\title[Azimuthal \feh\ variations]{Azimuthal metallicity variations, spiral structure, and the failure of radial actions based on assuming axisymmetry}

\author[Debattista et al.]{Victor P. Debattista$^1$\thanks{E-mail:
    vpdebattista@gmail.com}, Tigran Khachaturyants$^2$, Jo\~ao A. S. Amarante$^{2,3,1}$\thanks{E-mail: joaoant@gmail.com}\thanks{Visiting Fellow at UCLan}, \newauthor Christopher Carr$^4$, Leandro {Beraldo e Silva}$^{5,6}$ and Chervin F. P. Laporte$^{7,3,8}$ \\ 
$^{1}$ Jeremiah Horrocks Institute, University of Central
  Lancashire, Preston, PR1 2HE, UK \\
$^{2}$ Department of Astronomy, Shanghai Jiao Tong University, 800
  Dongchuan Road, Shanghai 200240, P.R. China \\
$^{3}$ Institut de Ci\`encies del Cosmos (ICCUB), Universitat de
  Barcelona (IEEC-UB), Mart\'i i Franqu\`es 1, E-08028 Barcelona,
  Spain \\
$^{4}$ Department of Astronomy, Columbia University, 550 West 120th
  Street, New York, NY, 10027, USA \\
$^{5}$ Department of Astronomy, University of Michigan, 500 Church
  St., Ann Arbor, MI, 48109, USA \\
$^{6}$ Department of Astronomy \& Steward Observatory, University of
  Arizona, Tucson, AZ, 85721, USA \\
$^{7}$ LIRA, Observatoire de Paris, PSL Research University, CNRS,
  Place Jules Janssen, 92195 Meudon, France \\
$^{8}$ Kavli IPMU (WPI), UTIAS, The University of Tokyo, Kashiwa,
  Chiba 277-8583, Japan
}
\begin{document}   

\date{{\it Draft version on \today}}
\pagerange{\pageref{firstpage}--\pageref{lastpage}} \pubyear{----}
\maketitle

\label{firstpage}

\begin{abstract}
  We study azimuthal variations in the mean stellar metallicity,
  \avfe, in a self-consistent, isolated simulation in which all stars
  form out of gas. We find \avfe\ variations comparable to those
  observed in the Milky Way and which are coincident with the spiral
  density waves. The azimuthal variations are present in young and old
  stars and therefore are not a result of recently formed
  stars. Similar variations are present in the mean age and
  \al-abundance. We measure the pattern speeds of the \avfe-variations
  and find that they match those of the spirals, indicating that
  spirals are the cause of the metallicity patterns. Because younger
  stellar populations are not just more \feh-rich and \al-poor but
  also dynamically cooler, we expect them to more strongly support
  spirals, which is indeed the case in the simulation. However, if we
  measure the radial action, \act{R}, using the St\"ackel axisymmetric
  approximation, we find that the spiral ridges are traced by regions
  of high \act{R}, contrary to expectations. Assuming that the passage
  of stars through the spirals leads to unphysical variations in the
  measured \act{R}, we obtain an improved estimate of \act{R} by
  averaging over a $1\Gyr$ time interval. This time-averaged
  \act{R}\ is a much better tracer of the spiral structure, with
  minima at the spiral ridges. We conclude that the errors incurred by
  the axisymmetric approximation introduce correlated deviations large
  enough to render the instantaneous radial actions inadequate for
  tracing spirals.
\end{abstract}


%

\section{Introduction}
\label{s:intro}


Observations of Milky Way (MW) disc stars have been finding azimuthal variations in the mean stellar metallicity at fixed radius. 
Early work using small samples of Cepheids, which are less than $1\Gyr$ old, found an azimuthal variation in the metallicity at the Solar radius \citep{luck+06}. On a larger scale, \citet{pedicelli+09} showed that the relative abundance of metal-rich and metal-poor Cepheids varies across the different quadrants. While \citet{luck+11} were unable to detect any azimuthal metallicity variations in a sample of 101 Cepheids, \citet{lepine+11} found azimuthal variations in both Cepheids and open clusters comparable to the radial gradient, and attributed them to the spiral structure.

Large surveys, such as APOGEE and the \gaia\ data releases \citep{apogee, GaiaDR2Kinematics} have significantly accelerated the characterisation of azimuthal stellar metallicity variations.
\citet{bovy+14} used a $\sim10^4$ red clump star sample from APOGEE, with an age distribution favouring $1-4\Gyr$, to search for azimuthal metallicity variations. They found azimuthal variations to be smaller than  $0.02$ dex within an azimuthal volume spanning $45\degrees$. 
\citet{poggio+22} used \gaia\ DR3 data to map metallicity variations within $4\kpc$ of the Sun. They showed that metal-rich stars trace the spiral structure. They also found that these variations are stronger for young stars, but are also present in old stellar populations.
\citet{hawkins23} used a sample of OBAF stars from LAMOST, and a separate sample of giant stars from \gaia\ DR3 to measure radial and vertical metallicity gradients. He also found azimuthal metallicity variations of order $\sim 0.1$ dex on average, which are co-located with the spiral structure in the \gaia\ data (but not in the LAMOST data).
\citet{imig+23} used 66K APOGEE DR17 red giants to suggest that the azimuthal metallicity variations in the Solar Neighbourhood do not seem to track the spiral structure.
Using APOGEE DR17, \citet{hackshaw+24} recently found azimuthal variations in \alfe\ of order $0.1$~dex; they also showed that azimuthal variations seem to be stronger in older populations.

Theoretical studies of azimuthal metallicity variations attribute them to spiral structure, via a number of different mechanisms.
The simplest idea is that the metallicity variations reflect the higher metallicity of recently formed stars, since the gas metallicity itself has an azimuthal variation of $\sim 0.1$~dex \citep{wenger+19}.
Comparable variations in gas metallicity have also been found in external galaxies
\citep[\eg][]{sanchez-menguiano+16, vogt+17, ho+18, kreckel+19, hwang+19}. \citet{spitoni+19} developed a chemical evolution model of the Galactic disc perturbed by a spiral. They found large metallicity variations near the corotation of the spiral.
\citet{solar+20} measured the metallicity of star-forming regions in EAGLE \citep{eagle} simulated galaxies finding variations of $\sim0.12$~dex.
\citet{bellardini+22} used a suite of FIRE-2 cosmological simulations \citep{Hopkins2018fire2} to show that the azimuthal variation in the stellar metallicity {\it at birth} decreases with time, mirroring the evolution of the gas phase metallicity \citep{bellardini+21}.

Alternatively, azimuthal variations of the mean stellar metallicity have been viewed as a signature of radial migration. 
Using an $N$-body simulation of a barred spiral galaxy, in which star particles had metallicity painted on based on their initial radius, \citet{pdimatteo+13} proposed that azimuthal metallicity variations are produced by the radial migration induced by the non-axisymmetric structure. Meanwhile, in a zoom-in cosmological simulation, \citet{grand+16b} found that the kinematic motions induced by the spirals drive an azimuthal variation in the metallicity, with metal-rich stars on the trailing side and metal-poor ones on the leading side, which they interpreted as a signature of migration of stars from a stellar disc with a radial metallicity gradient.
\citet{carr+22} noted that, in high-resolution $N$-body simulations \citep{laporte+18} of the Sagittarius dwarf's encounter with the Milky Way, azimuthal metallicity variations appear in the disc's impulsive response to the passage through the disc, during which migration is also enhanced, and then continuing within the spirals in the periods between passages.
\citet{sanchez-menguiano+16} observed an offset between the metallicity of the gas and the location of the spiral, which they interpreted as the result of gas migration.

A third hypothesis on the origin of azimuthal metallicity variations posits that they are produced by the different reactions of different stellar populations to a spiral perturbation.
\citet{khoperskov+18b} used pure $N$-body (no gas or star formation) simulations of compound stellar discs to show that azimuthal variations can be produced by the varying response of cool, warm and hot stellar populations to spirals. Even in the absence of a metallicity gradient in the disc, if the different populations have different metallicities, azimuthal variations result. They showed that, when migration dominates the azimuthal variations (because all populations had the same radial metallicity profile), the metallicity and density peaks are offset, whereas when the variations are driven by the different response of different populations then the density and metallicity peaks tend to align.

In this paper we explore the link between spiral structure and azimuthal metallicity variations. In agreement with \citet{khoperskov+18b}, we show, using a model with star formation and self-consistent chemical evolution, that the main driver of these variations are differences in how strongly a stellar population can trace the spiral structure based on its radial action. In Section~\ref{s:simulation} we describe the simulation used in this paper. In Section~\ref{s:chemistry} we describe the azimuthal chemical variations, both for \feh\ and \ofe. We find that variations in the average \feh\ closely track the spiral structure. Therefore in Section~\ref{s:omega} we construct a method to track the pattern speeds of the average metallicity variations, taking care to ensure that the average metallicity is not weighted by the density. Thus the pattern speeds of the metallicity variations we obtain are independent from those of the density. We show that there is such a close correspondence between the pattern speeds of the average metallicity and of the density in the disc region that it must be the spirals that are driving the metallicity variations. Section~\ref{s:age} explores the idea that the metallicity variations are due to young stars but finds that the azimuthal metallicity variations are present in all age populations, both very young and those older than $6\Gyr$. Then in Section~\ref{s:jr} we explore the dependence on the radial action. We find that the radial action provides a very robust explanation for the azimuthal metallicity variations, provided that the radial action is time averaged to reduce the errors introduced by the assumption of axisymmetry. We show that instantaneous values of the radial action instead provide almost exactly the wrong answer, in that stars with seemingly large radial action trace the spiral. 
In Section~\ref{s:migration} we explore whether migration is driving the azimuthal metallicity variations in the model. Multiple lines of evidence suggest that migration is not the dominant factor in setting up the azimuthal variations.
In Section~\ref{s:actionspace} we explore whether the temporal changes in the radial action are due to libration about resonances. We show that, even when there are signs of libration, there is also a large jitter in the time evolution of  the instantaneous radial action computed under the assumption of axisymmetry.
We present our conclusions in Section~\ref{s:discussion}.
The appendices present comparable analyses at other times, explore using medians instead of means, and track a sample of stars in action space chosen to favour those at resonances.


\section{Simulation description}
\label{s:simulation}

The model we use here was first described in \citet{fiteni+21}, who
referred to it as model M1\_c\_b.  The model evolves via the cooling of
a hot gas corona, in equilibrium within a spherical dark matter halo.
The dark matter halo has a virial mass $M_{200} = 10^{12} \Msun$ and a
virial radius $r_{200} \simeq 200~\kpc$, making it comparable to the
MW \citep[see the review of][and references therein]{bland-hawthorn_gerhard16}. The gas corona has the same profile but with an
overall mass that is $10\%$ that of the dark matter. The initial gas
corona, as well as the dark matter halo, is comprised of 5 million
particles, which means gas particles have an initial mass of $2.8
\times 10^4 \Msun$. The dark matter particles come in two mass species, one of mass $2 \times 10^5\Msun$ inside $r =
200\kpc$ and the other of mass $7.1 \times 10^5 \Msun$ outside. Gas particles are
softened with $\varepsilon = 50~\pc$, while the dark matter particles
have $\varepsilon = 100~\pc$.  A cylindrical spin is imparted to the
gas particles such that the gas corona has $\lambda \equiv
L|E|^{1/2}/(G M_{vir}^{5/2}) = 0.065$, where $L$ and $E$ are the total
angular momentum and energy of the gas, and $G$ is the
gravitational constant \citep{peebles69}. After setting up the system,
we evolve it adiabatically for $5\Gyr$ without gas cooling or star
formation to ensure that it is fully relaxed.

We then evolve these initial conditions with gas cooling and star
formation. As the gas cools, it settles into a rotationally supported
disc and ignites star formation. Thus all stars form directly out of
gas.  Individual star particles inherit their softening, $\varepsilon
= 50~\pc$, and chemistry from the gas particle from which they form.
Star formation, with an efficiency of 5 per cent, requires a gas particle to have
cooled below 15,000 K, exceeded a density of $0.1$ amu cm$^{-3}$ and
to be part of a convergent flow. Star particles are all born with the same
mass, $9.3 \times 10^3\Msun$, which is $1/3$ of the (initial) gas
particle mass. Gas particles continue to form stars until their mass
falls below $5.6 \times 10^3\Msun$ ($20\%$ of the initial gas particle
mass), at which point a gas particle is removed and its mass distributed to the nearest gas particles. After $13\Gyr$,
roughly $2.6$ million gas particles remain.

Once star formation begins, feedback from supernovae Ia and II, and from AGB
winds, couples $4\times 10^{51}$ erg per supernova to the interstellar
medium using the blastwave prescription of \citet{stinson+06}. Thermal
energy and chemical elements are mixed amongst gas particles using the
turbulent diffusion prescription of \citet{sshen+10} with a mixing
coefficient of 0.03.

The model is evolved for $13\Gyr$ with {\sc gasoline} \citep{gasoline,
  gasoline2}, an $N$-body$+$SPH code\footnote{{\sc gasoline} is
available at https://www.gasoline-code.com.}.  Gravity is solved using
a KD-tree; we employ a tree opening angle of $\theta = 0.7$ with a
base time-step of $\Delta t = 5\Myr$. Time-steps of individual
particles are then refined by multiples of 2 until they satisfy the
condition $\delta t \equiv \Delta t/2^n < \eta \sqrt{\varepsilon/a_g}$,
where $a_g$ is the gravitational acceleration of a particle, and we
set the refinement parameter $\eta = 0.175$. Time-steps for gas
particles are also refined in a similar manner to satisfy the
additional condition $\delta t_{gas} < \eta_{courant}h/[(1 + \alpha)c
  + \beta\mu_{max}]$, where $\eta_{courant} = 0.4$, $h$ is the SPH
smoothing length set over the nearest 32 particles, $\alpha=1$ is the
shear coefficient, $\beta=2$ is the viscosity coefficient, $c$ is the
sound speed and $\mu_{max}$ is the maximum viscous force between gas
particles \citep{gasoline}.  With these time stepping recipes,
typically 10 rungs (maximum $n=9$, corresponding to $\delta t =
9766\yr$) are required to move all the particles.

We save snapshots of the simulation every $5 \Myr$. This high cadence
gives a Nyquist frequency of $\sim 300 \kmsk$ (for $m=2$
perturbations), which is well above any frequency of interest in our
MW-sized galaxy.  By $t=13~\Gyr$, which is the main snapshot we
consider here, the simulation has formed 11,587,120 star particles
(2.9M in the radial range $5 \leq R/\kpc \leq 10$, where we focus our
attention), allowing us to dissect the model by stellar population
properties while ensuring that we have enough particles for reliable
study.

\citet{fiteni+21} showed that this model experiences an episode of
mild clump formation lasting to $\sim 2\Gyr$, which gives rise to a
small population of retrograde stars in the Solar Neighbourhood. A bar also forms at $t\simeq
4~\Gyr$, which subsequently weakens at $t\simeq 7\Gyr$ before
recovering. Meanwhile, \citet{khachaturyants+22} showed that weak
bending waves propagate through the disc, with peak amplitudes $\la
50~\pc$ in the Solar Neighbourhood. \citet{khachaturyants+22b}
computed the pattern speeds of the density perturbations
(bar$+$spirals), as well as those of the breathing and bending waves
in this model. They showed that the breathing waves have the same
pattern speeds as the spirals and bar, but very different pattern
speeds from the bending waves, from which they concluded that
breathing waves are driven by density perturbations. Finally
\citet{ghosh+22} used this model at $12\Gyr$ to show that breathing
motions induced by spirals get larger with distance from the
mid-plane, and are stronger for younger stars
than for older stars.

We focus most of our analysis on the single snapshot of this model at
$t=13\Gyr$. We have verified that results are broadly similar at other
times, and in Appendix~\ref{app:othertimes} we present the azimuthal
metallicity variations at other times.


\section{Chemical azimuthal variations}
\label{s:chemistry}

\subsection{The azimuthal variation of \avfe}

\begin{figure*}
\centerline{
\includegraphics[angle=0.,width=0.5\hsize]{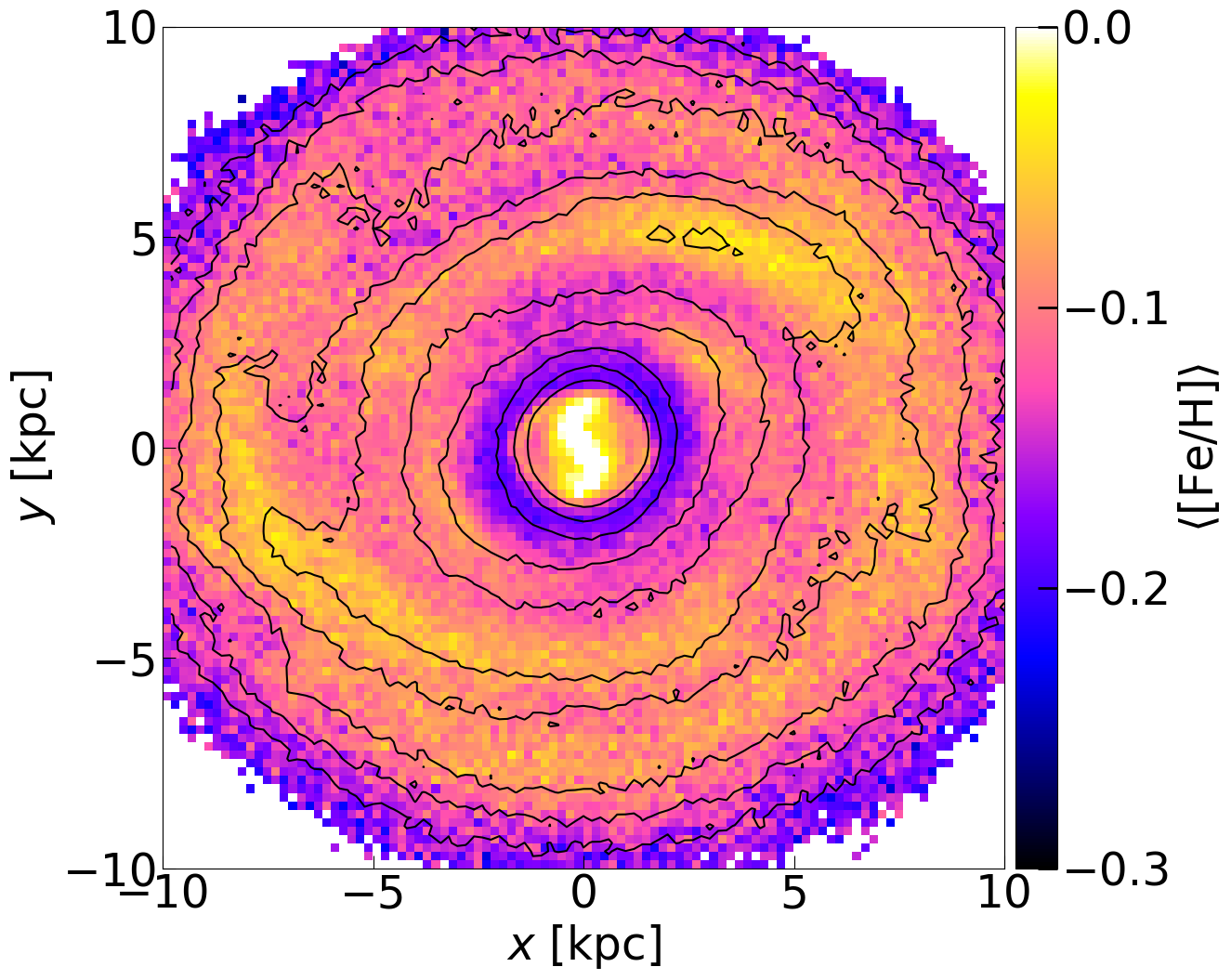}
\includegraphics[angle=0.,width=0.5\hsize]{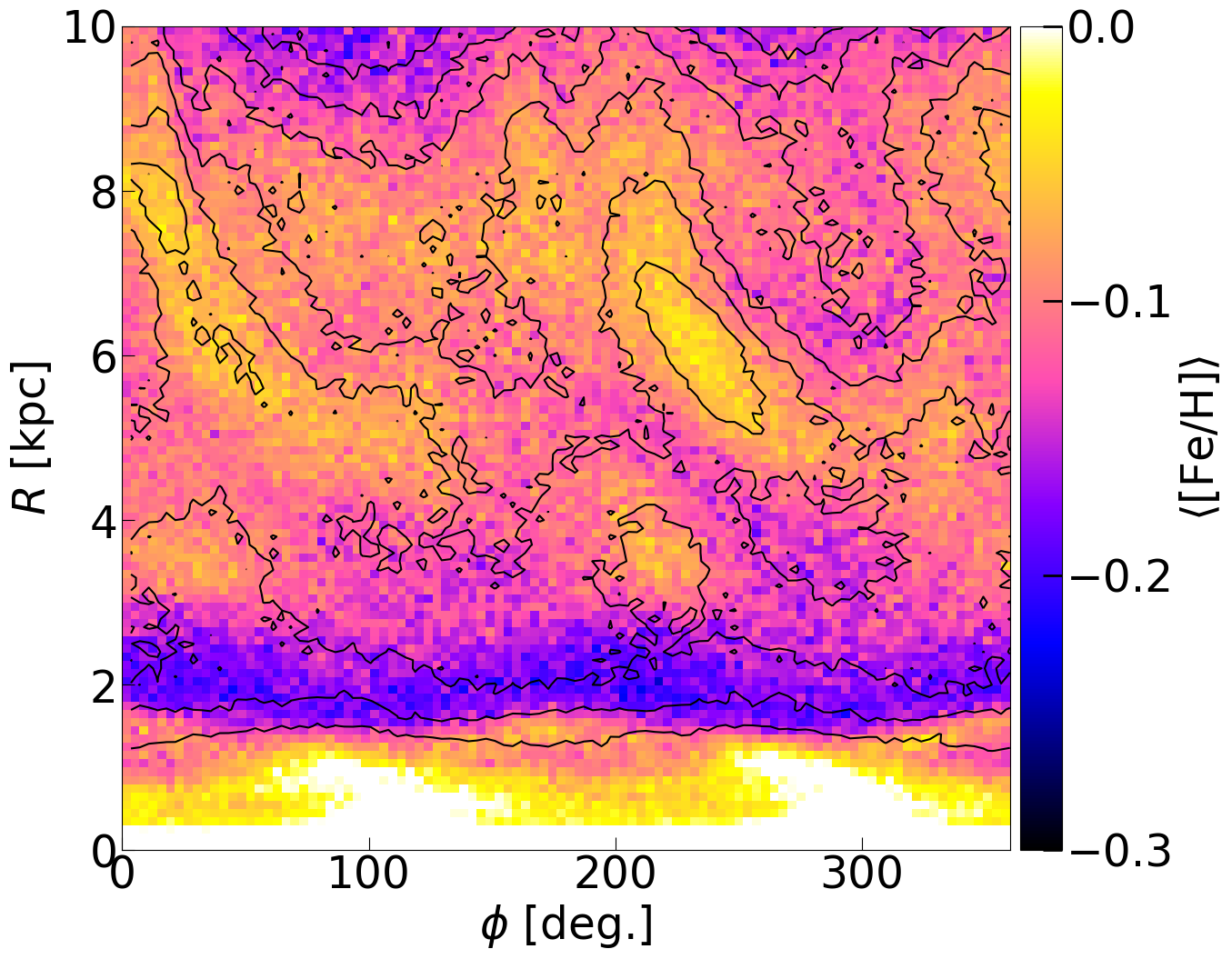}\
}
\centerline{
\includegraphics[angle=0.,width=0.5\hsize]{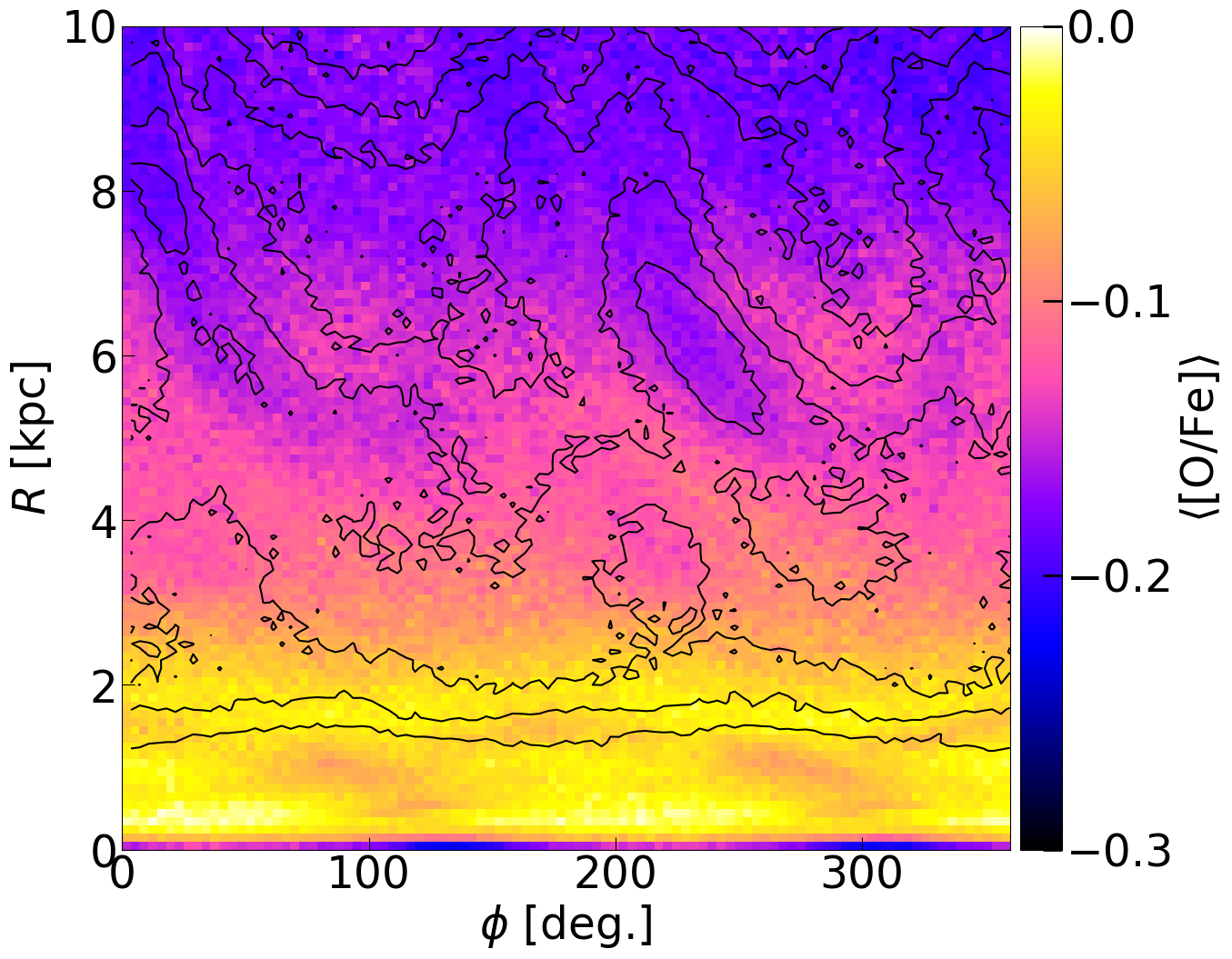}
\includegraphics[angle=0.,width=0.5\hsize]{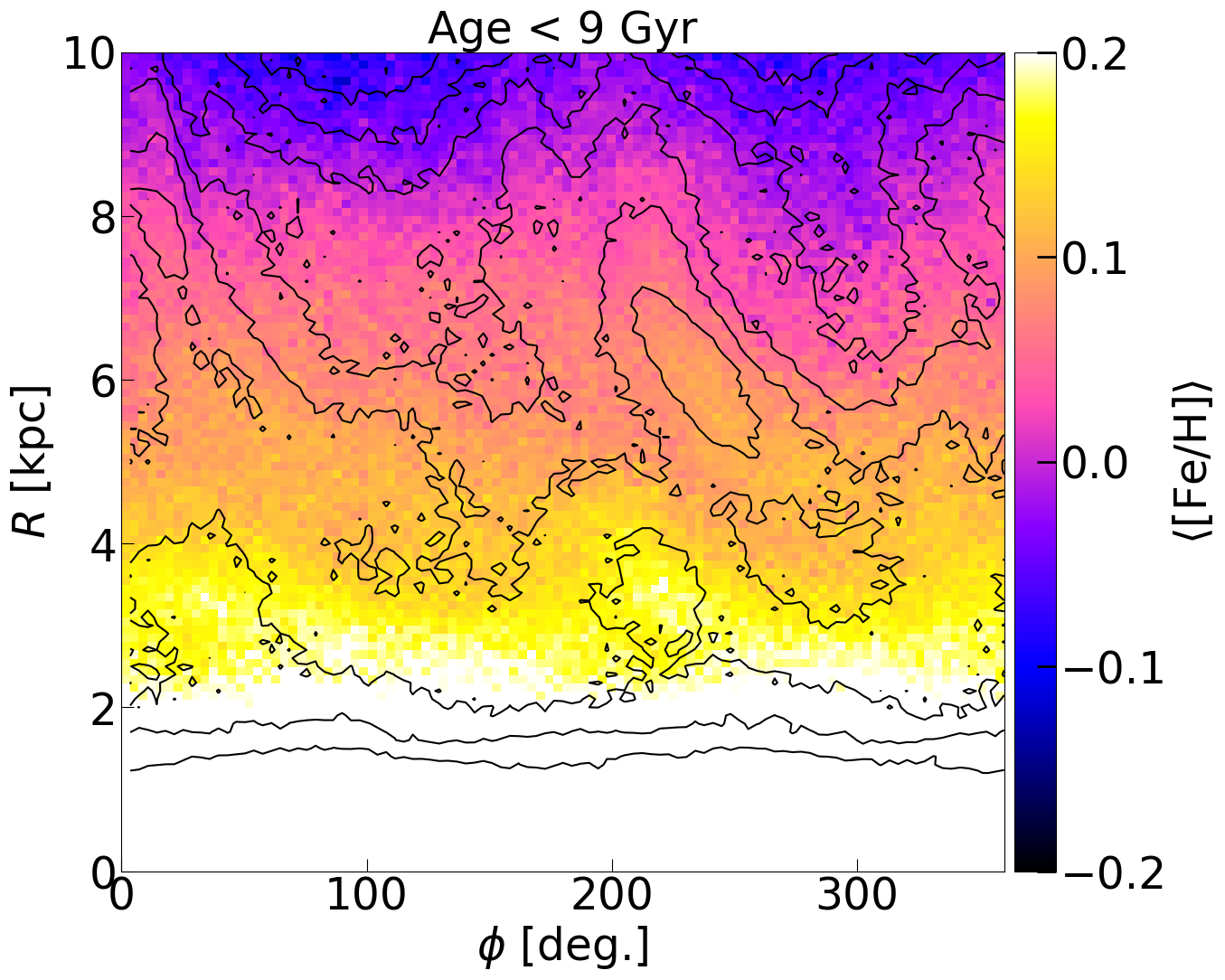}\
}
\caption{\avfe\ in Cartesian (top left) and cylindrical (top right)
  coordinates. The bottom left panel shows \avg{\ofe}\ in cylindrical
  coordinates while the bottom right panel shows \avfe\ excluding
  stars older than $9\Gyr$.  In all panels the contours correspond to
  the total stellar surface density. The disc is rotating in a counter-clockwise
  sense, \ie\ rotation is in the direction of increasing $\phi$. The
  model is shown at $13~\Gyr$.
\label{f:azimuthal}}
\end{figure*}

In Fig.~\ref{f:azimuthal} we present the mean metallicity, \avfe, of
the model at $13~\Gyr$ in both Cartesian (top left) and cylindrical
(top right) coordinates. Clear azimuthal variations in \avfe\ are
present. The density contours show a weak bar extending to $R\sim
2~\kpc$ and open spirals beyond. In the disc region ($R > 5\kpc$), the peak \avfe\ coincides with
regions of high density on the spirals.

\begin{figure*}
  \centerline{
    \includegraphics[angle=0.,width=0.5\hsize]{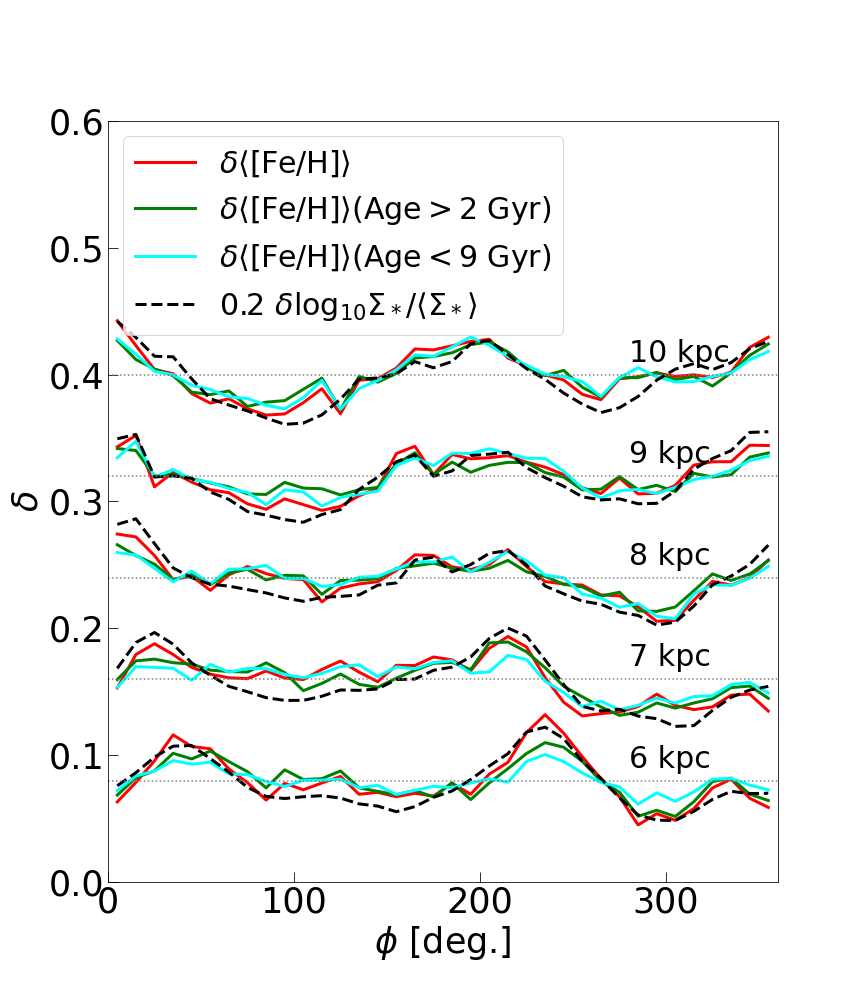}
    \includegraphics[angle=0.,width=0.5\hsize]{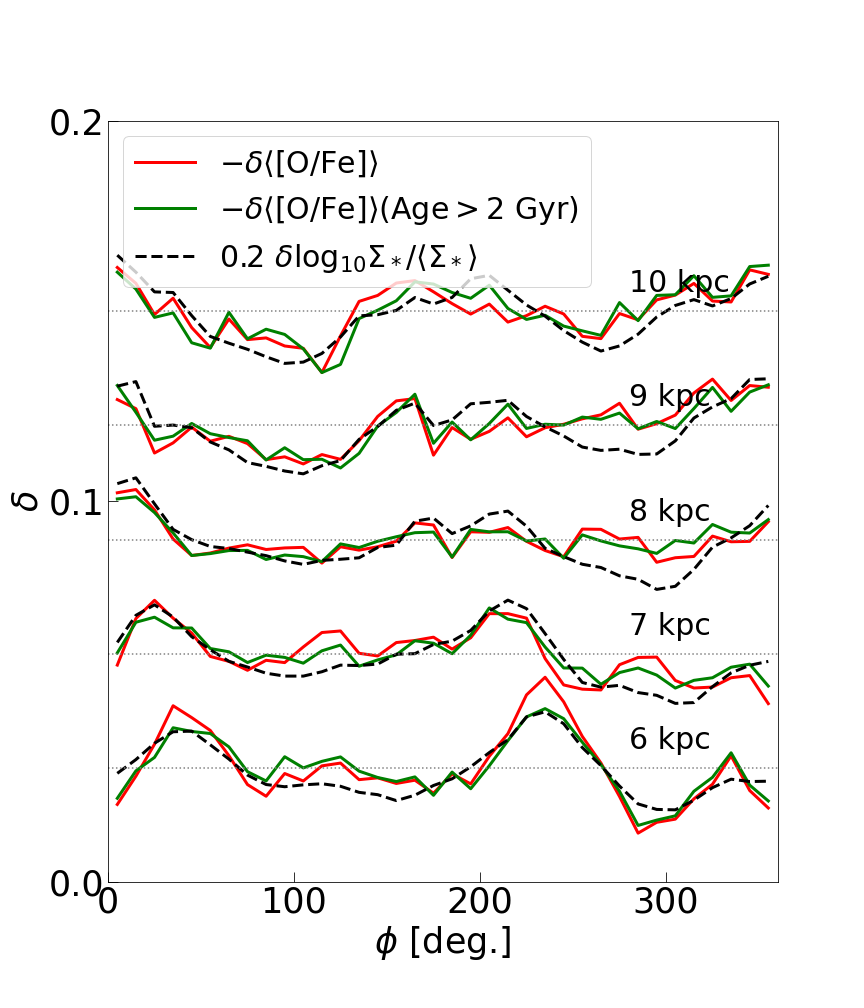}
  }
\caption{The variation of \avfe\ (left) and $-\avg{\ofe}$ (right)
  compared with the density variation at a series of $500~\pc$-wide
  annuli centred at $6\kpc$ to $10\kpc$, as indicated. The average values for each radius
  are subtracted from each azimuthal profile and the profiles are then
  vertically offset by a fixed amount to show the variation. The
  azimuthal profiles of the density have been scaled by the indicated
  factors for ease of comparison. The dotted horizontal lines indicate
  the zero for each radius.
\label{f:razimuthal}}
\end{figure*}

The left panel of Fig.~\ref{f:razimuthal} compares the azimuthal variations of the stellar
surface density, $\delta \Sigma/\avg{\Sigma}$, with those of the mean metallicity, $\delta \avfe$, in the disc region.
At these radii ($5.75 \leq R/\kpc \leq 10.25$), where the spirals
dominate, the variations in \avfe\ closely follow those of the density, with peaks and troughs at the same locations. This panel also shows that the $\delta 
\avfe$ of stars older than $2\Gyr$ behaves very similarly to that of the full distribution, with only slightly weaker
amplitudes, indicating that recent metal-rich star formation is not driving the \avfe\ variations.
The main variation in both the density and \avfe\ has an $m=2$
multiplicity. The maximum to minimum variations in \avfe\ are of order
$0.05-0.1$ dex outside the bar radius. In the MW, \citet{hawkins23} found metallicity variations of the same order.
The top panels of Figs.~\ref{f:medianmaps} and \ref{f:medianazimuthal} show maps and azimuthal profiles of the median \feh, finding a comparable behaviour as for the means.

\begin{figure*}
\centerline{
\includegraphics[angle=0.,width=0.5\hsize]{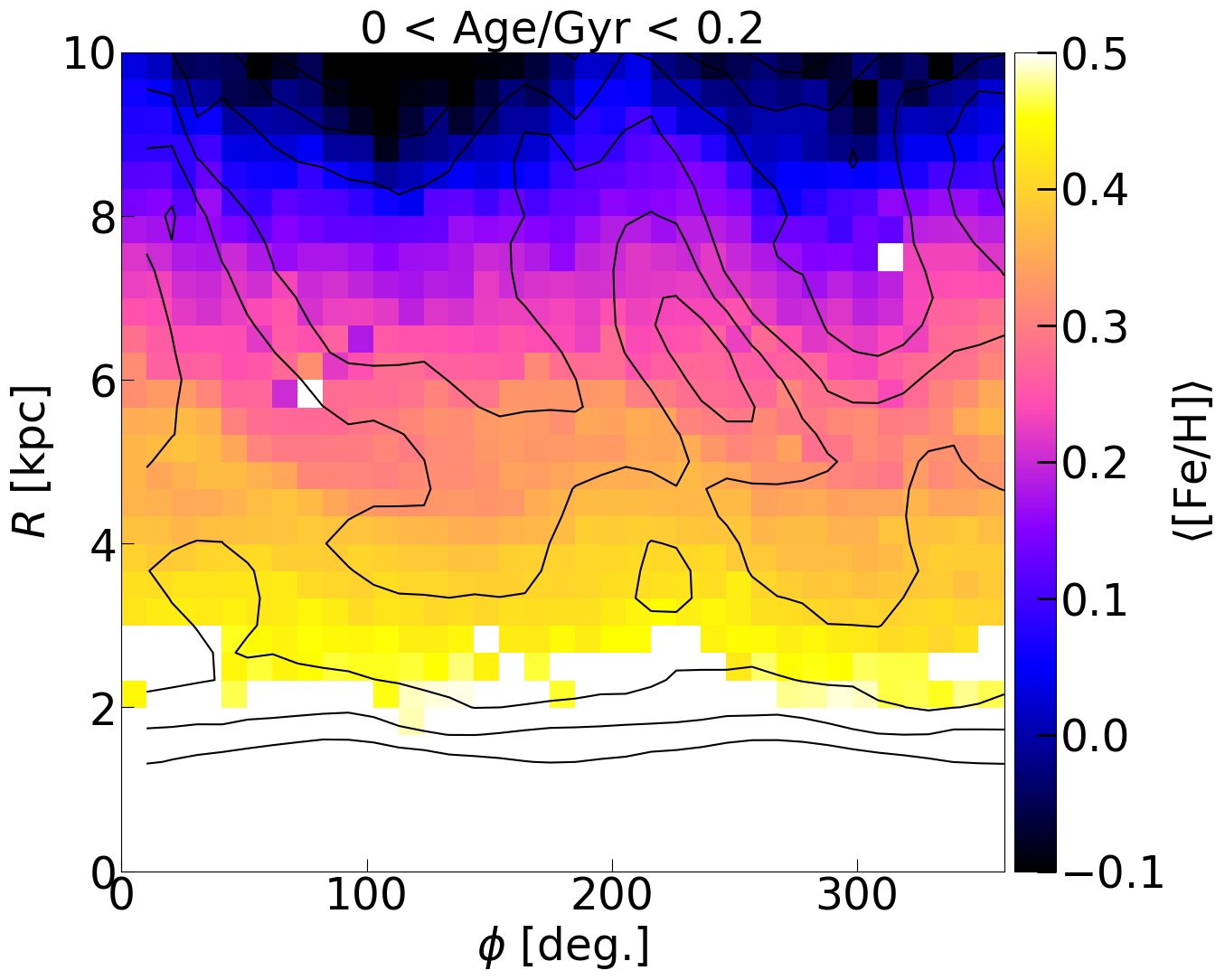}
\includegraphics[angle=0.,width=0.5\hsize]{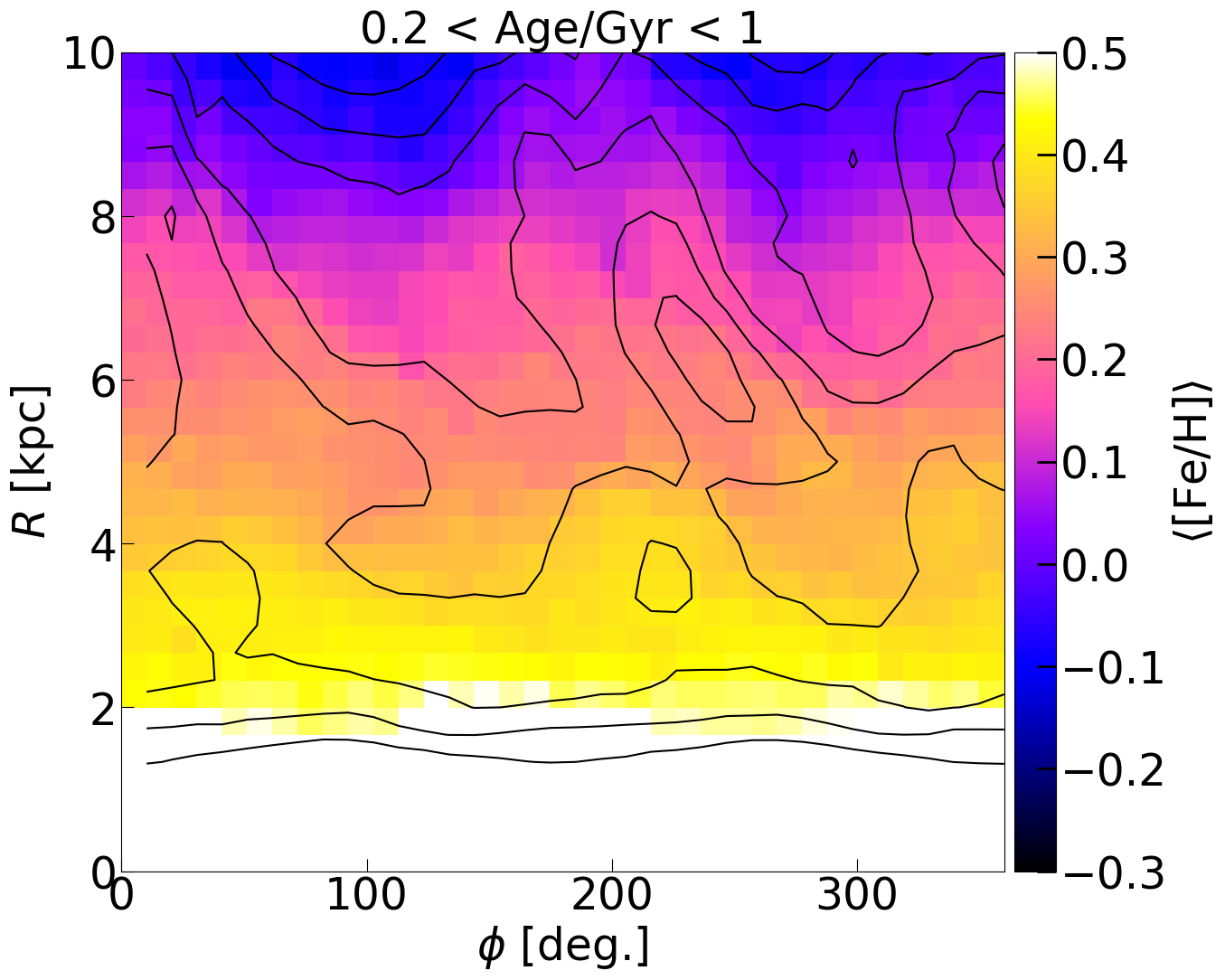}\
}
\centerline{
\includegraphics[angle=0.,width=0.5\hsize]{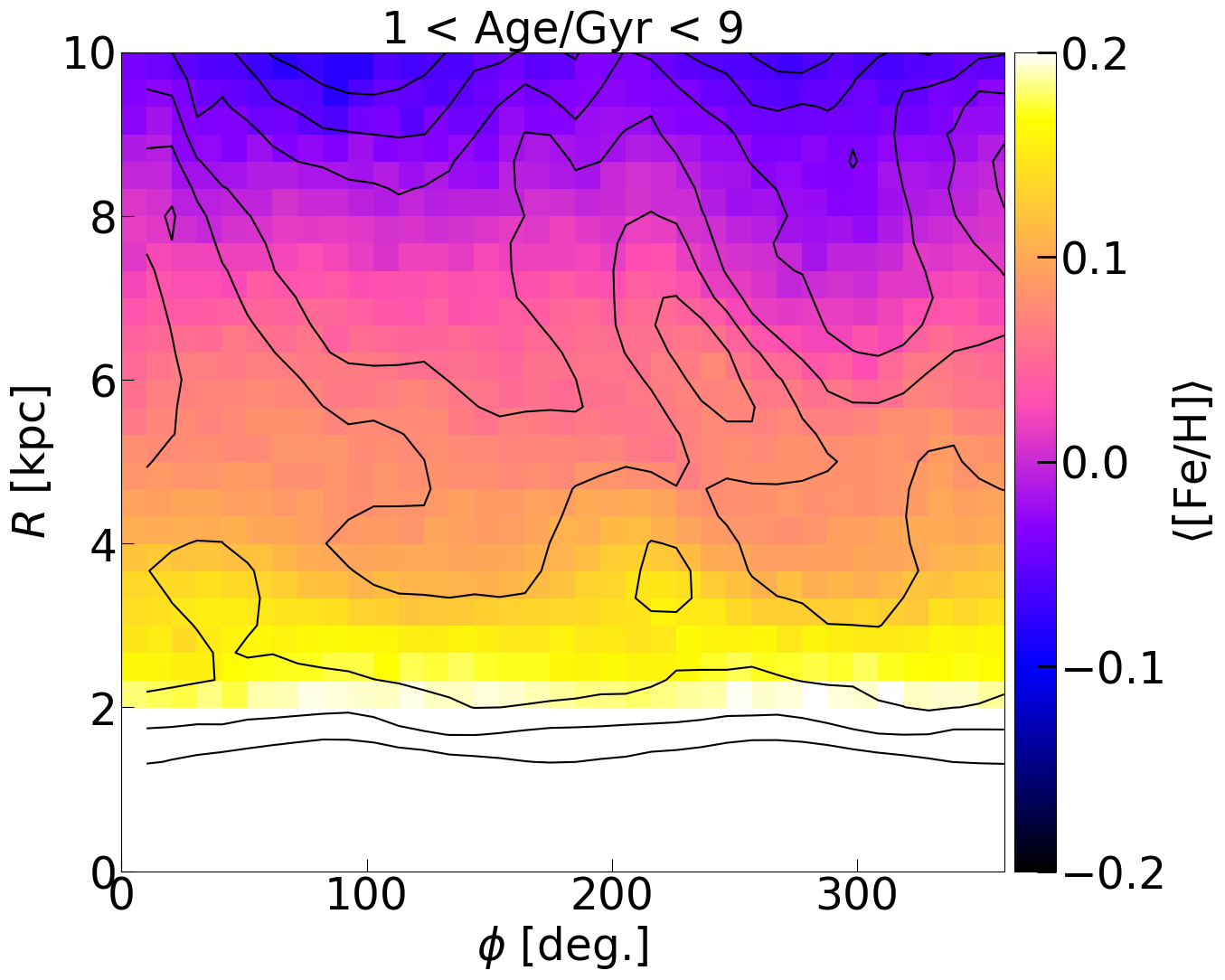}
\includegraphics[angle=0.,width=0.5\hsize]{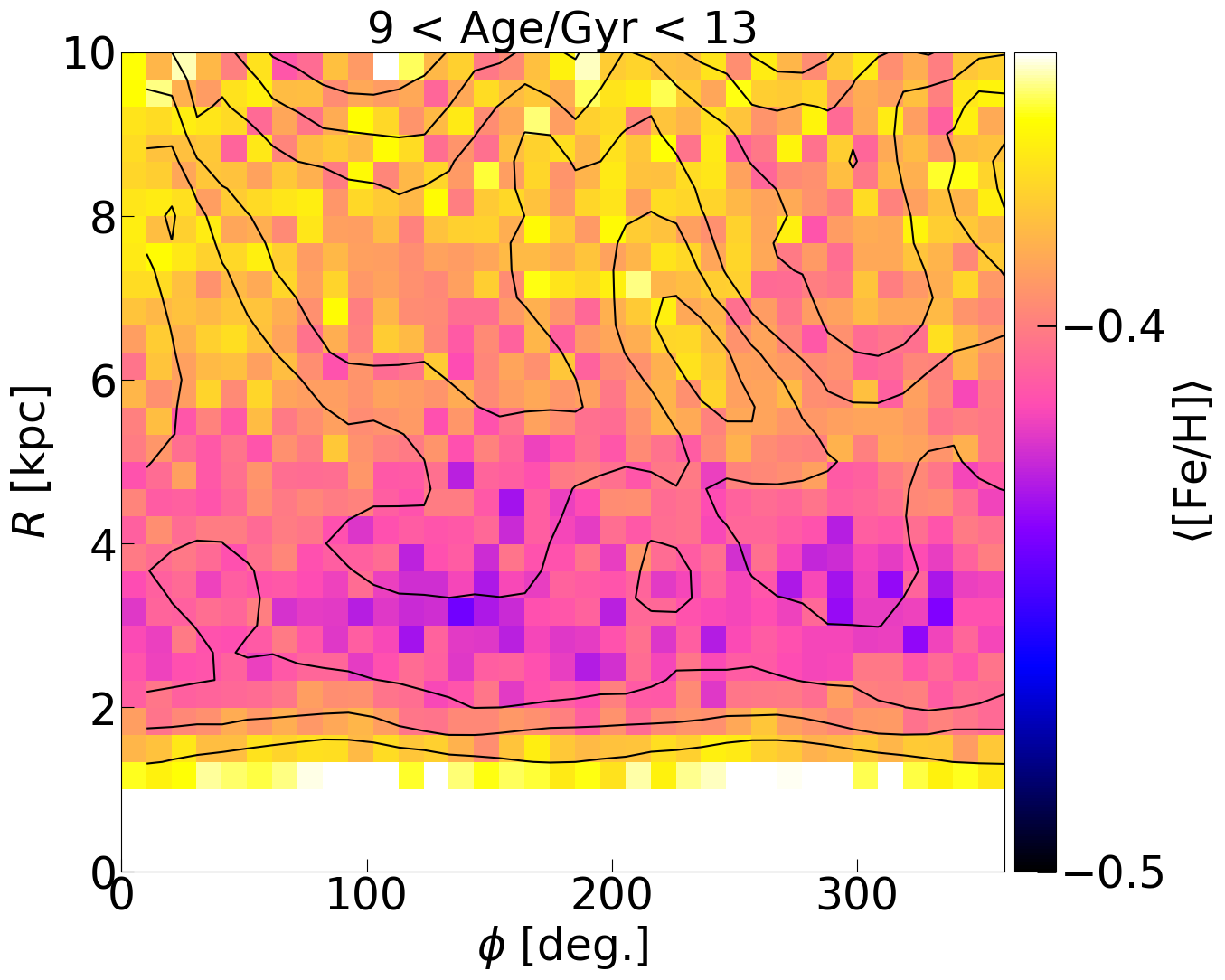}\
}
\caption{The distribution of stellar \avfe\ at $13 \Gyr$ split by age,
  as indicated. White represents bins with \avfe\ off the scale
  (inside $\sim 2\kpc$) or with less than 10 particles (elsewhere). In
  all panels the identical contours correspond to the total surface
  density. The disc is rotating in a counter-clockwise sense,
  \ie\ rotation is in the direction of increasing $\phi$.
\label{f:azimbyage}}
\end{figure*}

Since the azimuthal variations in \avfe\ are not limited to the young stars, we proceed by exploring the \avfe-azimuthal variations of different age populations.  Fig.~\ref{f:azimbyage} shows \avfe\ in
cylindrical coordinates for four age groups. A clear azimuthal
variation is present in the very young ($\age \leq 200$ \Myr) stars,
as seen in the MW \citep{luck+06, lemasle+08, pedicelli+09}. A similar pattern
is seen in slightly older stars, up to $1\Gyr$ old. This behaviour
might be expected if the spirals are the locii of metal-rich star
formation. But a similar azimuthal variation in \avfe\ is evident in
the intermediate age stars ($1 < \age/\Gyr < 6$) and even, to a lesser
extent, in the old stars ($6 < \age/\Gyr < 13$). Thus the azimuthal
variation of \avg{\feh}\ is not a result of recent star formation but
must be inherent to spiral structure.

\begin{figure*}
\centerline{
\includegraphics[angle=0.,width=0.5\hsize]{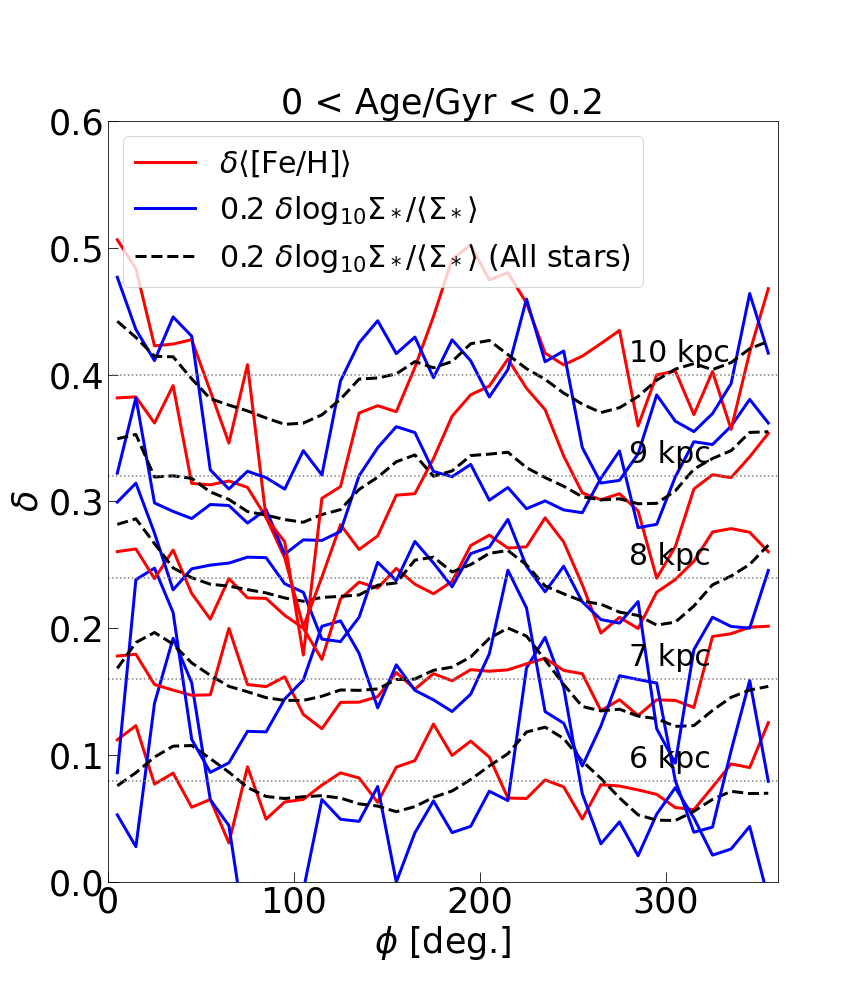}
\includegraphics[angle=0.,width=0.5\hsize]{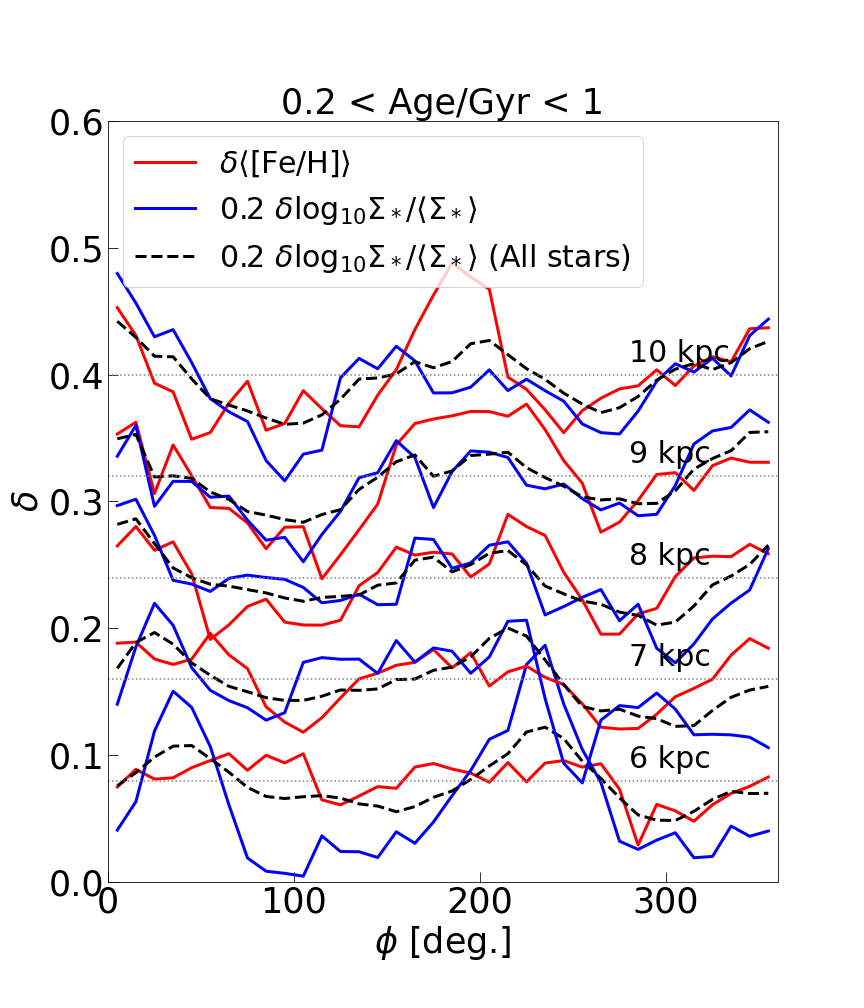}\
}
\centerline{
\includegraphics[angle=0.,width=0.5\hsize]{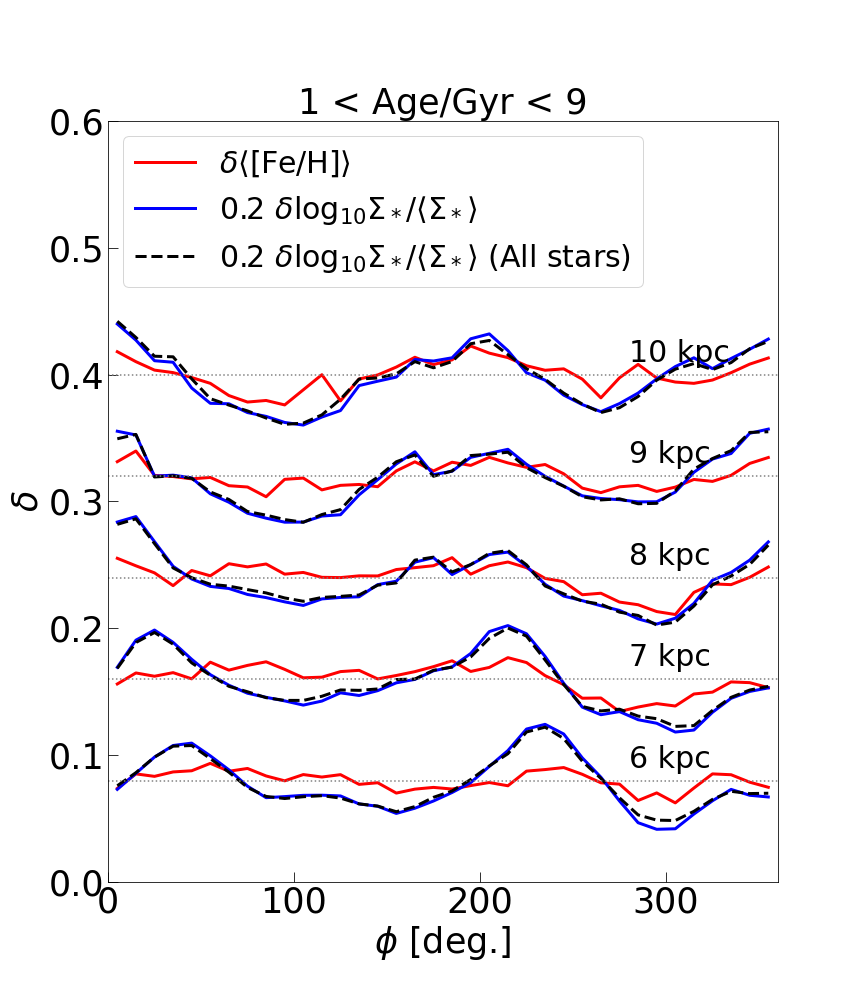}
\includegraphics[angle=0.,width=0.5\hsize]{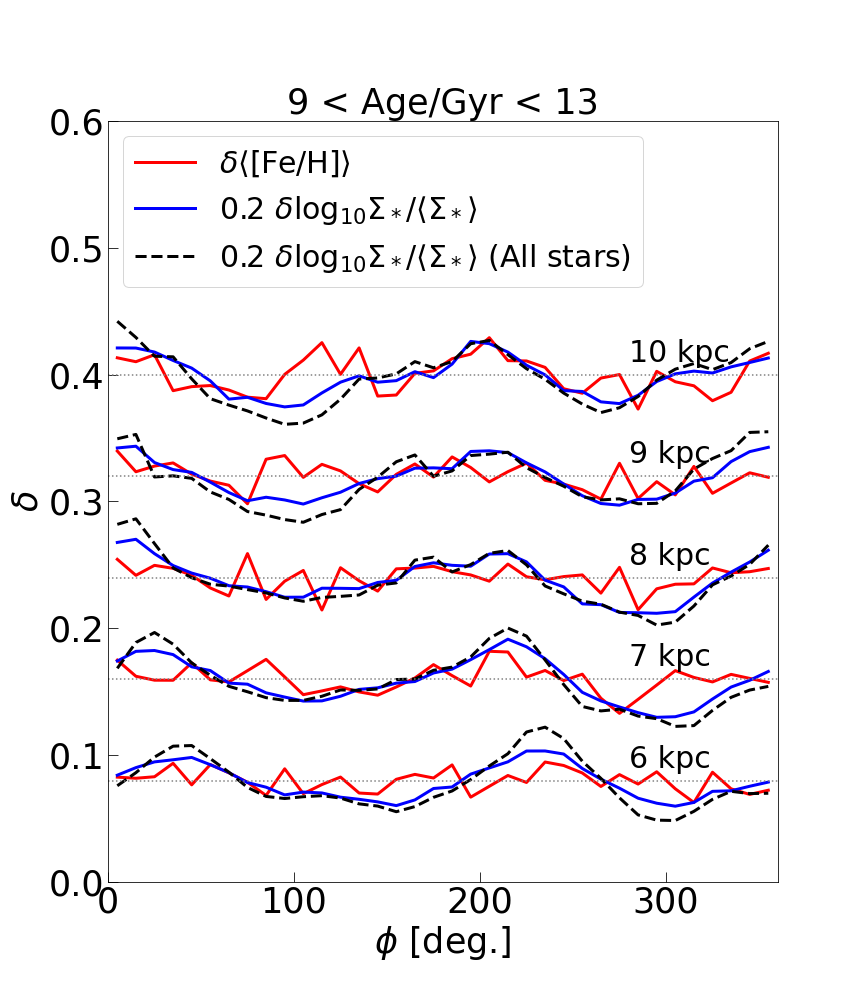}\
}
\caption{The variation of \avfe\ 
compared with the density variation, split by age for the same populations as in Fig.~\ref{f:azimbyage}, at a series of $500~\pc$-wide annuli centred at $6\kpc$ to $10\kpc$, as indicated. In each panel the solid lines refer to \avfe\ (red) and population density (blue) while the dashed solid lines refer to the density of all stars. The average values for each radius are subtracted from each azimuthal profile and the profiles are then vertically offset by a fixed amount to show the variation. The azimuthal profiles of the density have been scaled by the same factor of 0.2 for ease of comparison. The dotted horizontal lines indicate the zero for each radius.
\label{f:azimprofbyage}}
\end{figure*}

In Fig.~\ref{f:azimprofbyage} we compare the azimuthal variations of
\avfe\ and the density for each of the age populations and with the
full density distribution. The density variations in each population
(solid blue lines) track those of the full distribution (dashed black
lines); the comparison is noisy in the youngest population (there is a
factor of $100$ more stars in the oldest population than in the
youngest) but seems to hold for this population too. While the
amplitude of the density variations become smaller with age, we find
no significant evidence of a phase shift between the azimuthal
variation of the density in populations of different ages. In general
the peak \avfe\ also seems to trace the peak density, particularly at
large radii, but the density minima are not always traced by prominent
minima in \avfe.

From the observations that the \avfe\ peaks are radially extended but coincident with the density peaks, we conclude that the cause of
the \avfe\ variations is unlikely to be a resonant phenomenon, since azimuthal variations
due to stars librating about a
resonance might be expected to have \feh\ peaks azimuthally displaced relative to
the perturbation \citep[\eg][]{grand+16b, khoperskov+18b}.

\subsection{The azimuthal variation of \avg{\ofe}}
\label{ss:ofe}

The simulation also tracks the evolution of \ofe; the bottom left panel of
Fig.~\ref{f:azimuthal} shows the azimuthal variation of
\avg{\ofe}\ at $13 \Gyr$. We find very similar behaviour as for \avfe,
but since \feh\ and \ofe\ are anti-correlated, the minima in
\avg{\ofe}\ are at the density peaks.

The right panel of Fig.~\ref{f:razimuthal} compares the azimuthal
variations of \avg{\ofe}\ with those of the density (with a minus sign
on \avg{\ofe}\ so maxima and minima are aligned). The peak to trough
variation is of order $0.03$ dex at small radii, and declining at
larger radii. Overall a match in location of the peaks and troughs is
present but the correspondence is somewhat less exact than between
\avfe\ and $\delta\Sigma/\avg{\Sigma}$.


\section{Pattern speeds}
\label{s:omega}

Having shown the close correspondence between spirals and \avfe\ variations at one time, we explore their co-evolution across time by treating the \avfe\ variations as waves, and comparing their pattern speeds with those of the spirals.
We use the spectral analysis code of \citet{khachaturyants+22b}, which
uses binned data to measure pattern speeds. We bin the mass and
\feh\ data in polar coordinates with $R \in [0, 12] \kpc$ and $\Delta
R = 0.5 \kpc$, and $\varphi \in [0,360]\degrees$ and $\Delta\varphi =
10\degrees$. Thus our binning has $N_R \times N_\phi = 24 \times 36$
bins. 
In order to ensure that our pattern speeds for \feh\ are not produced
by young stars recently born in the spiral arms, which would trivially
result in the same pattern speed as the spirals, we exclude all stars
younger than $2\Gyr$ from all the pattern speed analysis.
We compute the stellar mass and \avfe\ in the binned
space and produce a 2D polar array at each snapshot. The resulting 2D
arrays are $M(R_i,\phi_j,t_k)$ and $f_e(R_i,\phi_j,t_k)$, where $R_i$
and $\phi_j$ are the coordinates of the centres of each bin in the
array, and $t_k$ is the time of the snapshot, which are spaced
at intervals of $\Delta t = 5\Myr$.  Following the notation of
\citet{khachaturyants+22} and \citet{khachaturyants+22b}, we define
the density Fourier coefficients as:
\begin{equation}
  c_m(R_i,t_k) = \frac{1}{\sum_j M(R_i,\phi_j,t_k)} \sum_j \left[ M(R_i,\phi_j,t_k) e^{i m \phi_j} \right],
\end{equation}
and similarly for \avfe:
\begin{equation}
  \gamma_m(R_i,t_k) = N_\phi^{-1} \sum_j \left[ f_e(R_i,\phi_j,t_k) e^{i m \phi_j} \right],
\end{equation}
where $m=1$ to $4$ is the azimuthal multiplicity of the Fourier
term. We then use a discrete Fourier transform
\begin{equation}
C_{m,n}(R_i) = \sum_{k=0}^{S-1} c_m(R_i,t_k) w_k e^{-2\pi i k n/S},
\end{equation}
with $n = 0, ..., S-1$, where $S$ is the number of snapshots in each
$1 \Gyr$ baseline (we use $S=200$), and the $w_k$ are the weights of a window
function, which we set to Gaussian:
\begin{equation}
w_k = e^{-(k-S/2)^2/(S/4)^2}.
\end{equation}
The frequency associated with each $n$ is given by $\Omega_n = 2 \pi
n/(m S \Delta t)$. Thus given we are using time baselines of $1\Gyr
\equiv S \Delta t$, the frequency resolution of our analysis is
$\Delta \Omega = 0.98\times 2 \pi/m \kmsk$ (where the factor $0.98$ is
the conversion factor to \kmsk).  The power spectrum, defined as the
power in each radial bin as a function of frequency, is given by
\begin{equation}
P_m(R_i,\Omega_n) = W^{-1} \left[ \left| C_{m,n}(R_i) \right|^2 + \left| C_{m,S-n}(R_i) \right|^2 \right],
\end{equation}
where $W = S\sum_{k=0}^{S-1} w_k^2$. 

In order to compute the total (radially integrated) power spectrum for a given $m$,
from which we recover the most important pattern speeds, we compute
the sum of the mass-squared-weighted power in each radial bin:
\begin{equation}
P_R(\Omega_n) = \frac{\sum_i M^2(R_i) P_m(R_i,\Omega_n)}{\sum_i M^2(R_i)}
\end{equation}
(where we ignore the multiplicity index $m$ on the notation since this will be clear). This mass weighting ensures that the power better reflects the overall
importance of a wave, rather than having large relative amplitude
perturbations in low density regions dominating the signal.  As we did
in \citet{khachaturyants+22b}, we focus on the positive pattern speeds
and use the method described in \citet{roskar+12} to obtain the
pattern speeds by fitting Gaussians to 
$P_R(\Omega_n)$, and subtracting successive peaks.

\begin{figure*}
\centerline{
\includegraphics[angle=0.,width=\hsize]{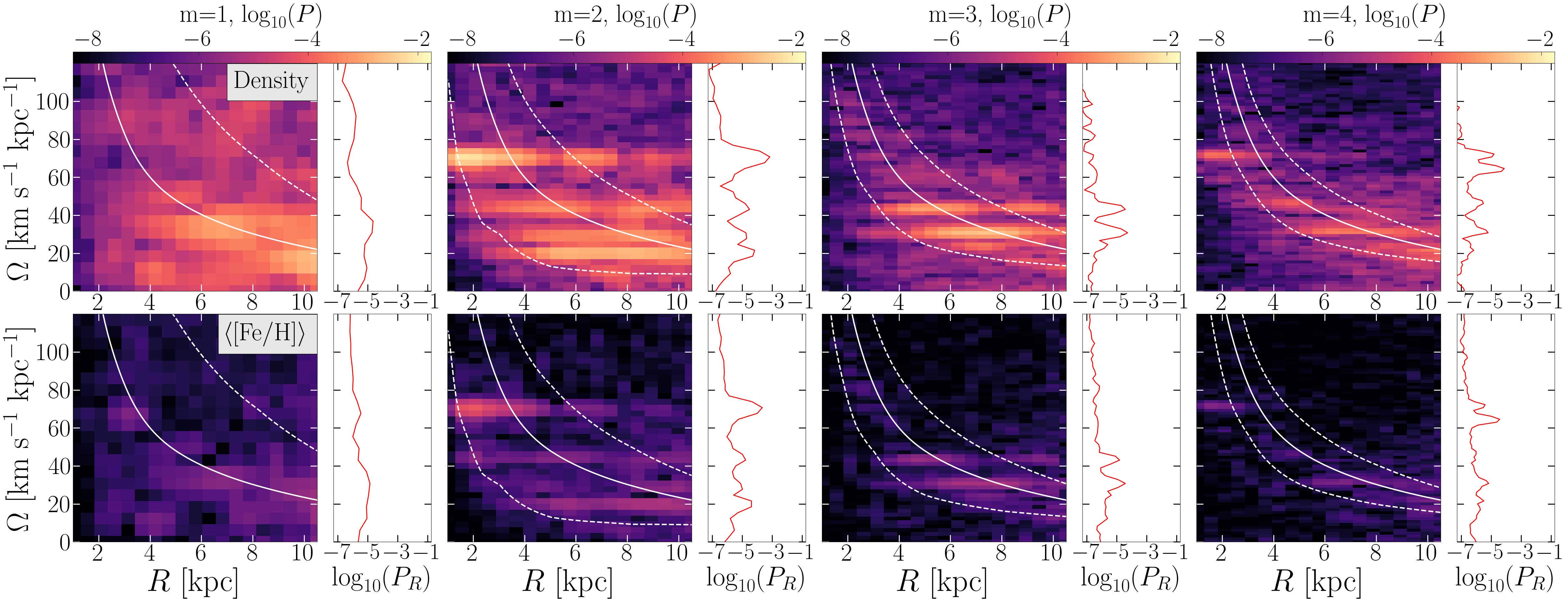}\
}
\caption{Power spectra (spectrograms) for density (top) and
  \avfe\ (bottom) waves of different multiplicities (columns) at
  $t=12-13\Gyr$. The solid lines show the rotation curve, $\Omega(R)$,
  evaluated at the central time of the interval using an interpolated
  potential. Dashed lines indicate $\Omega\pm\kappa/m$. To the right
  of each spectrogram we show the radius-integrated power. The colour
  scale limits are the same for each wave type over all multiplicities
  and models. The pattern speeds of $m>1$ density and \avfe\ waves
  match each other. Only stars older than $2\Gyr$ are considered.
\label{f:mspectrograms}}
\end{figure*}

We adopt a similar analysis for the \avfe\ pattern speeds, replacing
$c_m(R_i,t_k)$ by $\gamma_m(R_i,t_k)$.
The spectral analysis of the density (\ie\ the measurement of the pattern
speeds of the bar and spirals) has already been presented by
\citet{khachaturyants+22} (see their figures 6 and 7). They found that
the density Fourier expansion is dominated by $m=2$, with $m=1$
significantly weaker. In Fig.~\ref{f:mspectrograms} we plot
spectrograms of the density (top row) and \avfe\ (bottom row) distributions
for the time interval $t=12-13\Gyr$, for $m=1$ to $m=4$.  Each block
contains a spectrogram (left) and the radially integrated total power,
$P_R(\Omega)$ (right) of a given wave.  For $m\geq2$, the spectrograms
and total power spectra of the density and \avfe\ waves are in good
agreement with each other, with the strongest power in $m=2$ (note the
different scales of $P_R$ for different $m$ in
Fig.~\ref{f:mspectrograms}).  We therefore restrict our attention to
$m=2$ from here on.

\begin{figure}
\centerline{
\includegraphics[angle=0.,width=\hsize]{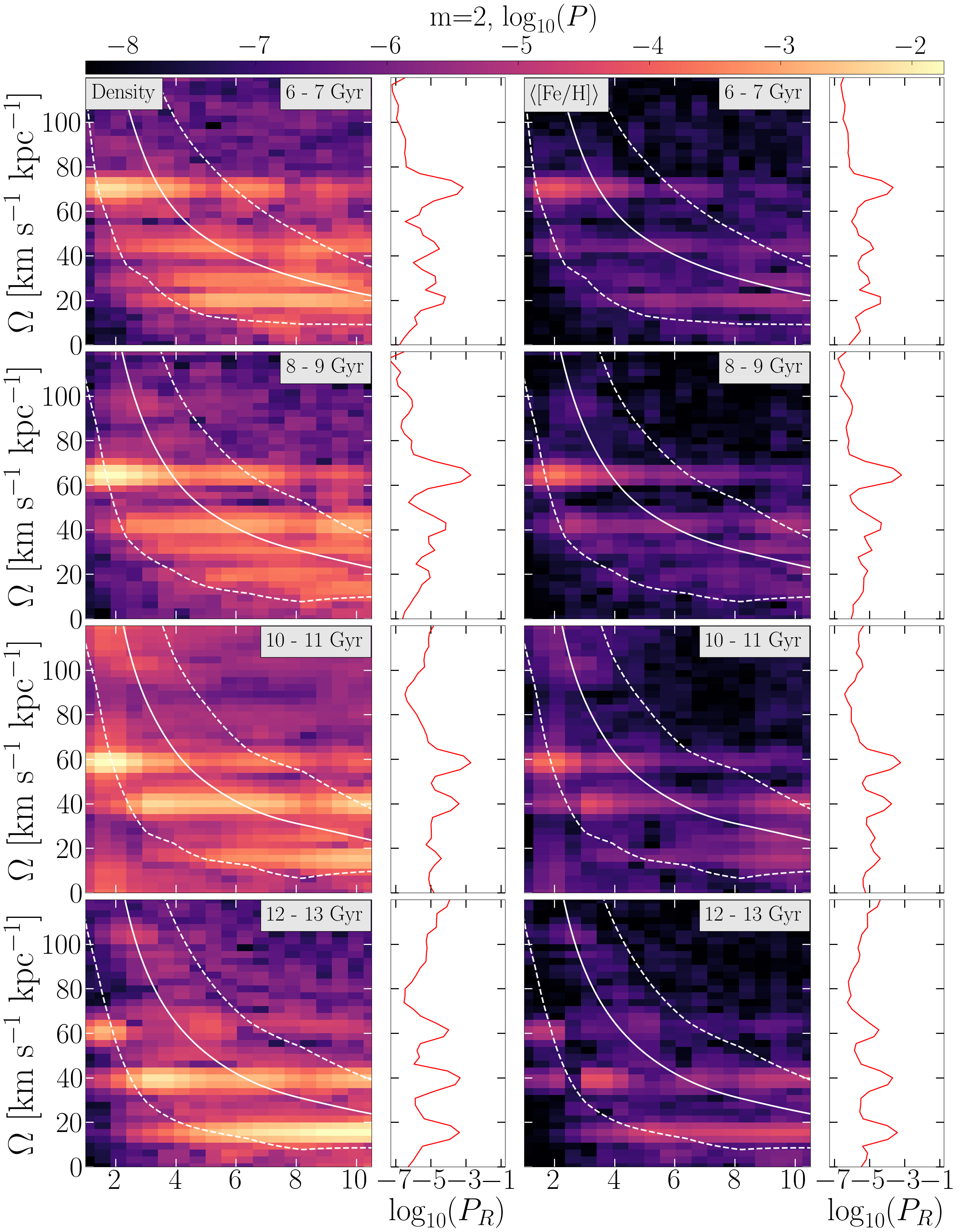}\
}
\caption{Similar to Fig.~\ref{f:mspectrograms}, but for the power
  spectra of density (left) and \avfe\ (right) $m=2$ waves at the
  different time intervals (rows) indicated at the top right of each
  panel. Only stars older than $2\Gyr$ are considered.
\label{f:tspectrograms}}
\end{figure}

Fig.~\ref{f:tspectrograms} therefore shows spectrograms of $m=2$ for
the density (left column) and \avfe\ (right column) at different times in the
evolution of the model. A clear correspondence between the
pattern speeds in the density and in \avfe\ is evident at each
time. When a peak in the power is present in the density we generally
find a corresponding peak for \avfe. At the earlier times the disc
hosts a larger number of pattern speeds; the number declines to about
3 pattern speeds at later times. The growth and slowdown of the bar from $6\Gyr$ to
$\sim 9 \Gyr$ and its subsequent weakening by $12\Gyr$ are also
evident in the strong feature at $60-80\kmsk$.

\begin{figure*}
\centerline{
\includegraphics[angle=0.,width=\hsize]{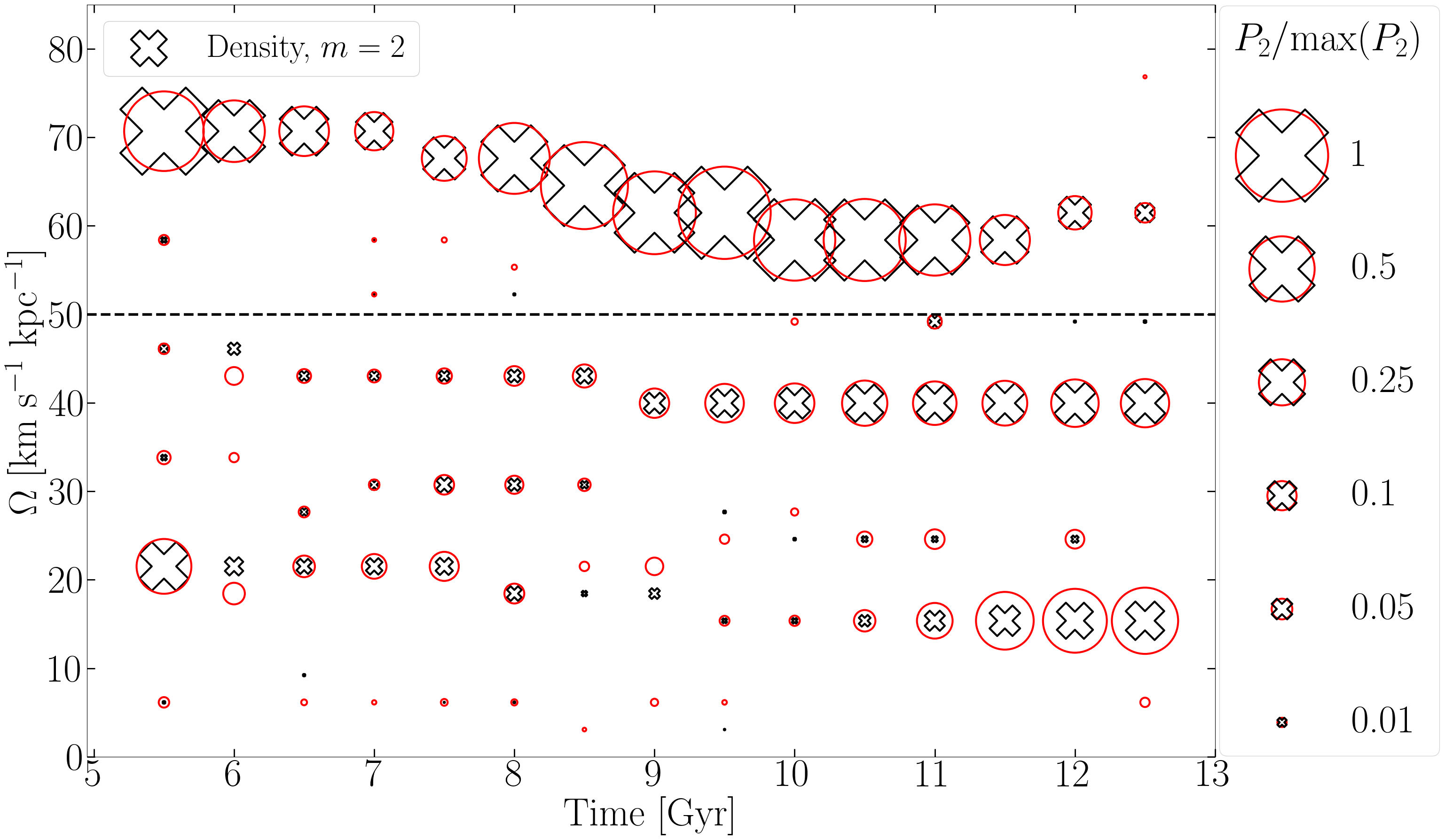}\
}
\caption{ The most prominent pattern speeds for the $m=2$ density
  (black crosses) and \avfe\ (red circles) waves. The marker size is
  indicative of the radius-integrated power at a given pattern
  speed. The peaks are scaled by the highest peak in the particular
  type of wave (density or \avfe) over all baselines,
  $P_2/\rm{max}(P_2)$ with the legend showing the relationship between
  the marker size and $P_{2}/\rm{max}(P_{2})$. Only pattern speeds
  with $P_{2}/\rm{max}(P_{2})\geq0.01$ are shown. We observe a
  correspondence between the $m=2$ density and \avfe\ perturbations:
  for the most part, the pattern speeds of the two match, except for
  the weakest patterns. The horizontal dashed line shows $\Omega =
  49\kmsk$; the region above this frequency is dominated by the
  bar. Each marker is shown at the mid-point of the corresponding
  Fourier time series; since these are $1\Gyr$ long, consecutive
  points are correlated (data points in the second half of each time
  series are the first half of the next time series).  Only stars
  older than $2\Gyr$ are considered.
\label{f:bubbleplot}}
\end{figure*}

The temporal evolution of the $m=2$ pattern speeds in the density
(black crosses) and \avfe\ (red circles) can be seen in
Fig.~\ref{f:bubbleplot}. The most prominent density perturbations,
including the bar, but also the slower
spirals, have pattern speeds that are well matched by pattern speeds
in the metallicity. Conversely, in almost all cases, the pattern
speeds identified in \avfe\ are matched by pattern speeds in the
density.  The size of the markers represents the power normalised by
the maximum power value each pattern reaches over the entire time
interval of this analysis, \ie\ $5-13~\Gyr$. The density and \avfe\ patterns fail to match only for the
weakest waves (smallest markers).

\begin{figure}
\centerline{
\includegraphics[angle=0.,width=\hsize]{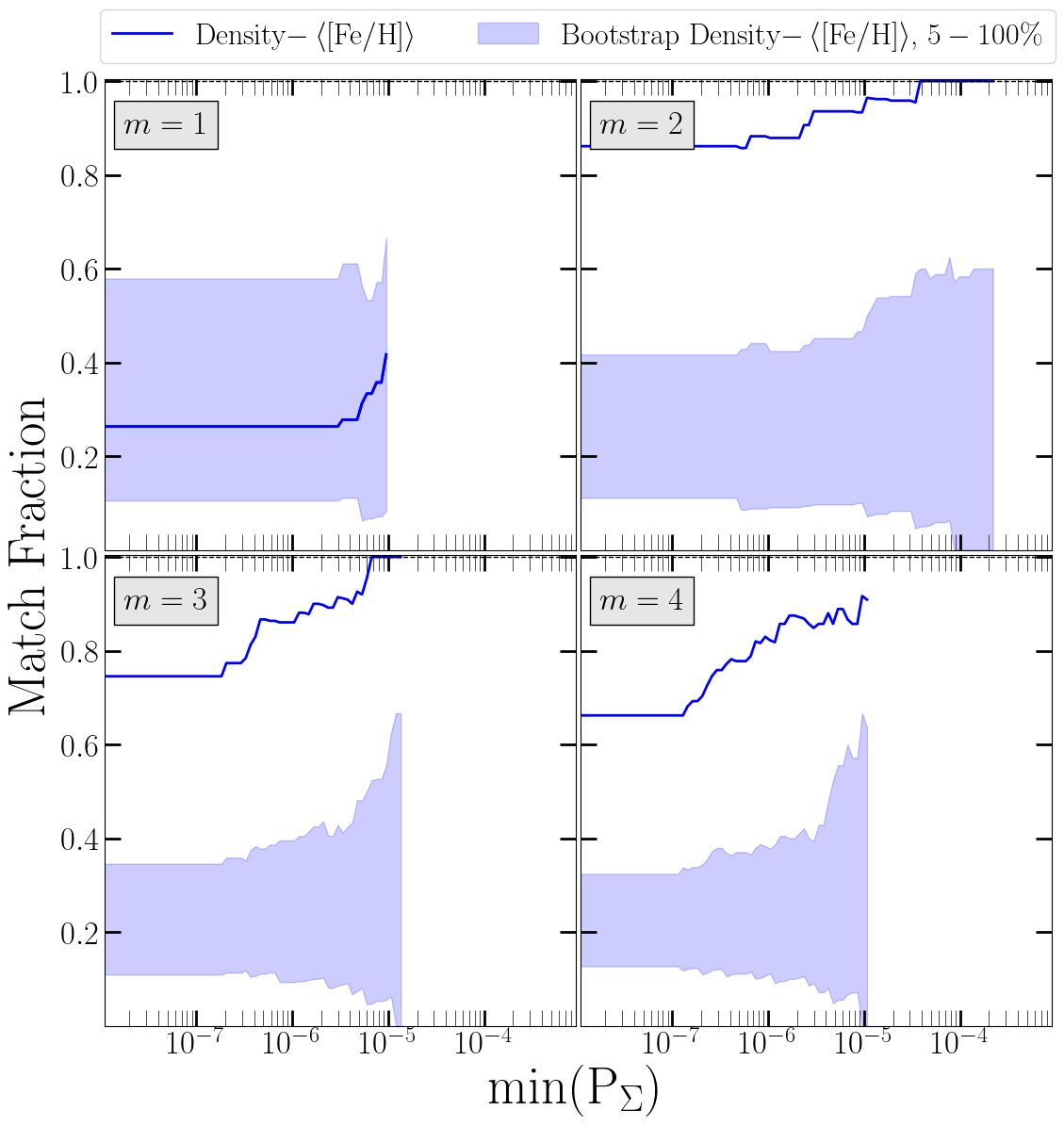}\
}
\caption{ Match fraction between the density and \avfe\ waves, given by Eqn.~\ref{e:matchfraction}, versus
  the minimum power of the density waves, $\min(P_\Sigma)$. The shaded region indicates the $5\%$ to $100\%$
  distribution of the match fractions resulting from a bootstrap resampling
  of random pattern speeds of \avfe\ waves. All random bootstrap samples have match fraction lower than in the simulation for $m>1$, implying the matches are not random.
  For this analysis we only consider pattern speeds $\Omega \leq
  49\kmsk$ (the dashed horizontal line in Fig.~\ref{f:bubbleplot}), to
  exclude the bar. We do not show $\min(P_\Sigma)$ with less than 10
  identified pattern speeds across all times. For $m>1$, we note an
  increase in match fraction with increasing $\min(P_\Sigma)$ that
  substantially exceeds the random bootstrap-sampled match
  fractions. Only stars older than $2\Gyr$ are considered in this
  analysis.
\label{f:matches}}
\end{figure}

\citet{khachaturyants+22b} showed that there is a strong connection
between spirals and breathing waves by showing that the pattern speeds
of strong breathing waves are always matched by pattern speeds of
density waves, and vice versa. They argued therefore that the
breathing motions must be part of the nature of spirals.
As in \citet{khachaturyants+22b}, we quantify the significance of the
match between the density and the \avfe\ waves in Fig.~\ref{f:matches}
by showing the fraction of identified density pattern speeds that are
matched by pattern speeds in \avfe\ (solid lines). The match fraction is defined
as the ratio of the number of density pattern speeds with metallicity matches to the
total number of density pattern speeds, \ie:
\begin{equation}
f = \frac{N(\Omega_{\rm{dens+met}})}{N(\Omega_{\rm{dens}})},
\label{e:matchfraction}
\end{equation}
where $N(\Omega_{\rm{dens}})$ ($N(\Omega_{\rm{dens+met}})$) is the number of density pattern speeds above the threshold (the number of density pattern speeds above the threshold that are matched by a metallicity pattern speed) for the full time interval $5-13~\Gyr$ (not just a single time).
In order to avoid including the bar, we consider only $\Omega_p < 49\kmsk$ for each multiplicity and
baseline.
This fraction
is shown for different minimum power thresholds of the density pattern
speeds, $\min(P_{\Sigma})$. The different panels show the matches for
different wave multiplicities.  Better matches between the density
and \avfe\ waves occur for $m>1$. We
check if these match fractions are compatible with random matches via a bootstrap algorithm where we
select 2500 iterations of random pattern speeds, with a time distribution exactly matching those of the \avfe\ waves, and compute the match fraction with the density pattern
speeds as before. Fig.~\ref{f:matches}
presents the $5-100\%$ probability intervals (shaded
regions) of the match fraction for these random samples. Only in
the case of the $m=1$ waves are the match fractions compatible with
the random matches, while $m>1$ match fractions are clearly higher than random
at all $\min(P_{\Sigma})$.
Since we have binned the \avfe\ field, and computed the Fourier time
series only of the binned values, the metallicity time series is not
contaminated by the density. As a result, the pattern speeds of the two would
be independent of each other if the two were unrelated. The high match
probability we find therefore indicates that the \avfe\ variations are
due to the spirals.  Thus we conclude that the \avfe\ waves are
driven by density waves and, in the disc, we must identify them with
spirals.  It remains to be understood how spirals cause the \avfe\ variations.


\section{Dissection by age}
\label{s:age}

\begin{figure}
\centerline{
\includegraphics[angle=0.,width=\hsize]{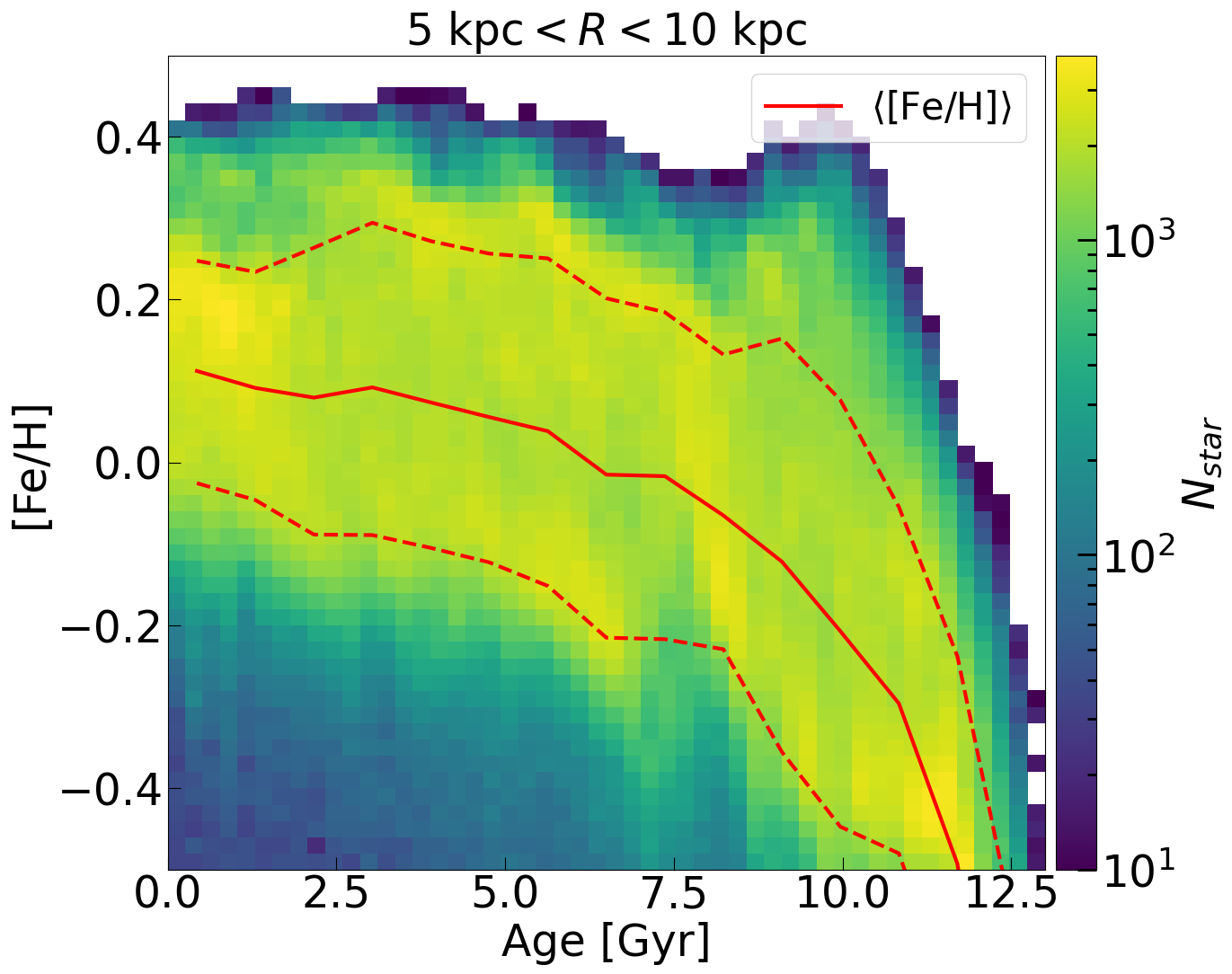}\
}
\caption{The age-metallicity relation for star particles at $5 \leq
  R/\kpc \leq 10$ at $13~\Gyr$. The solid line shows \avfe\ while the
  dashed lines show the $16\%$ and $84\%$ distributions.
\label{f:amr}}
\end{figure}

Having shown that the \avfe\ variations are part of the spirals, and do not result from recent star formation, we next explore the physical mechanism of the \avfe\ variations. 

First we consider whether \avfe\ variations reflect variations in the mean age.
Fig.~\ref{f:amr} shows the age-metallicity relation (AMR) of stars
between $5\kpc$ and $10\kpc$. A quite shallow relation is evident for
stars younger than $\sim 8\Gyr$, getting steeper for older stars. The
scatter in \feh\ increases with age. We have also checked that the AMR
of different subsections of this radial range look reasonably
similar. The AMR of the model for stars younger than $\sim 5\Gyr$ is
shallow but not perfectly flat. The AMR in the Solar Neighbourhood is
perhaps similar \citep{feuillet+19, anders+23, johnson+24}, although
some studies find a flatter AMR \citep{mackereth+17, delgadomena+19}.
The existence of an AMR implies that, if populations of different age
have spirals of different amplitude, then azimuthal variations in
\avfe\ would arise from the azimuthal variation of the spiral density
\citep[\eg][]{khoperskov+18b}.
\begin{figure}
\centerline{
\includegraphics[angle=0.,width=\hsize]{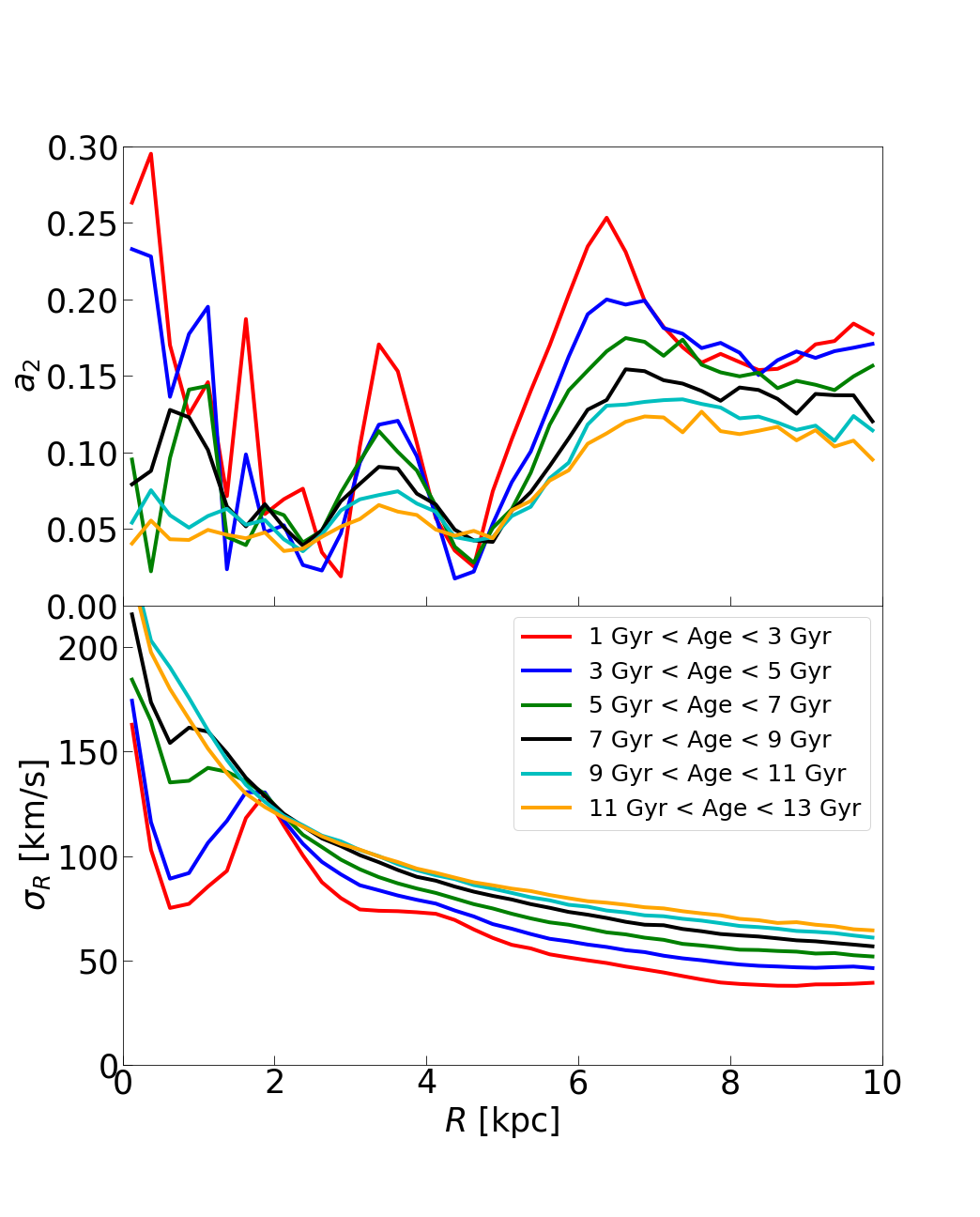}\
}
\caption{Top: The radial profiles of the amplitude of the $m=2$
  density Fourier moments for different age populations, as indicated in the legend. Outside
  $5\kpc$, spiral structure is weaker in old populations than in
  younger ones. Bottom: The radial velocity dispersion, \sig{R},
  profiles of the same stellar populations. Stars younger than $1\Gyr$
  are not plotted.
\label{f:agedispspiralamp}}
\end{figure}
The top panel of Fig.~\ref{f:agedispspiralamp} plots the radial
profiles of the $m=2$ amplitude of the density; it shows that spirals at $R>5\kpc$
are present in all age populations, although they are weaker in older
populations, as seen already in Fig.~\ref{f:azimbyage} \citep[see also][]{ghosh+22}. The bottom panel shows that the older populations support
weaker spirals because their radial velocity dispersion, \sig{R}, is
higher.

\begin{figure}
\centerline{
\includegraphics[angle=0.,width=\hsize]{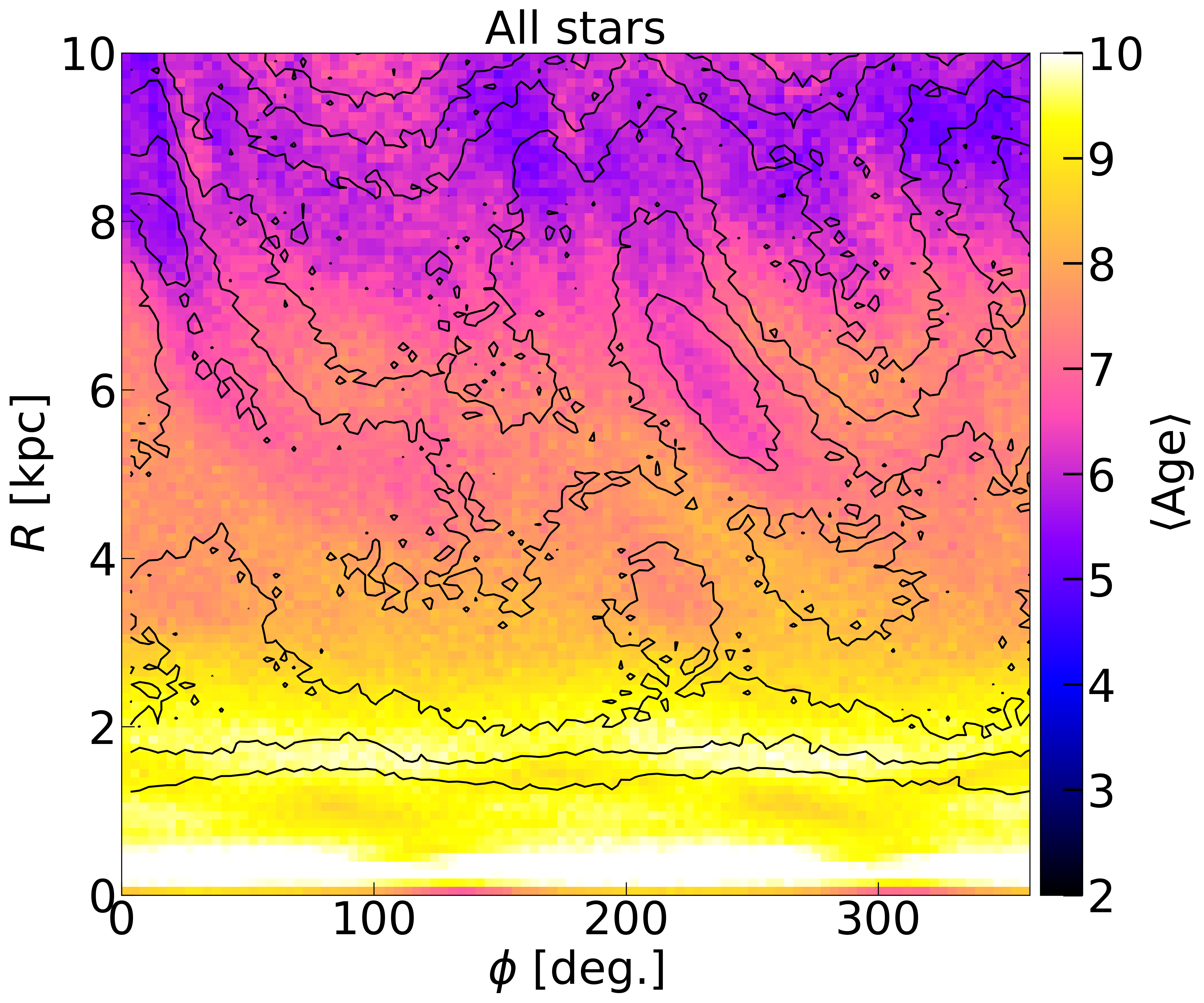}\
}
\centerline{
\includegraphics[angle=0.,width=\hsize]{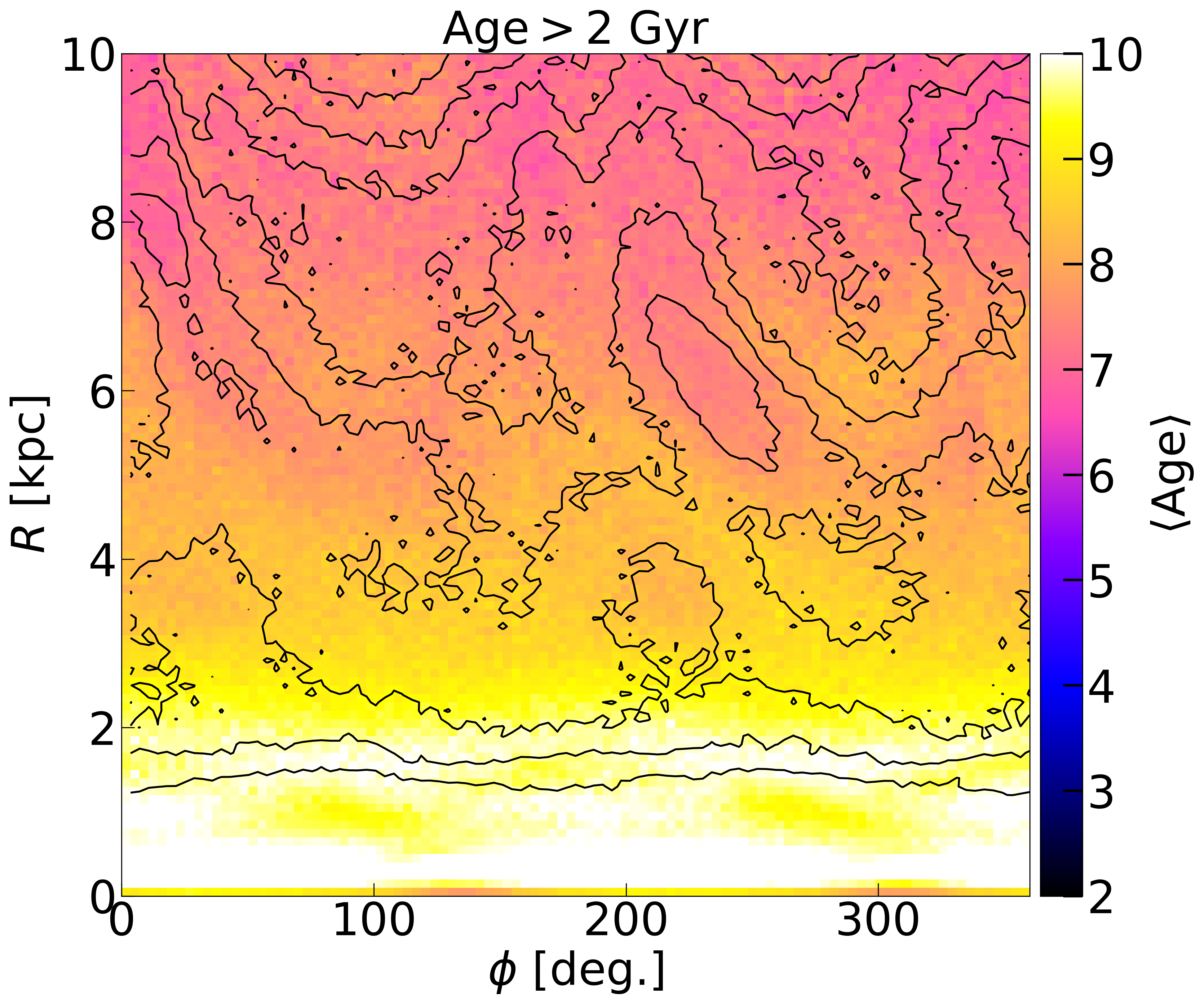}\
}
\caption{Mean age, \avag, in cylindrical coordinates. Top: for all
  stars. Bottom: for stars older than $2~\Gyr$. The contours
  correspond to the surface density. The disc is rotating in a
  counter-clockwise sense, \ie\ rotation is in the direction of
  increasing $\phi$. The model is shown at $13~\Gyr$.
\label{f:meanagemap}}
\end{figure}

Fig.~\ref{f:meanagemap} shows maps of the mean age, \avag. The top panel, which includes all stars, shows age variations, which have similar spatial distribution as the
metallicity variations of Fig.~\ref{f:azimuthal}. The
bottom panel shows \avag\ excluding stars younger than
$2~\Gyr$. The
variations in \avag\ retain a similar character to those of the full
population, showing again that recent star formation is not the cause
of the azimuthal variations. As expected, the spiral ridges are
regions of younger stars, which follows because these stars have lower
radial velocity dispersions and thus host stronger spirals
(Fig.~\ref{f:agedispspiralamp}).

\begin{figure}
\includegraphics[angle=0.,width=\hsize]{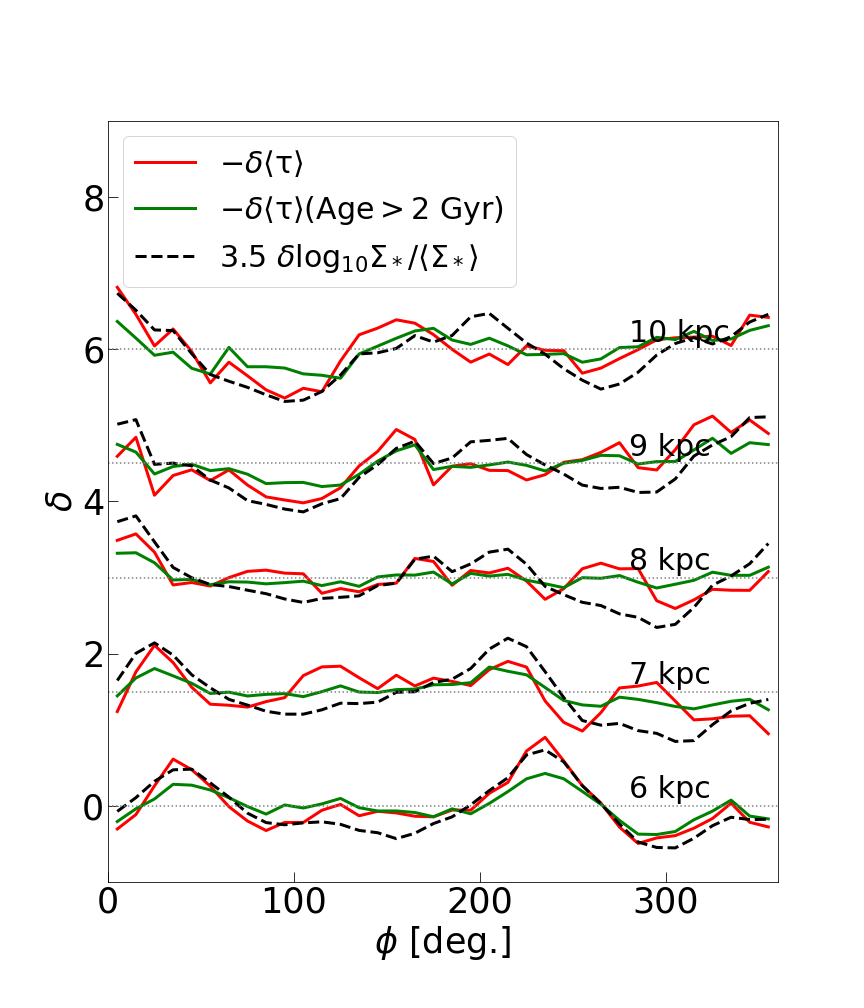}
\caption{The variation of \avag\ compared with the density variation
  at a series of $500~\pc$-wide annuli centred at $6\kpc$ to $10\kpc$, as indicated. (We use $-\avag$ to permit easier comparison with the density.) The average value at each radius is subtracted from each
  azimuthal profile and the profiles are then vertically offset by a
  fixed amount to show the variation. The azimuthal profiles of the
  density have been scaled by the factor $3.5$ for ease of
  comparison. The dotted horizontal lines indicate the zero for each
  radius. The red (green) lines show \avag\ profiles of all stars
  (stars older than $2~\Gyr$).
\label{f:tfrazimuthal}}
\end{figure}

Fig.~\ref{f:tfrazimuthal} compares the azimuthal variation of \avag\ and the density at different
radii. Age
variations of order $0.5-1\Gyr$ are present at all radii.  The green lines in
Fig.~\ref{f:tfrazimuthal} show the azimuthal profiles of \avag\ when
stars younger than $2\Gyr$ are excluded. The azimuthal variations are
now weaker, but still evident. Moreover these profiles largely track
those of all stars where the deviation from the radial average is
significant. Note that the density peaks correspond to locations where
the stars are, on average, younger.

While there is a reasonably good correspondence between the variation
of the density and mean age, comparison with Fig.~\ref{f:razimuthal}
shows that \avfe\ traces the density variations slightly better than
\avag\ does. This is particularly evident at larger
radii. \avag\ therefore is probably not directly the driver of the
\avfe\ variations. In any case, the variations in \avag\ are too small, given the AMR,
to be directly responsible for the variations in \avfe.


\section{Dependence on the radial action}
\label{s:jr}

We have shown above that older populations with larger \sig{R}\ have
weaker spirals. The radial motions of individual stars are best
quantified by the radial action, \act{R}, which is an integral of
motion when a system evolves adiabatically. We therefore expect to see
strong correlations between the azimuthal variations of \avfe\ and
those of \avg{\act{R}}. To test this idea, we compute the actions of
the model using the axisymmetric St\"ackel fudge of \citet{binney12},
as implemented in the {\sc agama} package \citep{agama}.

\subsection{Instantaneous radial actions}
\label{ss:jr}

\begin{figure}
\centerline{
\includegraphics[angle=0.,width=\hsize]{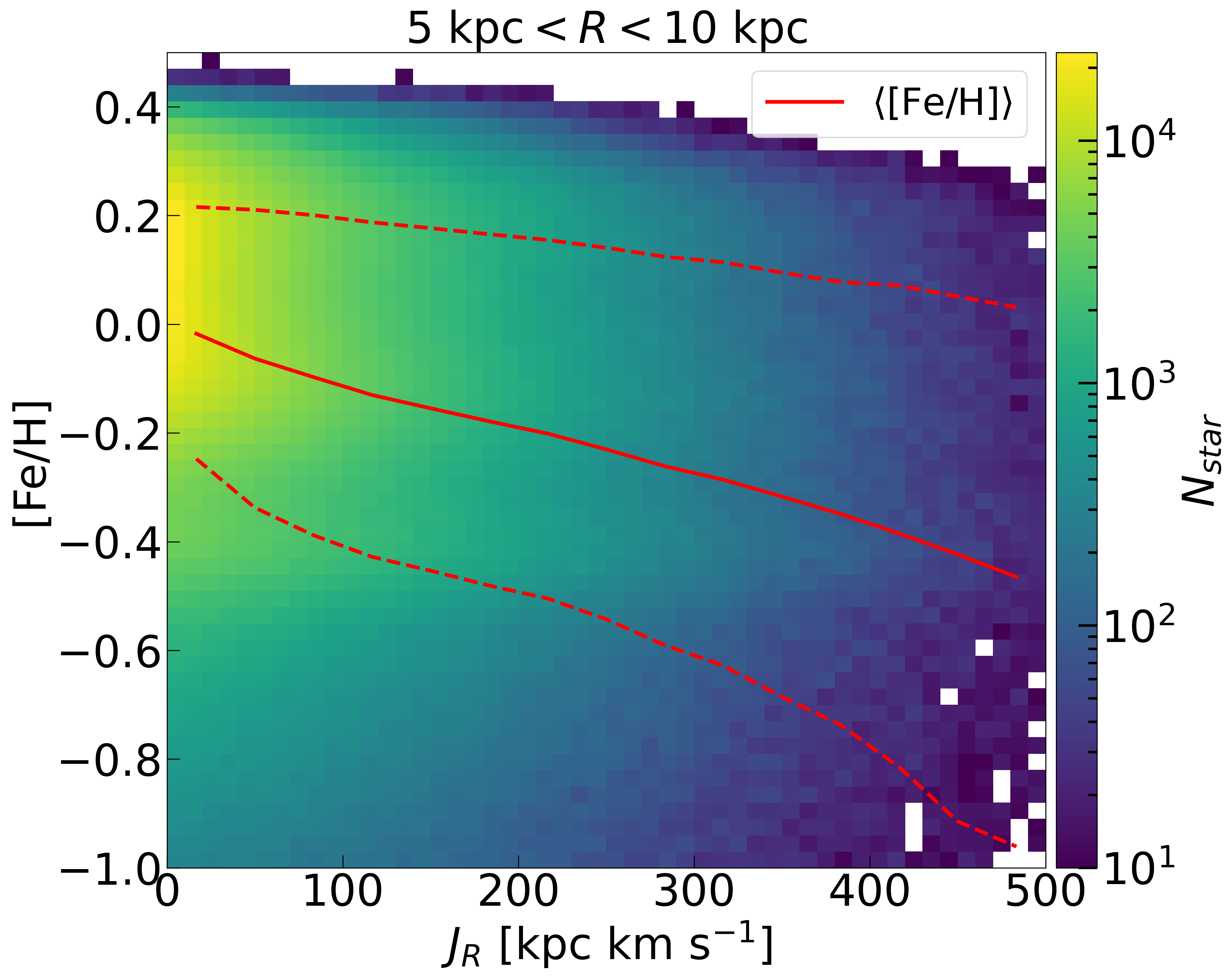}\
}
\centerline{
\includegraphics[angle=0.,width=\hsize]{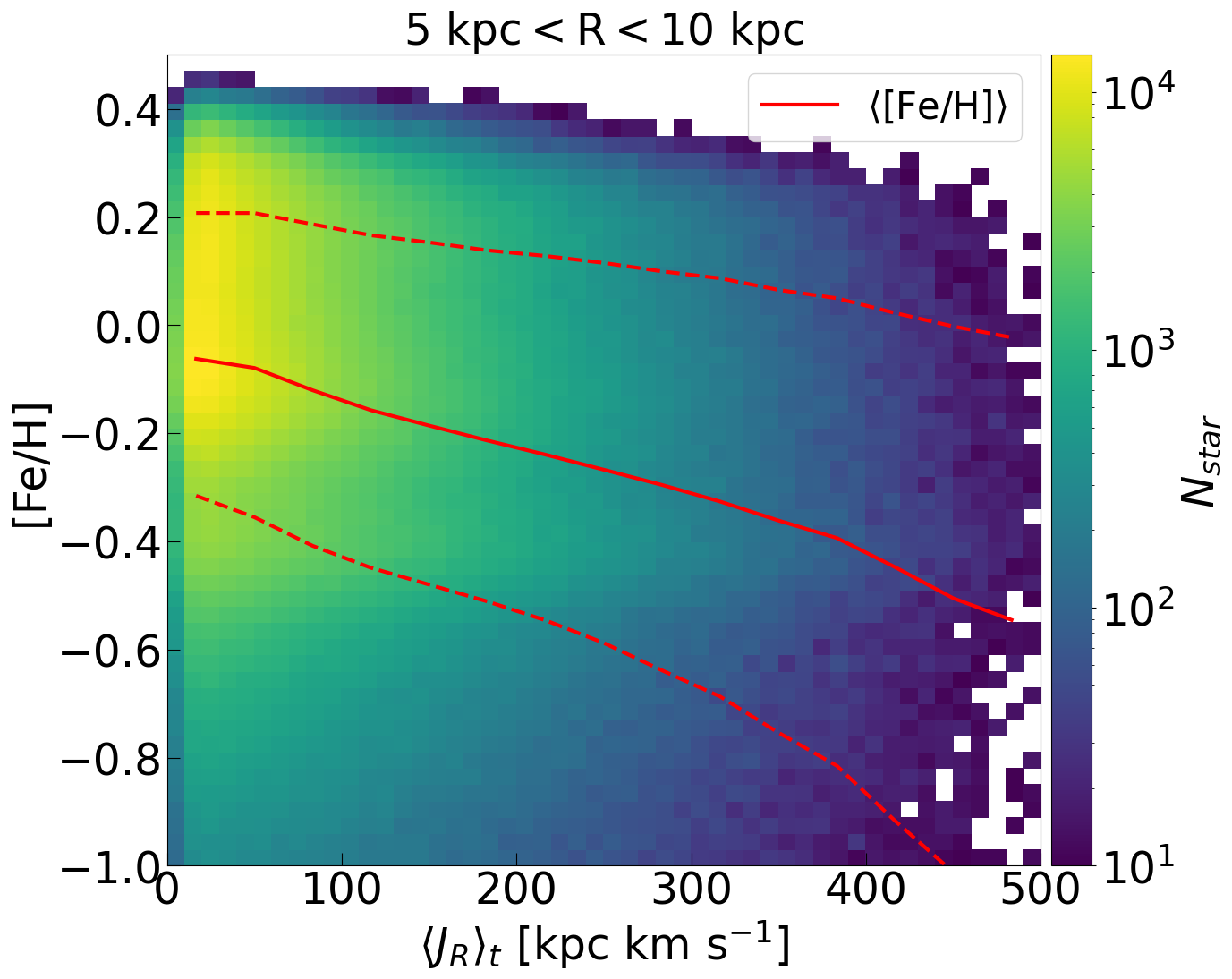}\
}
\centerline{
\includegraphics[angle=0.,width=\hsize]{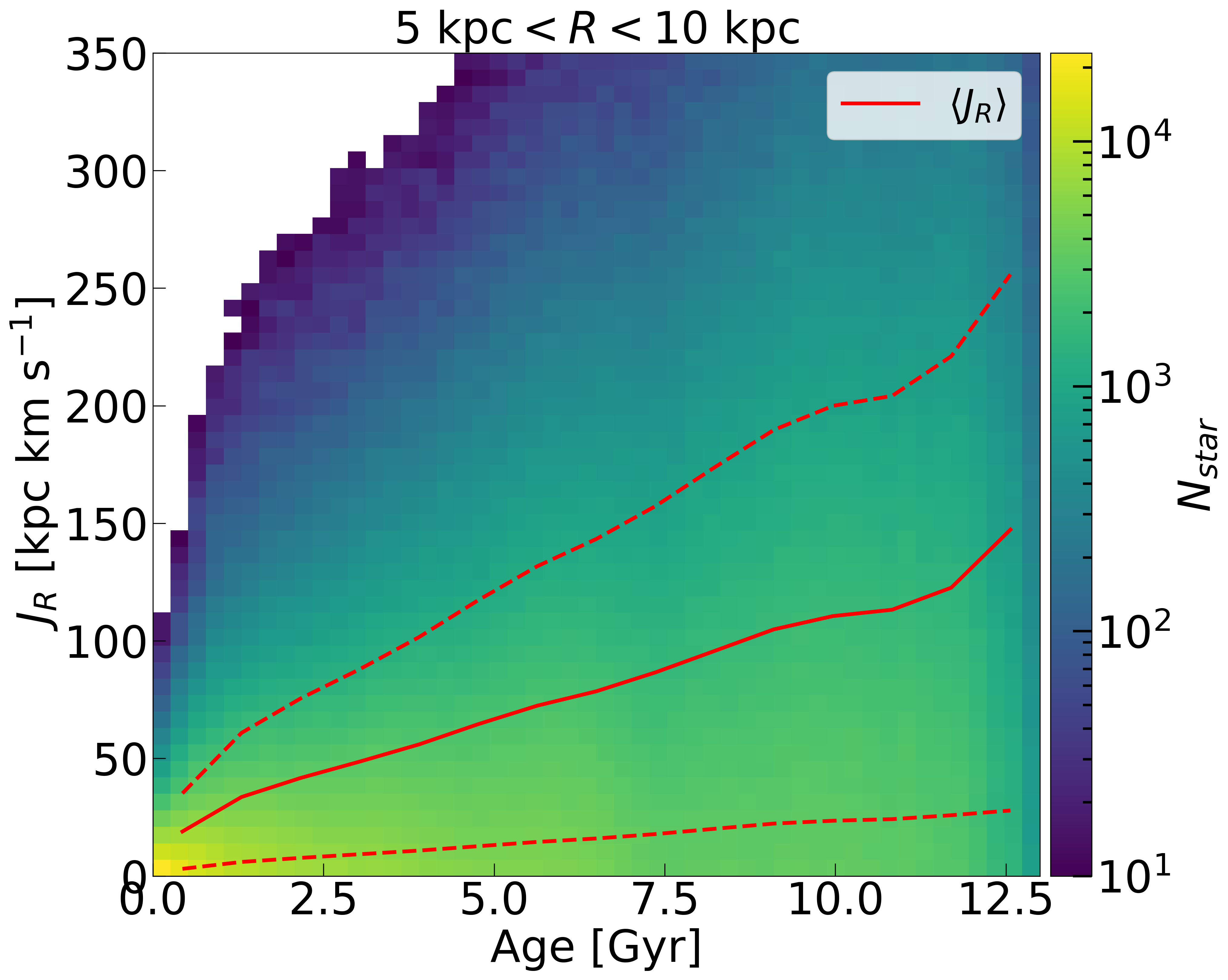}\
}
\caption{The \act{R}-\feh\ (top), \avgt{\act{R}}-\feh\ (middle) and
  \age-\act{R}\ relations (bottom) for star particles at $5 \leq
  R/\kpc \leq 10$ at $13~\Gyr$. The solid lines show \avfe\ (top and
  middle) and \avg{\act{R}} (bottom). In all panels the dashed curves
  show the $16\%$ and $84\%$ distributions. The middle panel excludes
  stars younger than $2\Gyr$, which results in the paucity of low
  \avgt{\act{R}}\ stars.
  \label{f:jrfehage}}
\end{figure}

The top panel of Fig.~\ref{f:jrfehage} shows the
\act{R}-\feh\ relation. Across the range in \act{R}\ the mean
metallicity varies significantly, and nearly linearly, although the
scatter about the mean is large. The bottom panel shows the
\age-\act{R} relation.  The average \act{R}\ grows linearly with age,
comparable to the trend seen in the MW \citep{beane+18,
  gandhi_ness19}, with \avg{\act{R}}\ at $10\Gyr$ roughly double that
measured in the MW. These relations suggest that the driver of the
azimuthal variations of stellar populations may be \act{R}, which both
determines how strong the spiral structure is in a given population,
and what the mean metallicity is.

\begin{figure}
\centerline{
\includegraphics[angle=0.,width=\hsize]{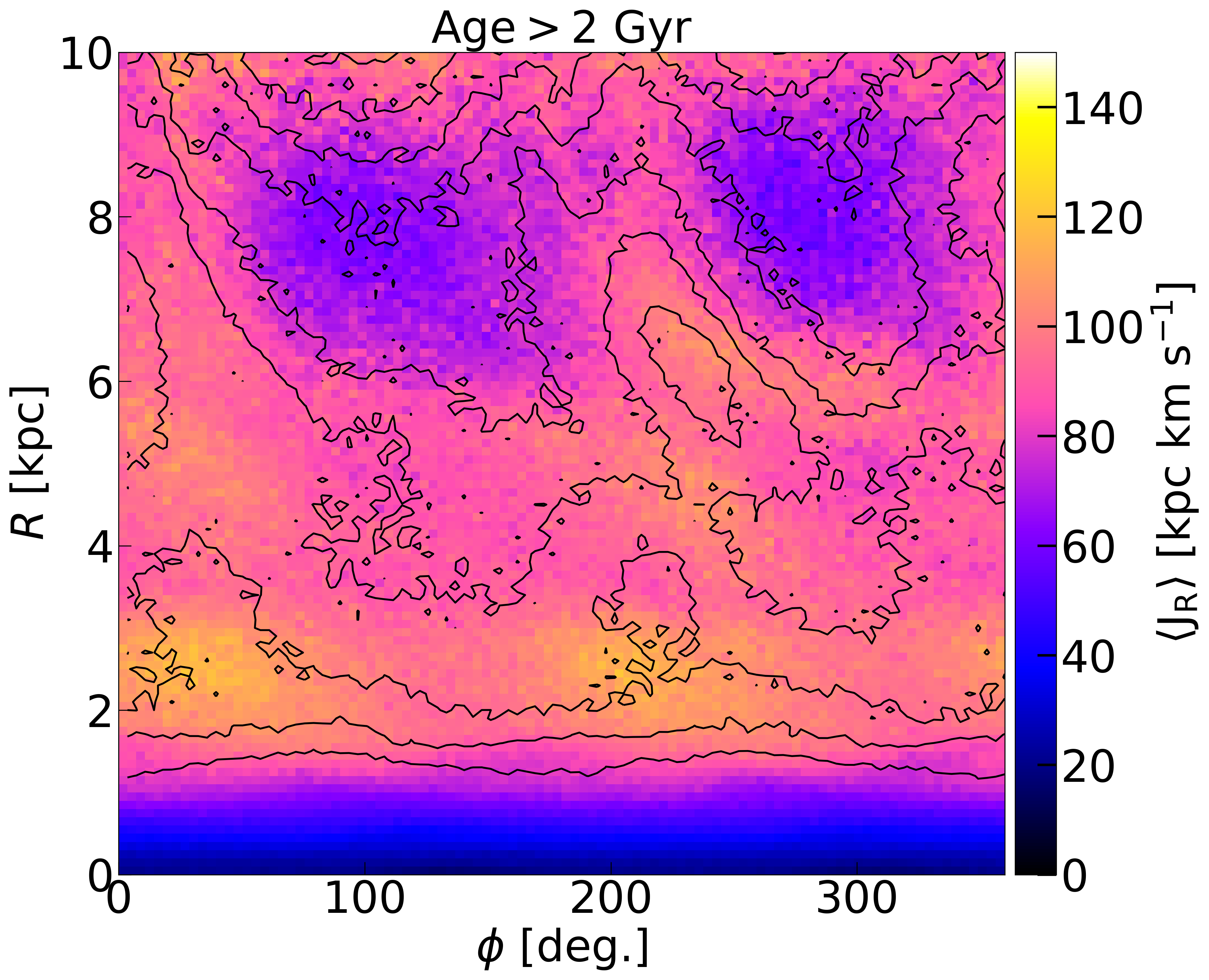}\
}
\centerline{
\includegraphics[angle=0.,width=\hsize]{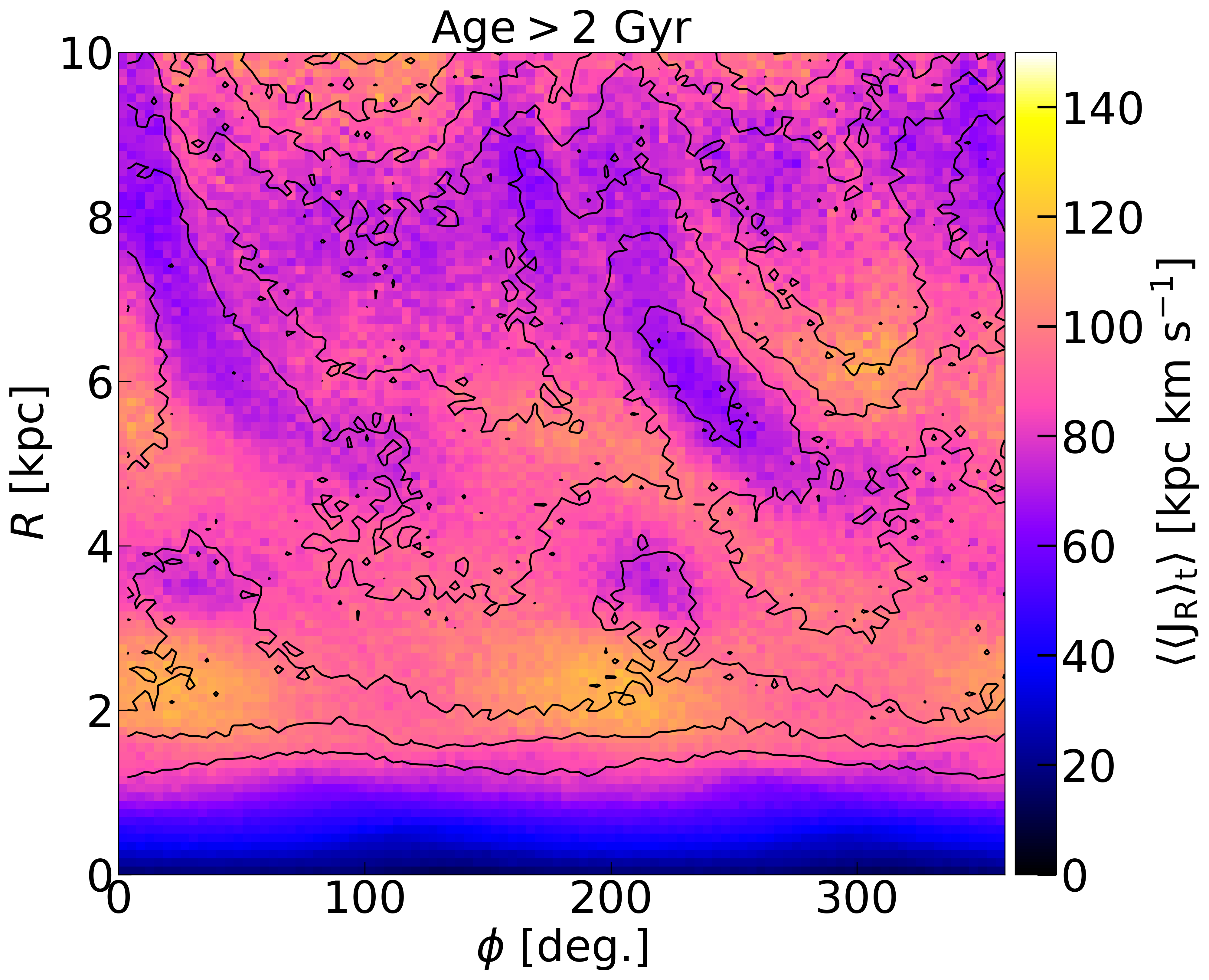}\
}
\caption{The mean instantaneous radial action, \avg{\act{R}}, (top) and the mean
  time-averaged radial action, \avg{\avgt{\act{R}}}, (bottom) in
  cylindrical coordinates. The contours correspond to the surface
  density. Only stars older than $2~\Gyr$ have been considered. 
  The disc is rotating in a counter-clockwise sense, \ie\ rotation is
  in the direction of increasing $\phi$. The model is shown at
  $13~\Gyr$.
\label{f:meanjrajrmaps}}
\end{figure}

\begin{figure*}
\centerline{
  \includegraphics[angle=0.,width=0.5\hsize]{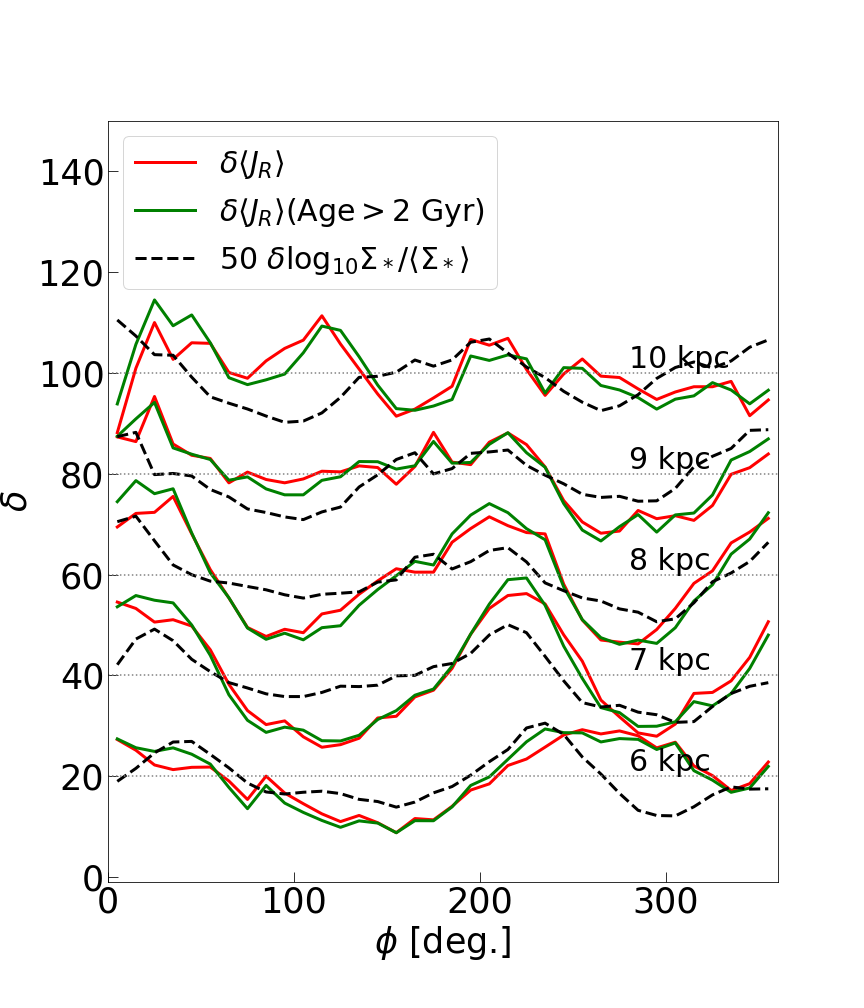}
  \includegraphics[angle=0.,width=0.5\hsize]{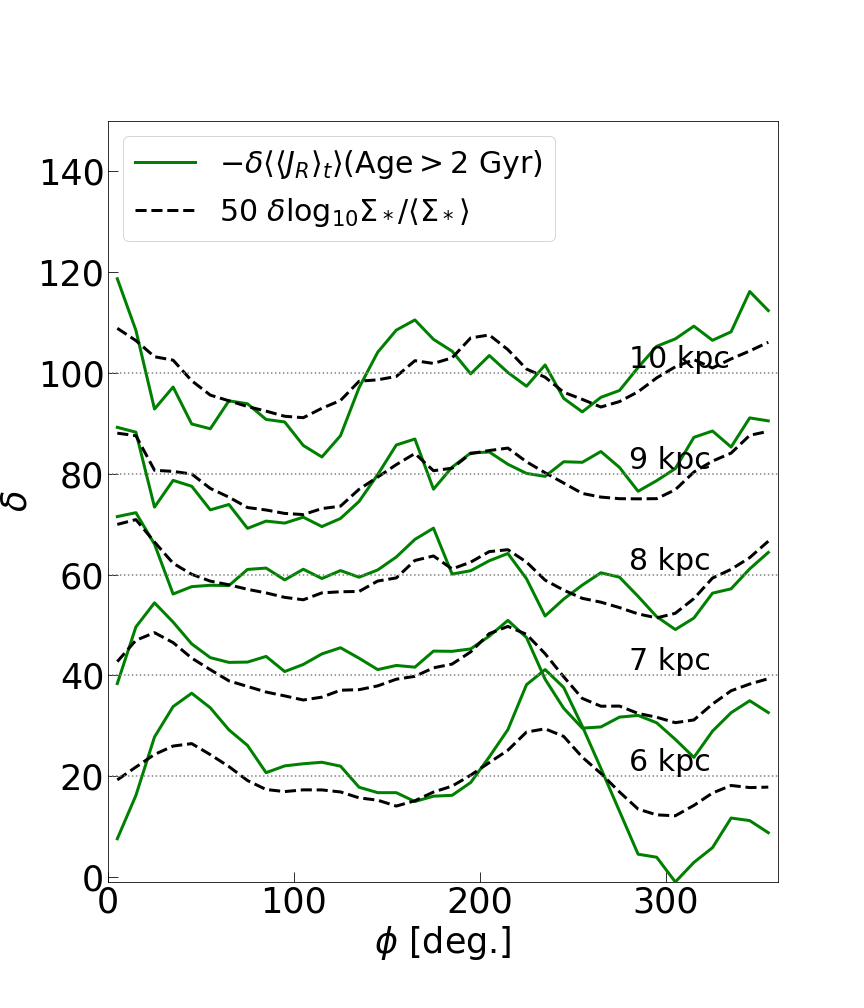}
}
\caption{The variation of \avg{\act{R}}\ (left) and $-\avg{\avgt{\act{R}}}$\ (note the sign difference) compared with the density
  variation at a series of $500~\pc$-wide annuli centred at $6\kpc$ to $10\kpc$, as indicated.  The average value
  at each radius is subtracted from each azimuthal profile and the
  profiles are then vertically offset by a fixed amount to show the
  variation. The azimuthal profiles of the density have been scaled by
  the factor $50$ for ease of comparison. The dotted horizontal lines
  indicate the zero for each radius. The red lines show
  \avg{\act{R}}\ profiles of all stars (left), while the green lines show stars older than $2~\Gyr$ only (both panels).
\label{f:jrazimuthal}}
\end{figure*}

The top panel of Fig.~\ref{f:meanjrajrmaps} maps the average radial action, \avg{\act{R}}, excluding stars younger than $2\Gyr$. Significant variations are evident, with
\avg{\act{R}}\ maxima near the locations of density peaks. However at
some radii (\eg\ $6.5~\kpc$) the location of the \avg{\act{R}}\ peaks
are clearly offset from the location of the density peaks.
The left panel of Fig.~\ref{f:jrazimuthal}, which compares the
azimuthal profiles of the density and of \avg{\act{R}}, also shows
that the peaks and troughs are offset relative to each other, with the
offset varying with radius.

Figs.~\ref{f:meanjrajrmaps} (top)  and \ref{f:jrazimuthal} (left) are
puzzling because they imply that, broadly speaking, the spiral density
peaks are better traced by stars with large \act{R}, which should be
older, and therefore represent \feh-poor stars. However we have seen that the \avfe-peaks are
located at the density peaks. Moreover, in
Fig.~\ref{f:agedispspiralamp} we showed that older populations are
hotter and host weaker spirals, contradicting the behaviour seen in
\act{R}. 

We resolve this difficulty in the next section by
showing that the inherent assumption of axisymmetry in computing
\act{R}\ gives rise to correlated errors.

\subsection{Time-averaged radial actions}
\label{ss:tjr}

\begin{figure}
\centerline{
\includegraphics[angle=0.,width=1.1\hsize]{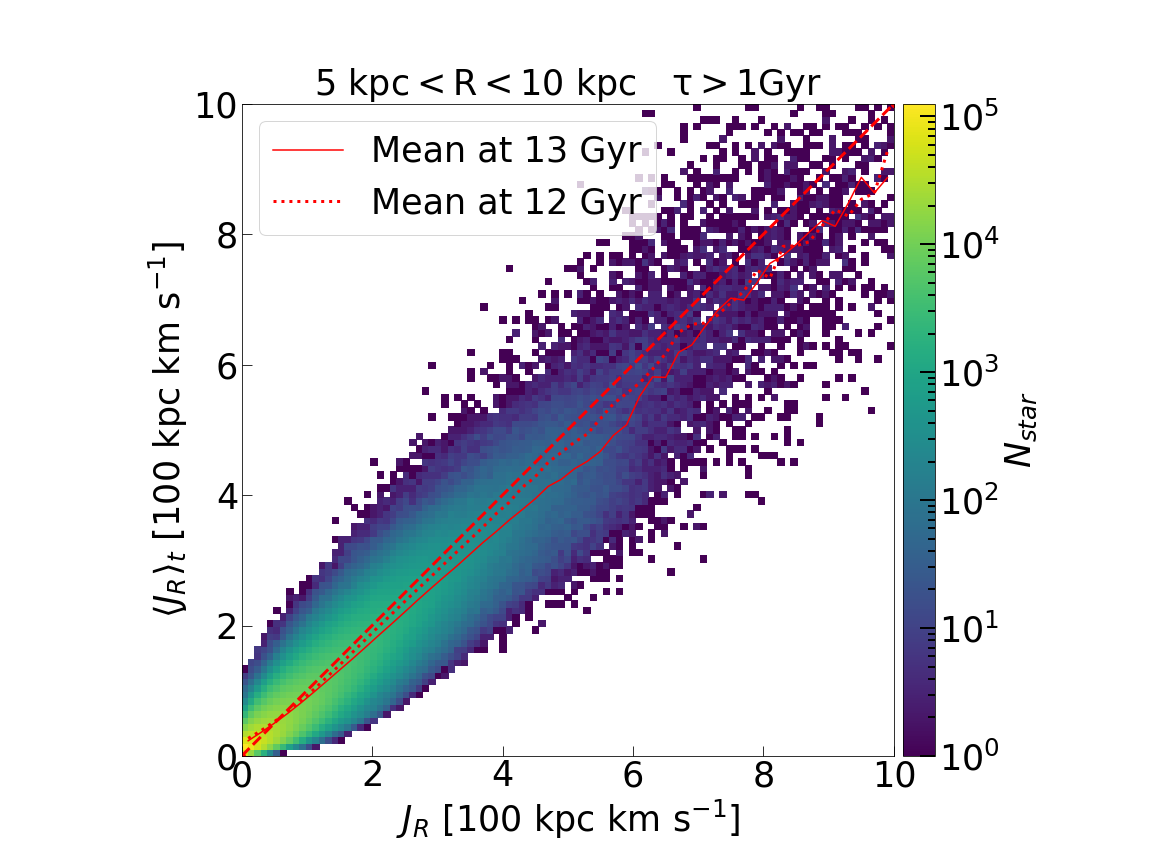}\
}
\centerline{
\includegraphics[angle=0.,width=1.1\hsize]{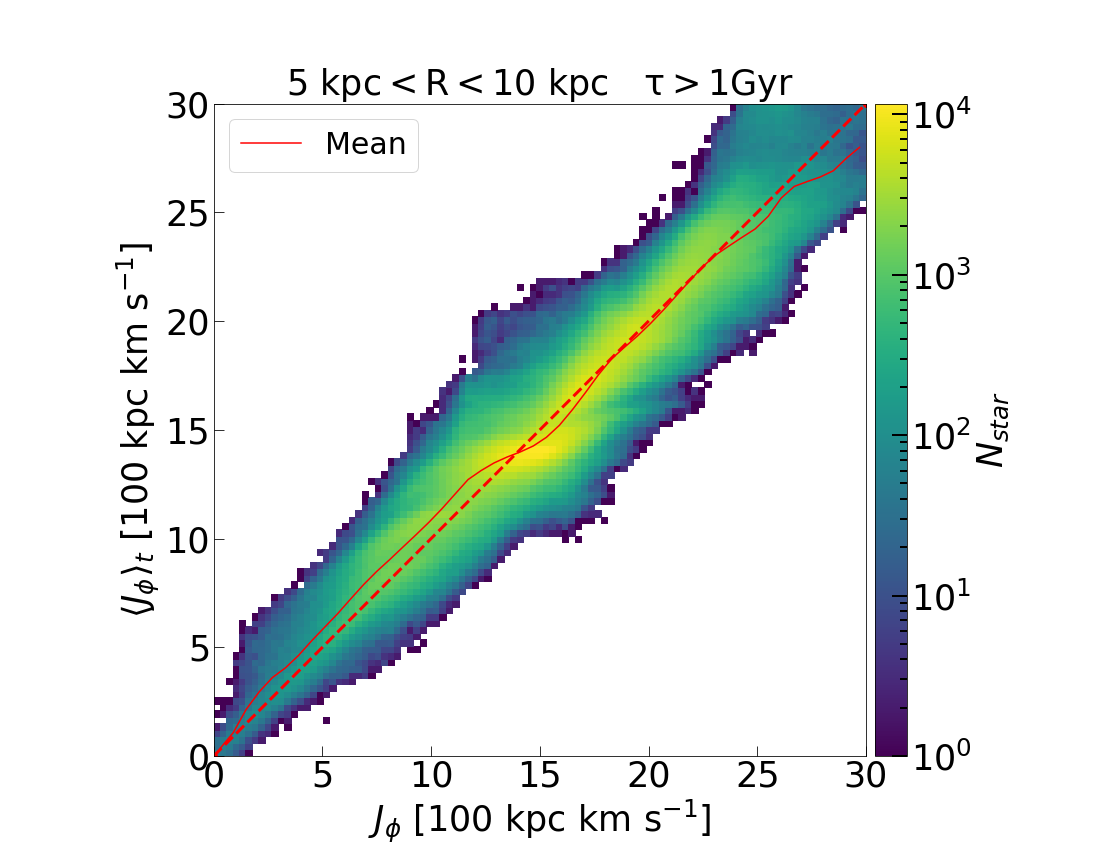}\
}
\caption{Comparison of \avgt{\act{R}}\ and \act{R}\ (top) and
  \avgt{\act{\phi}}\ and \act{\phi} (bottom) for stars which are at $5
  \leq R/\kpc \leq 10$ at $13 \Gyr$. The dashed lines show the
  diagonal, $\act{x} = \avgt{\act{x}}$ while the solid lines show the
  mean relation. The radial action shows the effect of modest heating
  while the angular momentum shows the effect of migration via
  resonances. Stars younger than $1\Gyr$ are not included. In the top
  panel the dotted line shows the mean \act{R}\ if we instead use
  \act{R}\ at $12\Gyr$ on the horizontal axis.
\label{f:avjrvsjr}}
\end{figure}

Because {\sc agama} assumes axisymmetry when computing actions, subtle
but spurious correlations between the azimuthal phase of a star with
respect to a spiral and its computed \act{R}\ may arise. In order to
mitigate, somewhat, such effects, we also compute the time average of
\act{R}, which we denote \avgt{\act{R}}. We carry out these time
averages over a  $1\Gyr$ time interval. This is long enough for
stars to have drifted across spirals several times. For those stars
trapped at the corotation resonance, $1\Gyr$ is long compared to the
lifetimes of transient spirals \citep[see][]{roskar+12}, so that they will
not have been trapped throughout this time. We compute the actions at $21$ snapshots, between $12~\Gyr$ and
$13~\Gyr$, inclusive, and average \act{R}\ to obtain
\avgt{\act{R}}. 

We compare the time-averaged radial action to the instantaneous one
(at $13\Gyr$) in the top panel of Fig.~\ref{f:avjrvsjr}. The mean of
the distribution (solid line) is slightly offset towards $\act{R} >
\avgt{\act{R}}$; this is possibly the effect of stellar orbit heating
during this time interval. We can indeed see, in the bottom panel of
Fig.~\ref{f:agedispspiralamp}, that stellar populations are slowly
heating.  The distribution is broadened, spreading across the 1:1
line, suggesting that the instantaneous action is varying
significantly over the $1\Gyr$ time interval. The most likely cause of
this variation is that the assumption of axisymmetry is introducing
errors in the measurements.

The distribution in the top panel of Fig.~\ref{f:avjrvsjr} appears featureless.  In contrast, the comparable plot for \act{\phi}-\avgt{\act{\phi}}\ (bottom panel) shows clear features over the same time interval. These features are likely produced by the radial migration driven by transient spirals \citep{sellwood_binney02}.

\begin{figure}
\centerline{
\includegraphics[angle=0.,width=\hsize]{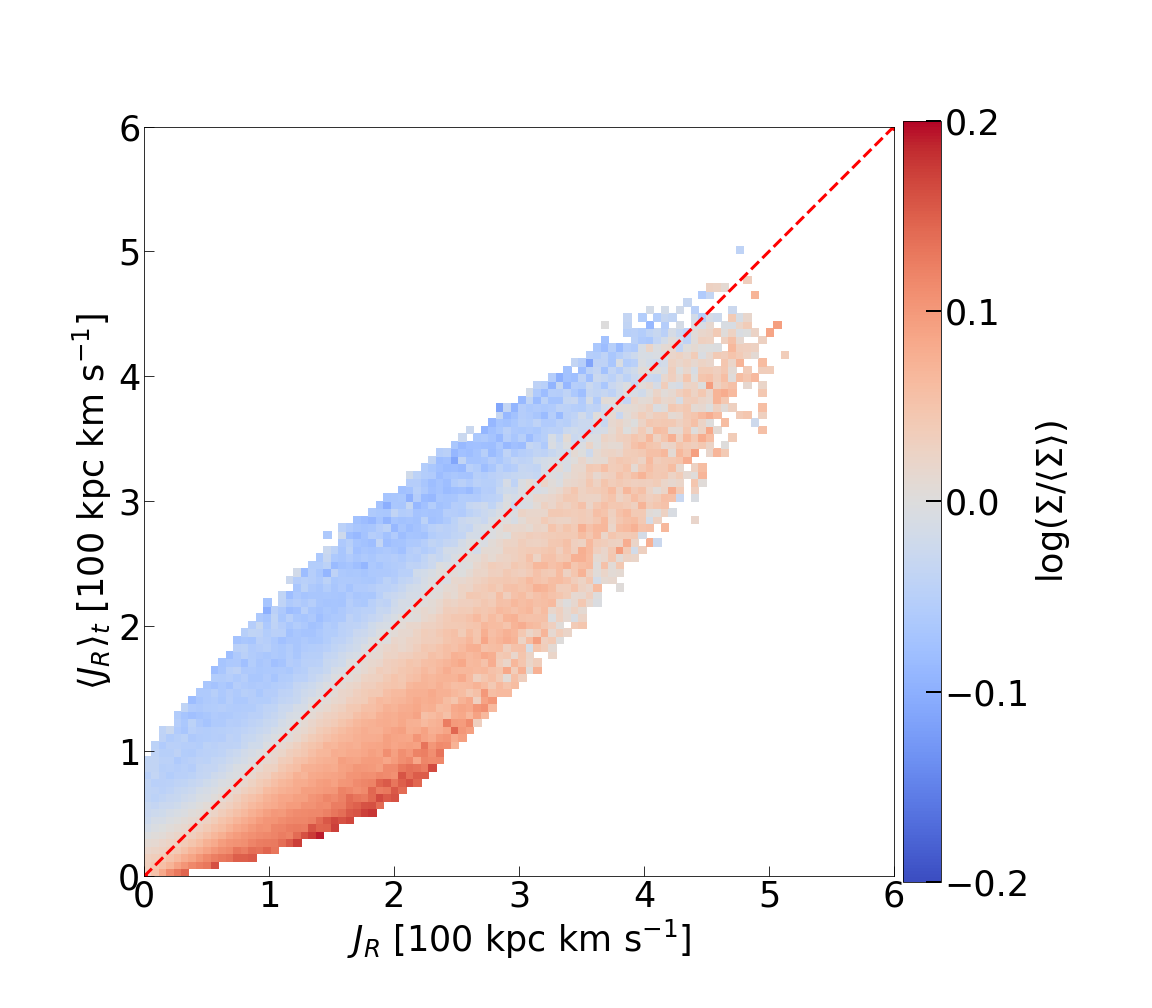}\
}
\caption{The density variation of individual stars at $t=13\Gyr$ in (\act{R},\avgt{\act{R}})-space. The deviation of \act{R}\ from the diagonal line is highly correlated with the over-density. Only bins with more than 20 particles are shown.
\label{f:residuals}}
\end{figure}

To confirm the hypothesis that the instantaneous value of \act{R}\ is
being perturbed by the spirals, we measure the overdensity of
individual particles, $\Sigma/\left<\Sigma\right>$, where
$\left<\Sigma\right>$ is the azimuthally averaged density. We measure
the local density at each particle by constructing a KD tree of the
star particle distribution based on the Euclidean 2D mid-plane
distances using the {\sc scipy.spatial.KDTree} package. For each
particle we measure the distance to the nearest 32 particles and
compute the density over these particles, and then compute the ratio
of the particle's density to the azimuthal average. We plot the
\avgt{\act{R}}-\act{R}\ distribution coloured by this density excess
in Fig.~\ref{f:residuals}. The deviation of the instantaneous
\act{R}\ from the diagonal is clearly very well correlated with the
excess density, with stars having $\act{R} > \avgt{\act{R}}$ ($\act{R}
< \avgt{\act{R}}$) in high (low) density regions, indicating that
\act{R}\ has correlated errors.

\begin{figure}
\centerline{
\includegraphics[angle=0.,width=\hsize]{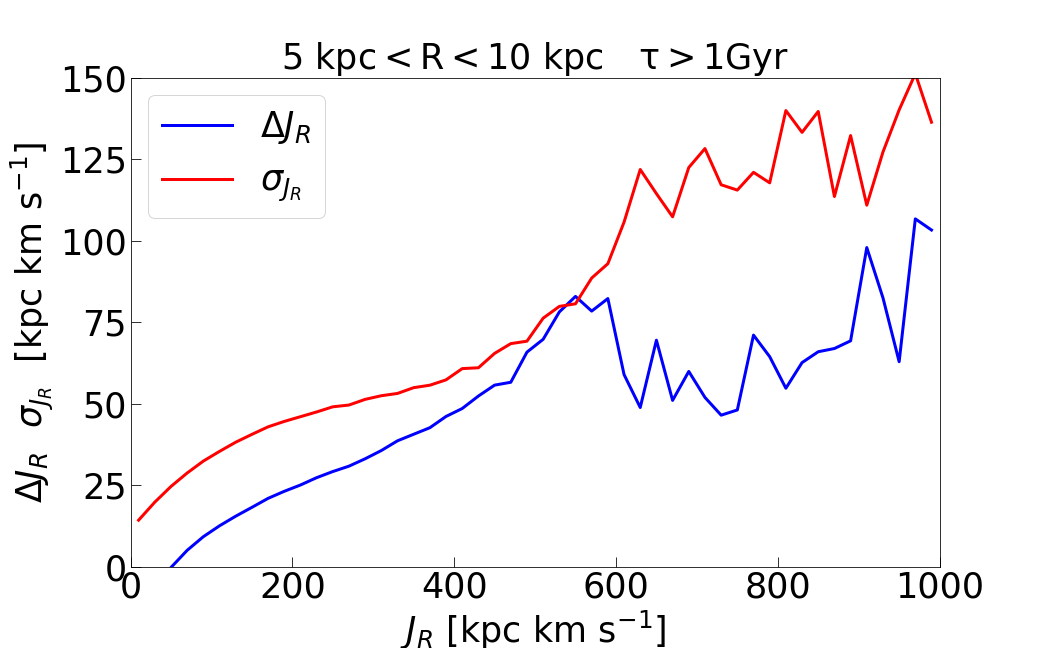}\
}
\caption{The heating, $\Delta \act{R} = \avg{\act{R} -
    \avgt{\act{R}}}$ (blue), and the instantaneous radial action
  measurement uncertainty, \sig{\act{R}}, computed as the standard
  deviation of $(\act{R} -
    \avgt{\act{R}})$ (red). Only stars older than $1\Gyr$ in the disc are considered. See the text for why $\Delta\act{R}$ over-estimates the heating.
\label{f:error}}
\end{figure}

In order to compare the measurement uncertainty with the heating rate, we first bin particles by \act{R}\ at $13\Gyr$. For each bin, we estimate the uncertainty in the measurement of \act{R}\ as the 
the standard deviation in \avgt{\act{R}}\ at a given \act{R}, which we term $\sigma_{\act{R}}$. We estimate the heating rate as the difference between \act{R}\ and the average \avgt{\act{R}}\ in each bin, \ie\ $\Delta \act{R} = \avg{\act{R} - \avgt{\act{R}}}$. 
Fig.~\ref{f:error} shows our measurements of $\Delta \act{R}$ and
$\sigma_{\act{R}}$. Over this $1\Gyr$ time interval,
$\sigma_{\act{R}}$ is smaller than $\Delta \act{R}$, \ie\ the
uncertainty in measuring the instantaneous \act{R}\ introduced by the
assumption of axisymmetry is larger than the radial heating.
In reality we have over-estimated the heating, since this is based on
the assumption that all the difference between \avg{\act{R}}\ and
\avg{\avgt{\act{R}}} (corresponding to the difference between the
$1:1$ (dashed) line and the mean \act{R}\ (solid) line in the top
panel of Fig.~\ref{f:avjrvsjr}) is due entirely to heating. However,
the dotted line in the same panel shows the relation between the same
\avgt{\act{R}}\ and \act{R}\ now computed at $12\Gyr$; this line is
also below the diagonal. The offset between the diagonal and the solid
line, therefore, is partly due to the intrinsically skewed
distribution of \avgt{\act{R}}, which means that we have
over-estimated the heating rate. We have also estimated the heating
rate by averaging \act{R}\ between $12-12.5\Gyr$ and between
$12.5-13\Gyr$, which also shows that $\Delta \act{R}$ overestimates
the heating rate. Crucially then, we are not mistaking actual
(physical) heating, from whatever source, for the (unphysical) scatter
arising from the assumption of axisymmetry.

The middle panel of Fig.~\ref{f:jrfehage} shows the
\avgt{\act{R}}-\feh\ relation; this is very similar to the
\act{R}-\feh\ relation, since the errors in \act{R}\ should be
independent of \feh. Thus the azimuthal variations of \avfe\ may be
driven by those of \avgt{\act{R}}.
The bottom panel of Fig.~\ref{f:meanjrajrmaps} maps \avg{\avgt{\act{R}}} in cylindrical coordinates. The spiral arms are very well delineated by \avg{\avgt{\act{R}}}, much better than by
\avg{\act{R}} (top panel). The differences between the two maps are quite striking
with the peak densities being associated with minima in
\avg{\avgt{\act{R}}}\ but closer to maxima in \avg{\act{R}}. That low \avgt{\act{R}}\ should track the spirals is natural since radially cool stars support the spiral structure better. At the same time, the larger
radial perturbations these stars experience means that when the
actions are computed they will appear to be radially hotter, leading to apparently larger \act{R}. 

The right panel of Fig.~\ref{f:jrazimuthal} plots the azimuthal variations
of \avg{\avgt{\act{R}}}\ and compares them with those of the
density. Compared with \avg{\act{R}}\ in the left panel, a number of important
differences are evident. First we confirm that, while the variations
in \avg{\act{R}}\ somewhat correlate with the density variations,
those in \avg{\avgt{\act{R}}}\ anti-correlate with the density
variations, as expected. Furthermore, while the correlation with
\avg{\act{R}}\ is relatively poor, that with $-\avg{\avgt{\act{R}}}$\ is
quite strong. 
Combined with the \avgt{\act{R}}-\feh\ relation of
Fig.~\ref{f:jrfehage} \avgt{\act{R}}\ is able to explain the strong correlation between the
density and \avfe\ variations.

\begin{figure}
\centerline{
\includegraphics[angle=0.,width=\hsize]{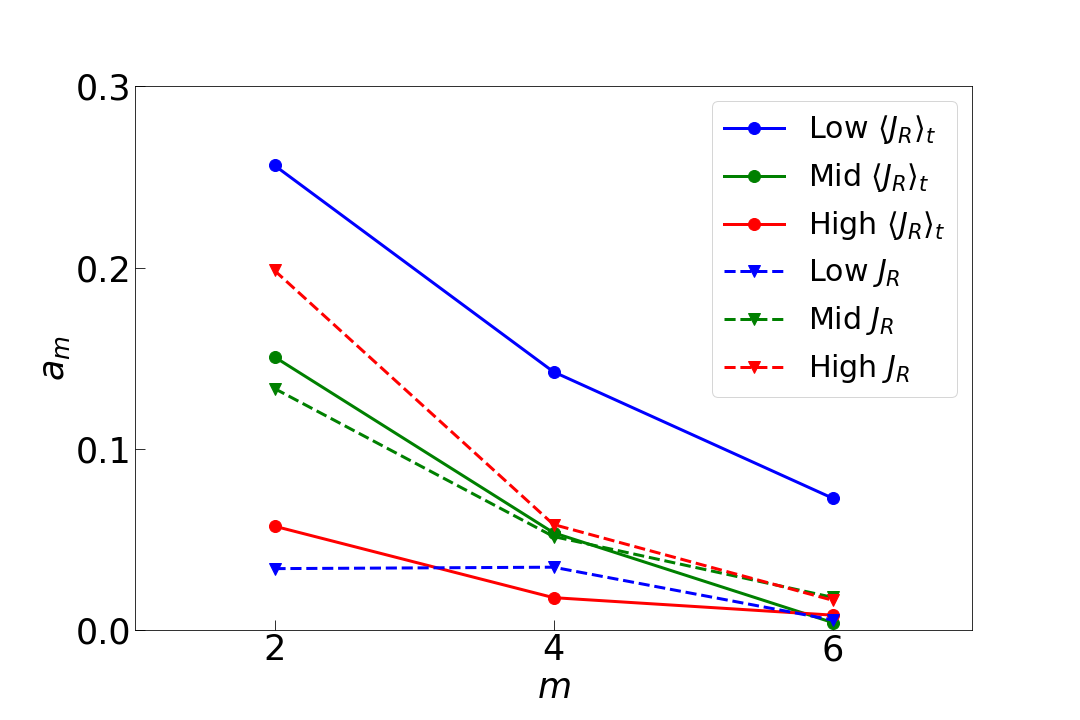}\
}
\caption{The $m=2$, $4$ and $6$ Fourier amplitudes of the density distribution at $5.5<R/\kpc<6.5$ separated into equal-sized low, medium and high populations of \avgt{\act{R}}\ (solid lines) and \act{R}\ (dashed lines). 
\label{f:aJRJRm246amp}}
\end{figure}

Fig.~\ref{f:aJRJRm246amp} shows the Fourier amplitudes of $m=2,~4$ and
$6$ of the stellar surface density at $R= 6\kpc$ (where the spirals
are strong) separated into 3 roughly equal-sized populations of low,
medium and high \act{R}\ and \avgt{\act{R}}. The low
\avgt{\act{R}}\ population has the strongest amplitudes, including at
$m=6$, indicating that this population is able to support sharp
features in the spiral structure. The mid and high
\avgt{\act{R}}\ have progressively weaker amplitudes. Instead, in
\act{R}, the strongest amplitudes are in the high-\act{R}\ population,
although these amplitudes are smaller than the ones in the
low-\avgt{\act{R}}\ population. The amplitudes decrease as
\act{R}\ decreases, the opposite of what is expected. Moreover, the
$m=6$ amplitude is nearly zero for all \act{R}\ populations,
indicating that sharp features in physical space cannot be represented
by \act{R}\ populations. Clearly \act{R}\ does a poorer job of tracing
the spiral structure than does \avgt{\act{R}}. The middle and bottom
panels of Fig.~\ref{f:medianmaps} show that the median
\avgt{\act{R}}\ also traces the spirals better than the median
\avg{\act{R}}. The bottom panel of Fig.~\ref{f:medianazimuthal} shows
that the azimuthal variations of the median \avgt{\act{R}}\ track
those of the density very well.

Thus, while the time-averaged actions cannot be the true actions at
any given time, they are a sufficiently close approximation to be
useful for studying the effects of spiral structure.

\subsection{Correlations}
\label{ss:correlations}

We compute the correlations between the variations in the density, and
those of the population properties for stars in the disc region. In
other words we quantify the correlations seen in
Figure~\ref{f:razimuthal}, and others like it. From the cylindrical
maps (in the radial range $5 \leq R/\kpc \leq 10$, which covers
$N_\phi = 101$ and $N_R = 51$ bins, giving $N_\phi \times N_R = 5151$
pixels), we subtract the radial profile of the azimuthal average in
each of the quantities of interest ($\Sigma$, \feh, \ofe, \age,
\act{R}, and \avgt{\act{R}}). We then compute the correlation
coefficient between all the pixel values for the variations in the
density and each of the other variables in turn.

\begin{figure}
\centerline{
\includegraphics[angle=0.,width=\hsize]{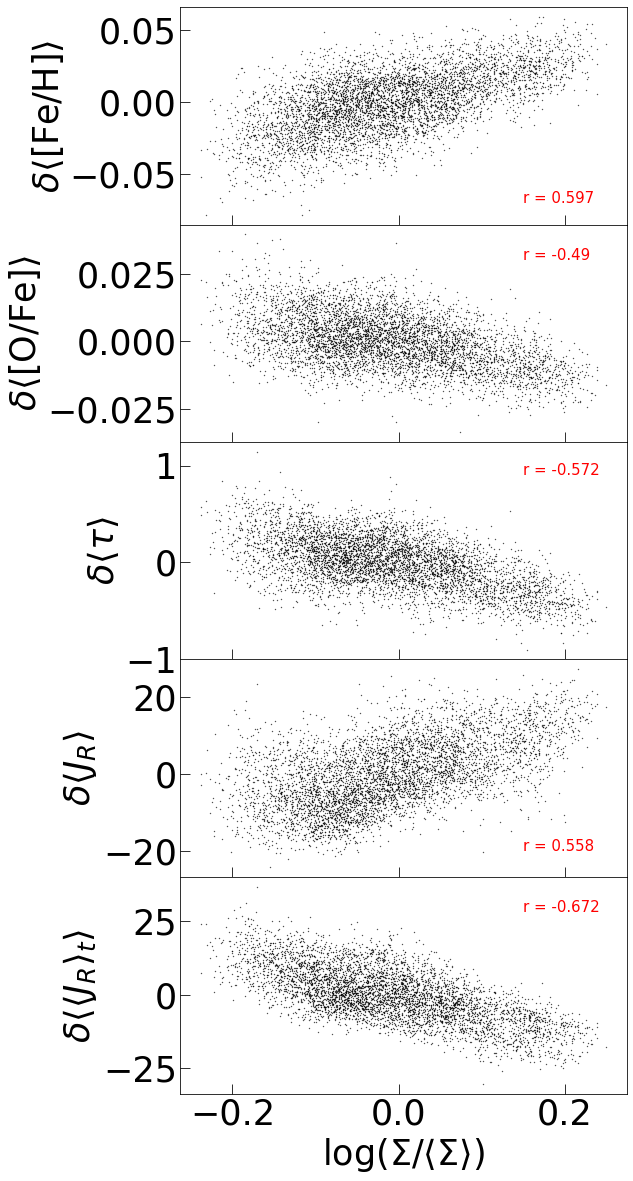}\
}
\caption{The dependence of azimuthal variations in \feh\ (top row),
  \ofe\ (second row), \age\ (third row), \act{R}\ (fourth row) and
  \avgt{\act{R}}\ (bottom row) on the azimuthal variations in
  density. The Pearson $r$ value of each correlation is indicated in
  each panel. The strongest correlation is between the density variations and those in the
  \avgt{\act{R}}. Note the change in the sign of $r$ for the correlation between the
  density variations with $\delta\avg{\act{R}}$ and with $\delta\avg{\avgt{\act{R}}}$.
  Populations younger than $2~\Gyr$ are excluded from this plot. The units of $\delta\avag$ are \Gyr, while the units of $\delta\avg{\act{R}}$ and $\delta\avg{\avgt{\act{R}}}$ are \kkms.
\label{f:denscorrs}}
\end{figure}

\begin{figure*}
\centerline{
\includegraphics[angle=0.,width=\hsize]{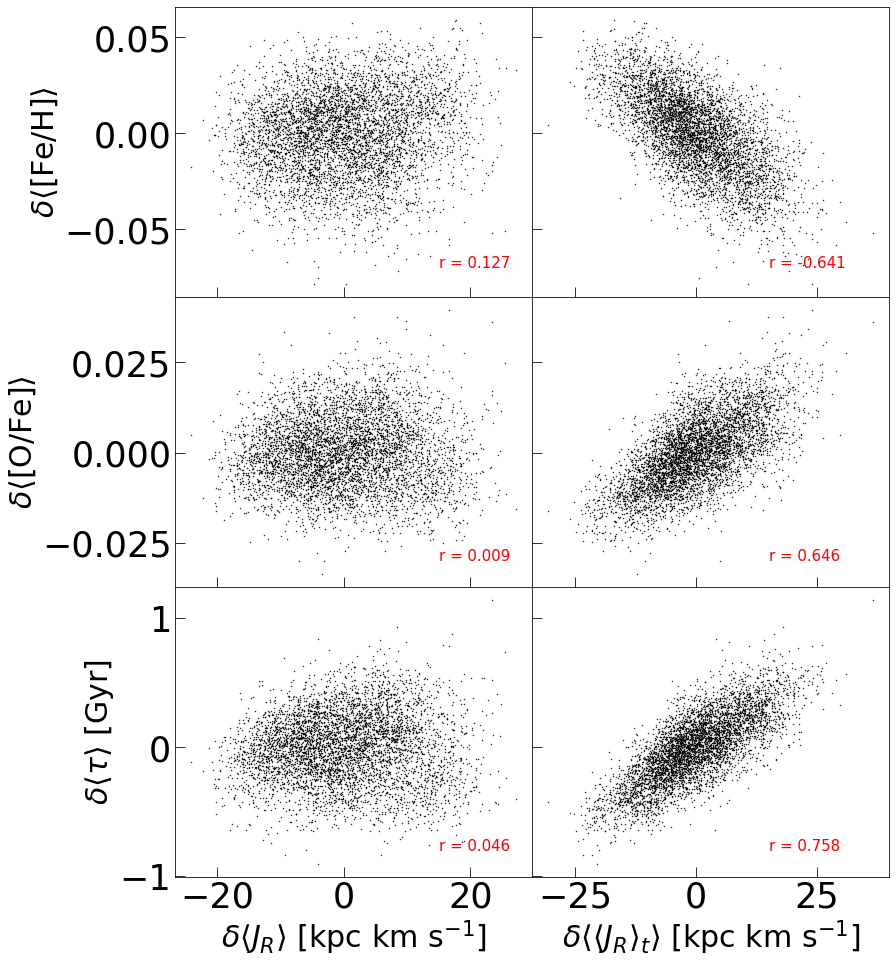}\
}
\caption{The dependence of variations in stellar population
  properties, $\delta\avfe$ (top), $\delta\avg{\ofe}$ (middle) and
  $\delta \avag$ (bottom) on the azimuthal variations of the radial
  action computed instantaneously, $\delta\avg{\act{R}}$, (left) and
  time-averaged, $\delta\avg{\avgt{\act{R}}}$,
  (right). The Pearson $r$ value of each correlation in indicated in
  each panel. The population variations are strongly correlated
  with those of \avgt{\act{R}}, but very poorly correlated with those
  of \act{R}.
\label{f:jrajrcorrs}}
\end{figure*}

Fig.~\ref{f:denscorrs} shows how stellar population properties vary with density variations. The spiral ridges
correspond to
high $\delta\avfe$, and low $\delta\avg{\ofe}$ and $\delta\avag$. It also shows that the
correlation between density variations and $\delta\avg{\avgt{\act{R}}}$
is stronger than that between density variations and
$\delta\avg{\act{R}}$. This latter relation is, in fact, even weaker than
the correlation between the density variations and $\delta\avfe$ and
$\delta\avag$. Even more strikingly, the correlation between the
density variations and $\delta \avg{\avgt{\act{R}}}$ is negative (\ie\ high density regions have small time-averaged radial action) while the correlation is positive for the instantaneous
one.
Fig.~\ref{f:jrajrcorrs} shows how the variations in the stellar
population properties (age, metallicity and \al-abundance) correlate
with the variations in \act{R}\ and \avgt{\act{R}}. The stellar population variations are poorly correlated with $\delta\avg{\act{R}}$ but strongly correlated
with $\delta\avg{\avgt{\act{R}}}$.
Together, Figs.~\ref{f:denscorrs} and \ref{f:jrajrcorrs} demonstrate
that the errors associated with the axisymmetric approximation in
computing the actions are sufficiently large to mask the actual
dependence of the population variations on the radial action.

We have also explored the vertical action, \act{z}\ and \avgt{\act{z}} (defined similar to \avgt{\act{R}}). We find that \act{z}\ behaves more similar to \avgt{\act{z}}, with spiral structure becoming progressively weaker as \act{z}\ increases. This indicates that the vertical motion is less affected by the spirals than the radial motion, consistent with the idea that in-plane and vertical motions are only weakly coupled.

\subsection{Synthesis}

In summary, the azimuthal \avfe\ variations appear to be the result of
the spirals being more strongly supported by stars with low intrinsic
\act{R}. Such stars are, on average, metal-rich, and relatively young,
which means that the locations of the spiral density peaks (troughs)
are also where these stars are over (under) represented, giving rise
to the mean metallicity peaks (troughs). However recovering this
behaviour requires that the radial actions are corrected for the error
from the assumption of axisymmetry; here we have partially done this
by time averaging \act{R}.


\section{The role of migration}
\label{s:migration}

Trapping at the corotation resonance of transient spirals drives
stellar migration \citep{sellwood_binney02, roskar+12,
  grand+12a}. Azimuthal variations in \avfe\ have been proposed to
result from radially migrating populations.
We show that migration is not a leading cause of the azimuthal
\avfe\ variations by showing that populations with a stronger radial
gradient do not produce an azimuthal offset, and that non-migrating
populations also exhibit azimuthal variations. Additional support
comes from considering the phase difference between the \avfe\ and
density waves.

\subsection{Populations with strong radial gradients}
\label{ss:radialgradient}

\begin{figure}
\includegraphics[angle=0.,width=\hsize]{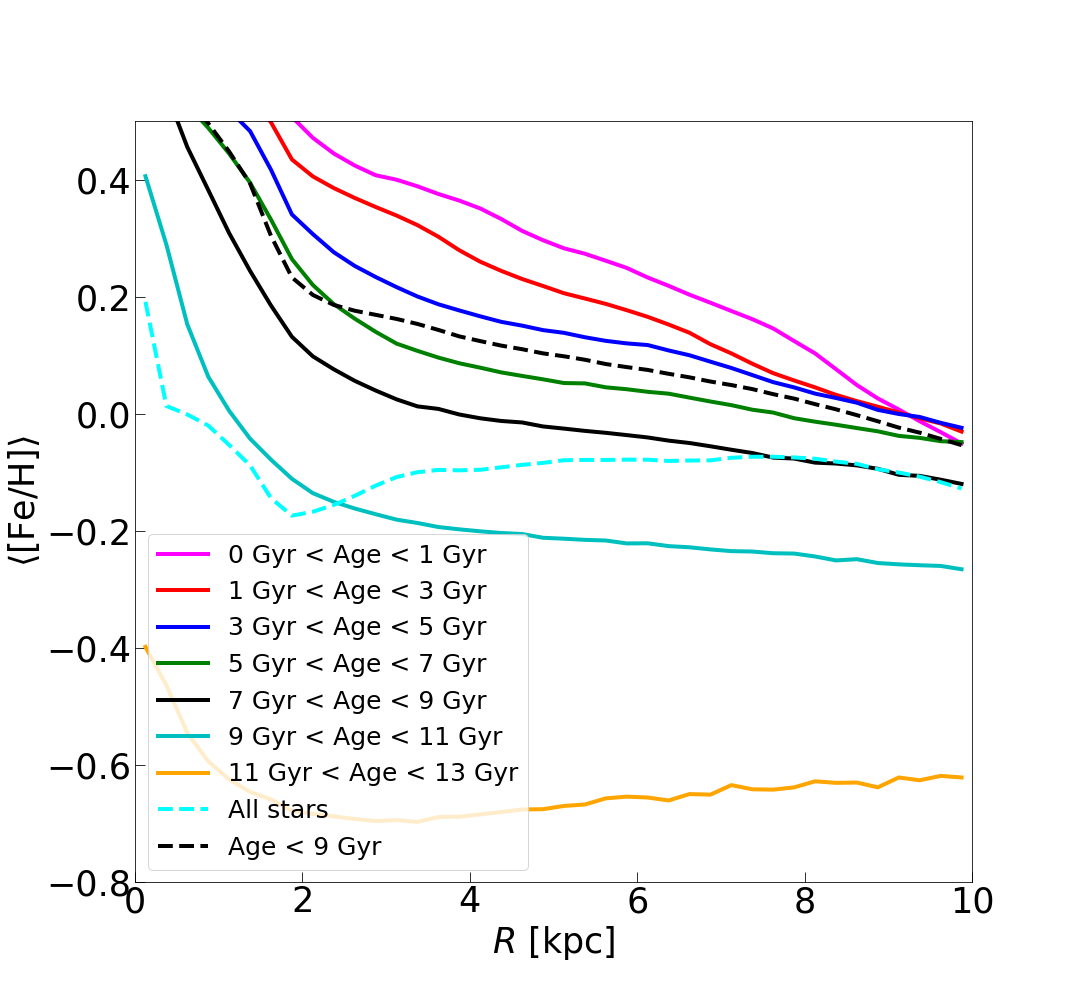}
\caption{\avfe\ profiles at $t=13\Gyr$ for different age populations
  as indicated. Outside the inner $5\kpc$, the profiles become
  progressively flatter for older populations. The overall
  \avfe\ profile (dashed cyan line) is flat. Excluding populations
  older than $9\Gyr$ (dashed black line) gives a gradient $\nabla
  \avfe \simeq 0.03$~dex.~$\kpc^{-1}$. }
\label{f:fehprofiles}
\end{figure}

\citet{khoperskov+18b} emphasise that migration-dominated
\avfe\ variations require a radial metallicity
gradient. Fig.~\ref{f:fehprofiles} shows the \avfe\ profiles of
different age populations in the model. This shows the usual influence
of radial migration, in that metallicity profiles become flatter for
older populations, which is the opposite trend expected for a disc
growing inside-out. A comparable behaviour has been observed in the MW
\citep{nordstrom+04, casagrande+11, bergemann+14, vickers+21,
  anders+23, willett+23}, including the slightly positive gradient of
the oldest stars \citep{nordstrom+04, casagrande+11, anders+23}. The
overall \avfe\ profile (dashed cyan line) is quite flat over $5 \leq
R/\kpc \leq 10$, which might explain why migration has not seemed to
matter thus far, as implied by the coincidence of the $\delta\avfe$
and density peaks. However, if we consider only stars younger than
$9\Gyr$, to match the bulk of the MW's thin disc, we find a
metallicity gradient of $\nabla \avfe \simeq 0.03$~dex~$\kpc^{-1}$
(dashed black line). This gradient, although roughly half that of the
MW's thin disc \citep[\eg][]{vickers+21, imig+23, hackshaw+24}, allows
us to test whether stellar migration of populations with a radial
metallicity gradient may dominate azimuthal \avfe\ variations. In the
bottom right panel of Fig.~\ref{f:azimuthal} we map \avfe\ for stars
younger than $9\Gyr$, superimposing the density map of all stars. No
obvious offsets between the peak \avfe\ and peak density are evident,
and the overall metallicity map looks broadly consistent with that for
all stars, apart from the different \avfe\ values. The same result can
be seen comparing the cyan line with the red or green lines in the
left panel of Fig.~\ref{f:razimuthal}, suggesting that migration does
not dominate the azimuthal \avfe\ variations.

\subsection{Non-migrating stars}
\label{ss:nonmigrators}

\begin{figure}
\centerline{
\includegraphics[angle=0.,width=1.1\hsize]{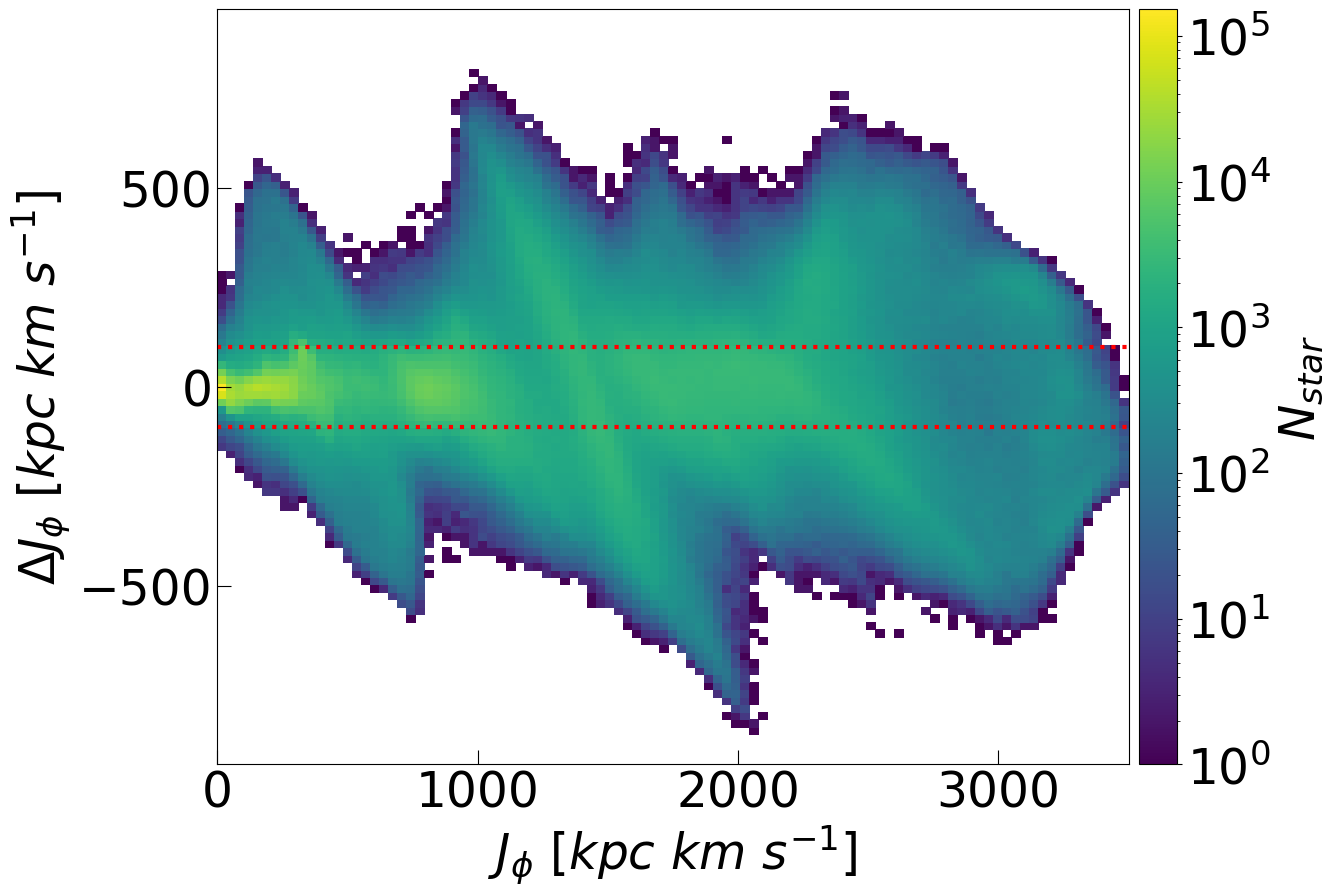}\
}
\caption{The distribution of stars in the $\Delta
  \act{\phi}-\act{\phi}$ plane.  Prominent ridges are evident,
  indicative of migration driven by spirals. The horizontal lines
  delineate the non-migrators, defined as those having $|\Delta
  \act{\phi}| < 100$~\kkms. The horizontal axis shows \act{\phi}\ at
  $12.65\Gyr$ while the vertical axis shows $\act{\phi}(13~\Gyr) -
  \act{\phi}(12.65~\Gyr)$. Only stars younger than $9\Gyr$ are
  included.}
\label{f:migration}
\end{figure}

We further explore whether migration plays a leading role in the
metallicity variations by studying a non-migrating population over the
time interval $12.65-13\Gyr$, during which a strong spiral forms.
We compute the change in angular momentum, $\Delta\act{\phi} =
\act{\phi}(12.65~\Gyr) - \act{\phi}(13~\Gyr)$. Fig.~\ref{f:migration}
plots the distribution of $\Delta\act{\phi}$ versus $\act{\phi}$. This
exhibits the usual stripes associated with spiral corotation migration
\citep{sellwood_binney02, roskar+12}. The dotted horizontal lines
delineate $|\Delta \act{\phi}| < 100\kkms$, which we take as a
non-migrating population (constituting $\sim 70\%$ of all stars).

\begin{figure}
\centerline{
\includegraphics[angle=0.,width=1.1\hsize]{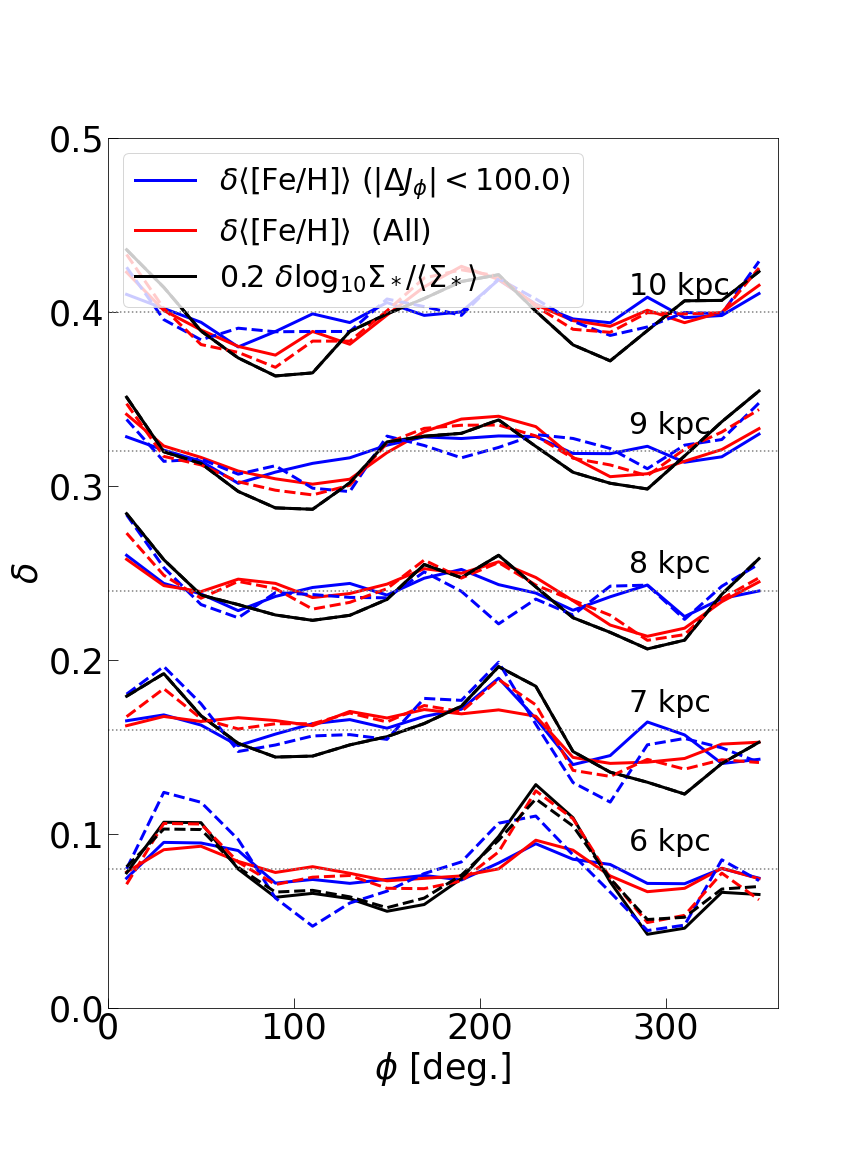}\
}
\caption{The effect of radial migration on \avfe\ variations.  The
  variations are shown for a series of 500 pc-wide annuli centred at
  $6\kpc$ to $10\kpc$, as indicated. The average values for each
  radius are subtracted from each azimuthal profile and the profiles
  are then vertically offset by a fixed amount to show the
  variation. The azimuthal profiles of the density have been scaled by
  the factor $0.2$, for ease of comparison. All profiles exclude stars
  formed after $12.65\Gyr$, for which we cannot compute $\Delta
  \act{\phi}$.  The red lines show the variations of stellar
  populations with no cuts on $\Delta \act{\phi}$, while the blue
  lines show the variations for the non-migrating populations,
  \ie\ those with $|\Delta \act{\phi}| < 100\kkms$. Dashed lines show
  all ages while solid lines show only populations younger than
  $9\Gyr$. Note that the density, shown by black lines, include both a
  solid line ($|\Delta \act{\phi}| < 100\kkms$) and a dashed line (no
  $|\Delta \act{\phi}|$ cut); other than at $6\kpc$, these overlap.
  The dotted horizontal lines indicate the zero for each radius. We
  show the $13\Gyr$ distribution.
\label{f:migration2}}
\end{figure}

Fig.~\ref{f:migration2} shows azimuthal profiles of $\delta\avfe$ for
all stars, and for the non-migrating population. We present this
twice, once for all ages (other than those that form after $12.65\Gyr$
for which we cannot compute $\Delta \act{\phi}$), which we show with
dashed lines, and excluding stars older than $9\Gyr$ (solid
lines). The \avfe~profiles of all stars (shown in red) follow the
density variations quite well, regardless of whether stars older than
$9\Gyr$ are excluded (solid lines) or included (dashed lines). When
non-migrating stars are considered, the profiles still somewhat track
those of the density but there is now considerably more noise in the
profiles, as attested by the point to point scatter. However for the
most part, the blue solid line follows the density variations, most
notably peaking at the same locations. If we include stars older than
$9\Gyr$, the non-migrating population (blue dashed line) still
generally peaks near the density peaks.

\subsection{Evolution of the phase difference}

\begin{figure}
\centerline{
\includegraphics[angle=0.,width=1.1\hsize]{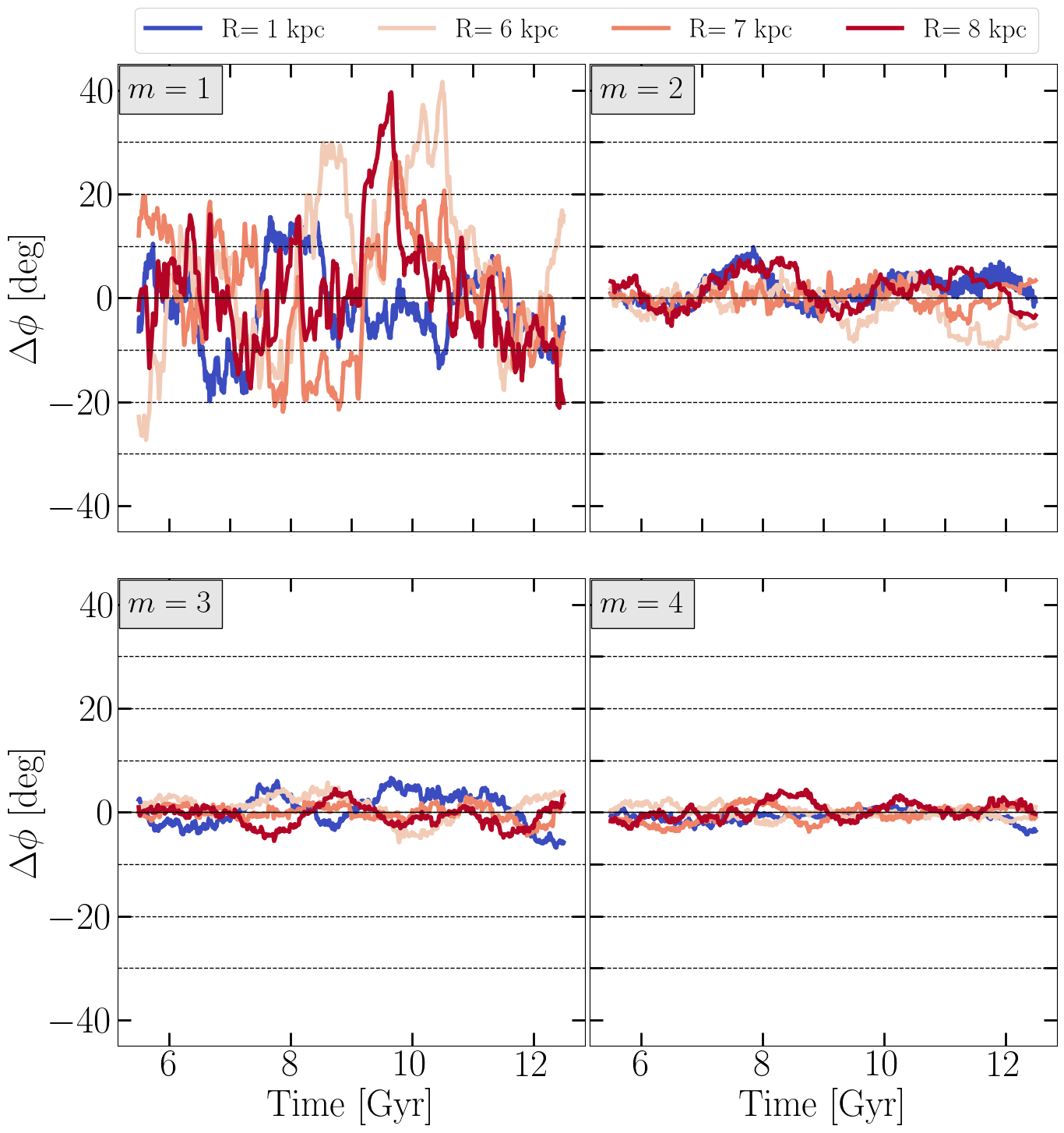}\
}
\caption{Moving average over $\Delta t = 1\Gyr$ intervals, of the
  phase shift, $\Delta \phi_m$, between the density and \avfe\ waves
  at different radii, as indicated by the colours, and harmonics
  (indicated at top left of each panel). $\Delta \phi_2$ appears to
  fluctuate about $0\degrees$ throughout the $7 \Gyr$ evolution with
  $\max(\Delta \phi_2)\leq10\degrees$. Only stars older than $2\Gyr$
  are considered in this analysis.}
\label{f:phases}
\end{figure}

Our final evidence that migration does not dominate the azimuthal
variations comes from considering the time evolution of the relative
phase of the density and metallicity. We compute the phases of the
waves using the Fourier analysis of Section~\ref{s:omega}. Since star
formation occurs preferentially on the leading (trailing) side of
spirals outside (inside) corotation, which leads to phase offsets for
young stars, as in Section~\ref{s:omega} we exclude stars younger than
$2\Gyr$ in this analysis. From the Fourier moments at each timestep,
we compute the phase shift between the density and \avfe\ waves,
$\Delta \phi_m = \phi_{\feh,m} - \phi_{\Sigma,m}$.
Because single phases are noisy, particularly when the spirals are
weak, or when the two spirals which overlap in the disc have similar
amplitudes, we average over $1\Gyr$ time
intervals. Fig.~\ref{f:phases} shows $\Delta \phi_m$ for $1 \leq m
\leq 4$. The phase shift for $m=1$ perturbations is large, but the
density waves of this multiplicity are weak. For $m = 2$, which has
the largest amplitudes, $|\Delta \phi_2|$ is everywhere less than
$10\degrees$. Fig.~\ref{f:phases} includes $\Delta \phi_m$ for $R =
1\kpc$, where the bar is resident, to show that even in this case
there is some deviation from $\Delta \phi_m = 0\degrees$, which
therefore indicates some of the uncertainty in the measurement
method. No sustained offset, $|\Delta \phi_2|$, is evident; the same
is true for the $m=3$ and $m=4$ moments. Thus we find no evidence that
the \avfe\ waves have a consistent phase shift relative to the
spirals, even though the pattern speeds of the main spirals are only
slowly evolving over this time. We have repeated this analysis
excluding stars younger than $2\Gyr$ and those born in the first
$4\Gyr$ (to ensure a significant disc \feh\ radial gradient) and still
find no consistent phase shifts.

We conclude therefore that, at least for this simulation, there is no
support for the idea that migration dominates the azimuthal
\feh\ variations. We discuss this result, and reasons why the model
might be underestimating the role of migration relative to the effect
of varying radial actions, in Section~\ref{s:discussion} below.


\section{The view in action space}
\label{s:actionspace}

Resonant trapping leads to \act{R}\ and \act{\phi} non-conservation,
with stars librating around the resonance when viewed in the space of
axisymmetric actions. In this section we examine whether the
\act{R}\ variations are due to such librations, rather than errors
arising from the axisymmetric approximation.
We consider the joint time-evolution of \act{R}\ and \act{\phi}, in order to show that the variations in \act{R}\ we have found are not due to resonances. 

Recall that when a resonance is present, stars are either trapped by
it, in which case their actions librate back and forth across the
resonant action, or they are untrapped, in which case their actions
vary periodically over their orbits. The variations of \act{R}\ and
\act{\phi}\ of both trapped and untrapped stars can be related by
noting that the Jacobi energy, $E_J \equiv E - \omp \act{\phi}$, is a
conserved quantity \citep[\eg][chapter 3.3.2]{binney_tremaine08},
where $E$ is a star's energy and $\omp$ the pattern speed of the
perturbation, provided that the system is nearly stationary in the
rotating frame of the perturber. Thus in the
$(\act{\phi},\act{R})$-plane, resonances cause stars to oscillate
tangential to the contours of constant $E_J$, even if they are not
resonantly trapped.

Furthermore a star is at a planar resonance when $N_R \omg{R} + N_\phi (\omg{\phi} - \omp) = 0$, where $N_R$ and $N_\phi$ are integers and \omg{\phi}\ and \omg{R}\ are a star's azimuthal and radial frequencies. The principal resonances are the corotation (CR; $N_R = 0$), and the outer and inner Lindblad resonances (OLR and ILR; $N_R = 1, ~N_\phi = \pm 2$). The conservation of $E_J$ leads to the condition
\begin{equation}
\Delta{\act{R}} = \frac{N_R}{N_\phi} \Delta{\act{\phi}}
\end{equation}
\citep[\eg][]{sellwood_binney02, chiba+21a}. Therefore, for the ILR, CR and OLR, the directions of oscillation of resonant orbits have slopes $-\frac{1}{2}$, $0$ and $+\frac{1}{2}$, respectively, in the (\act{\phi},\act{R})-plane. We can therefore test whether the variations in \act{R}\ and \act{\phi}\ follow these conditions by plotting their evolution in the $(\act{\phi},\act{R})$-plane.

\begin{figure*}
\centerline{
\includegraphics[angle=0.,width=0.9\hsize]{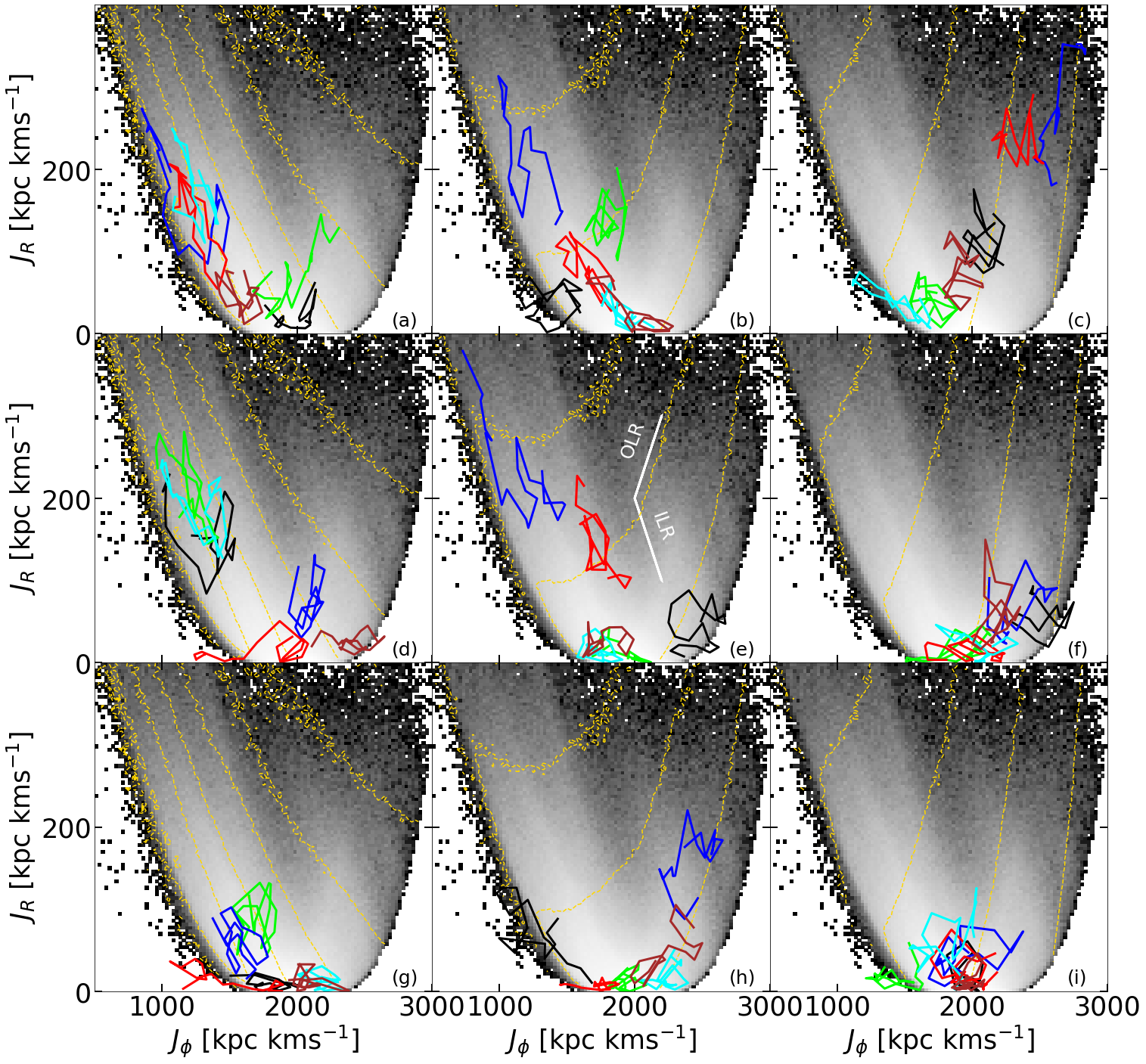}\
}
\caption{The evolution of a random set of 54 particles, selected as described in the text, in the action space (\act{\phi},\act{R}). In each panel we present 6 stars with different colours over a $1\Gyr$ time interval plotting every $50\Myr$. The plotted particles are ordered by age, with the oldest in the top row ($9.2\leq \age/\Gyr \leq 11.7$), the intermediate age ones in the middle row ($4.7\leq \age/\Gyr \leq 8.8$) and the youngest in the bottom row ($2.2\leq \age/\Gyr \leq 4.2$). The dashed yellow lines show contours of constant $E_J$ for $\omp = 15.4~\kmsk$ (left), $40.0~\kmsk$ (centre) and $61.5~\kmsk$ (right). The white lines in panel (e) indicate the libration vectors at the OLR (upper line) and the ILR (lower line).
The background shows a log scale representation of the number density of stars.
\label{f:jrjpevol}}
\end{figure*}

We carry out this test for a subsample of stars chosen at random. 
We begin by selecting stars older than $2\Gyr$ in the radial region $7 \leq R/\kpc \leq 9$ at $13\Gyr$. From the 940800 stars that satisfy these conditions, we select those with small vertical excursion through $\avgt{\act{z}} < 5\kkms$ (corresponding to an RMS height of $0.17\kpc$), which leaves 441931 stars.  
Fig.~\ref{f:jrjpevol} shows the orbital evolution of a sample of 54 such stars chosen at random in the (\act{\phi},\act{R})-plane over the time interval $12-13\Gyr$\footnote{We show a random distribution where we only avoid orbits that overlay the resonance arrows in panel (e).}. The age of the selected stars decreases from top to bottom. The three columns show $E_J$ contours for the 3 main pattern speeds at this time (including the bar's). It is immediately obvious that the computed actions of all the stars are varying. While some stars do appear to be at resonances (\eg\ ILR: (a) cyan and blue; CR: (g) black; OLR: (a) green) for the most part the stars do not appear to be librating about resonances. Nor do they faithfully follow the contours of $E_J$ for any of the main pattern speeds. In all cases, we find considerable jitter in the tracks, which we associate with measurement errors.

We conclude that the variations we find are not predominantly the
result of resonances. This does not mean of course that the actions
are not changing as a result of resonances, but the actions computed
assuming axisymmetry are not able to properly follow this
behaviour. While we have shown this for only a tiny fraction of stars,
we have verified that this behaviour is generic to other stars. In
Appendix~\ref{app:otherparticles} we present a choice of particles
which favours stars trapped at ILRs, and find the same pronounced
jitter in the instantaneous actions.

Nor does it seem likely that resonance overlap is driving the changes
in the actions, since such scattering leads to heating overall, which
we earlier found to be quite modest. We conclude that the variations
in \act{R}\ are largely the result of errors introduced by the
assumption of axisymmetry.


\section{Discussion} 
\label{s:discussion}

\subsection{The cause of azimuthal metallicity variations}
\label{ss:cause}

We have shown that the azimuthal metallicity variations are a result
of spiral structure. The peaks and troughs of the \avfe\ waves match
those of the density. Their pattern speeds also match those of the
spirals. The possibility that the \avfe\ variations are just a result
of young, recently formed stars is excluded by the simulation, and by
observations, which both show that \avfe\ variations occur also in
stars older than $2\Gyr$ \citep[\eg][]{bovy+14}, which will have had
sufficient time to move out of their natal spiral arms. Moreover,
\citet{roskar+12} found (their Fig.~6) that the lifetimes of spirals
is typically a few $100 \Myr$. We have shown that azimuthal
\avfe\ variations are caused by the stronger response to a spiral
perturbation of kinematically cooler (therefore younger and
metal-rich) stars compared with kinematically hotter (older and
metal-poorer) stars. As a result, the metal-rich stars trace the
density ridge of the spirals. For the same reason, the spirals are
traced by minima in \ofe\ and age (even when young stars are
excluded). Our conclusion with a fully self-consistent simulation
agrees with that of \citet{khoperskov+18b}. We do not find the offsets
between the peak metallicity and peak density characteristic of
metallicity variations produced primarily by migrating stars
\citep{grand+16, khoperskov+18b, khoperskov+23}.
We find that non-migrating populations also follow the overall trend
of the \feh\ variation. Because the population that is migrating at
any one time is a relatively small fraction of the total (consisting
of those currently resonantly trapped at the corotation of any given
spiral) this is not an efficient way to generate azimuthal variations,
because the majority of the stars do not partake in the
migration. Thus we propose that the azimuthal \avfe\ variations are
dominated by the weaker spiral response of populations with increasing
radial action and decreasing metallicity.
Our conclusion agrees with models 1 and 2 of \citet{khoperskov+18b},
where metallicity is a function of both radius (in model 2) and
stellar population temperature (both models), but differs from that of
model 3, in which, by design, the metallicity is a function only of
radius, and independent of \act{R}. As a result, azimuthal variations
in model 3 can only be driven by radial migration.  Since the age of
stars is related to both the metallicity and their dynamical
temperature, the conditions of model 3 are unlikely to arise in
nature.

The reason for the disagreement with the result of \citet{grand+16b}
is less clear.  These authors only considered stars younger than
$3\Gyr$; it may be that these populations were too similar in
\act{R}\ to generate much of a different spiral response. Another
possible cause for a strong effect of radial migration is that the
spiral structure in their model is corotating at all radii
\citep{grand+12a}, which may enhance migration-driven azimuthal
variations by migrating many stars over very large radial
ranges. Alternatively, migration is enhanced by satellite interactions
\citep{grand+16, carr+22} in their simulation, which are absent in our
model.

On the other hand, some important caveats need to be acknowledged
about our model, which may lead to underestimating the effect of
radial migration. The model may have a steeper age-metallicity
relation (AMR) in the disc than found in some studies of the MW
\citep[e.g.][]{mackereth+17, delgadomena+19}; this would imply that
migration is weaker in the model than in the MW. However other studies
find a steeper AMR in the MW \citep[e.g.][]{feuillet+19, anders+23,
  johnson+24}, consistent with the model. The AMR therefore does not
provide an unambiguous test of the model's migration history.

The model's metallicity gradient for stars younger than $9\Gyr$ is
lower than the MW's thin disc's by a factor of about 2. Since the
metallicity gradient depends on both the birth metallicity gradient
across time, as well as migration, it also does not test the model's
rate of migration. However, the lower gradient implies that migrating
populations generate a weaker azimuthal metallicity
variation. Moreover, the \act{R}-\feh\ relation in the model is
steeper than in the MW, which means that the spirals are bound to
produce stronger azimuthal variations because of \act{R}. All these
suggest that the relative effect of migration may be underestimated in
this model compared with the effect of varying \act{R}. Nevertheless,
the fact that only stars trapped at the spirals' corotation resonances
can produce the migration-induced azimuthal variations implies that
this is a less efficient mechanism than the \act{R}\ variations.

Since, at present, the MW's spiral structure is not fully known
\citep[see, for instance,][]{shen_zheng20}, the azimuthal metallicity
variations themselves may therefore be used to probe the spiral
structure. Unlike molecular gas clouds, the distance to stars within
$\sim 5\kpc$ can now be determined accurately with \gaia; furthermore
metallicity variations can be mapped away from the dusty mid-plane.

\subsection{Errors in the radial action}

That the axisymmetric approximation needed in computing actions using the St\"ackel approximation does not hold has long been obvious. The fact that in the St\"ackel approximation itself the errors incurred are less than $10\%$ \citep{binney12} has given rise to the expectation that other sources of error are equally small. Since results based on actions of Solar Neighbourhood disc stars computed using the axisymmetric St\"ackel approximation give physically reasonable results \citep[\eg][]{trick+19, monari+19, frankel+20, kawata+21, drimmel+23}, this has perhaps allayed concerns that the local spiral structure disturbs the computed actions much.
We have shown (Fig.~\ref{f:error}) that the presence of spiral structure causes errors of order $< 100\kkms$ in \act{R}\ through most of the disc. The error increases slowly with \act{R}\ such that the fractional error is larger at small \act{R}.

More worryingly, \act{R}\ computed using the axisymmetric
approximation must have {\it correlated} errors large enough to render
them compromised when employed to study spiral structure. In principle
this difficulty can be sufficiently circumvented in simulations by
replacing \act{R}\ by the time-averaged radial actions,
\avgt{\act{R}}, over a timescale sufficiently short to limit the
effect of resonant heating but longer than the finite lifetime of
spirals, to ensure any resonant stars are not trapped for the whole
time interval. Naturally this is not possible in the MW. It may be
possible to account for the perturbations from spirals to correct for
the radial action errors; the torus-mapping technique
\citep{binney_mcmillan11, binney16, binney18, binney20} is well-suited
to this problem \citep[see also][]{alkazwini+22}. However attempting
to map the spiral structure using just the radial actions based on an
axisymmetric St\"ackel approximation gives the wrong result. For
instance, \citet{palicio+23} found that the spirals traced by
\act{R}\ computed this way did not fully match with any of the current
models of the MW's spiral structure, although it should also be noted
that none of the models agree fully with any other.

Despite these errors, we note that the density distribution in Fig.~\ref{f:jrjpevol}, shown by the grey background, still exhibits multiple ridges in the space of instantaneous actions (\act{\phi},\act{R}). Similar ridges have been observed in the MW \citep{trick+19, monari+19}. Thus the errors are not so large, or the spirals so strong in the simulation, that this important observable is smeared out.

\subsection{Summary}

Our key results are the following:
\begin{enumerate}
\item We have shown that an isolated simulation develops azimuthal
  \avfe\ variations with peaks at the locations of the density peaks,
  corresponding to the spiral ridges. These azimuthal variations occur
  also when stars younger than $2\Gyr$ are excluded, indicating that
  high-metallicity star formation is not driving the
  \avfe\ variations. Indeed the \avfe\ variations occur in stars of
  all ages. The peak to trough variations of \avfe\ are of order 0.1
  dex. (See Section~\ref{s:chemistry}.)
\item We bin \feh\ on a cylindrical grid and use this binning to
  compute the pattern speeds of the \avfe\ variations. We compare
  these pattern speeds to those of the spiral density waves binned in
  the same way. Because we bin the average \feh, the density is
  factored out of the \avfe\ pattern speeds. We find that the pattern
  speeds of \avfe\ match those of the density waves, indicating that
  the \avfe\ waves are merely a manifestation of the spiral density
  waves. (See Section~\ref{s:omega}.)
\item Azimuthal variations occur also in \avg{\ofe}\ (peak to trough $\sim 0.025$ dex) and age, \avag, (peak to trough $\sim 1\Gyr$). The variations in \avag\  and \avg{\ofe}\ track density variations slightly less well than does \avfe, and both are anti-correlated with the density, as would be expected since \feh-rich stars should be younger and $\alpha$-poor. (See Sections~\ref{ss:ofe} and \ref{s:age}.)
\item The radial velocity dispersions, \sig{R}, of stars in the
  simulation increases with age, \age. As a consequence, the spirals
  are weaker in the old stars than in the younger stars. This
  variation in spiral strength with \age\ is why the \feh-rich stars
  are concentrated on the spiral density ridges. However, when we
  compute the radial action, \act{R}, and compare the azimuthal
  variation of \avg{\act{R}}, we find that they generally correlate
  with those of the density, the opposite of what is expected. (See
  Sections~\ref{s:age} and \ref{ss:jr}.)
\item We demonstrate that the axisymmetry approximation is introducing correlated
  errors in the computation of \act{R}. In order to explore this
  further, we measure \act{R}\ over $1\Gyr$, and compute the
  time-averaged radial action, \avgt{\act{R}}. We note both a heating
  of the stellar populations, and a comparable (but larger) scatter in
  \act{R}. We show that the azimuthal variations of
  \avgt{\act{R}}\ are anti-correlated with the variations of the
  density, as expected if the spirals are stronger in cooler, younger,
  metal-rich populations. We further show that the variations of
  metallicity, age and \al-abundance are poorly correlated with those
  of \avg{\act{R}}\ but very well correlated with those of
  \avgt{\act{R}}. (See Section~\ref{s:jr}.)
\item Thus the axisymmetric St\"ackel approximation over-estimates the
  radial action in high density regions and under-estimates it in low
  density regions. For studies that depend critically on the behaviour
  of one \act{R}\ population versus that of another, it is important
  to correct for this effect. We have done this by time averaging
  \act{R}\ over $1\Gyr$, during which time star particles will have
  drifted in and out of spiral arms. Such an approach is of course not
  possible in the MW; other ways of correcting for these errors are needed. (See Section~\ref{s:jr}.)
\end{enumerate}


\bigskip
\noindent
{\bf Acknowledgements.}

We thank the anonymous referee whose comments helped improve this
paper, and Pedro Palicio for useful discussions.
T.K. acknowledges support from the National Science Foundation of
China (NSFC) under grant No. 12303013 and support from the China
Postdoctoral Science Foundation under grant No. 2023M732250.
J.A. and C.L. acknowledge funding from the European Research Council
(ERC) under the European Union’s Horizon 2020 research and innovation
programme (grant agreement No. 852839). J.A. also acknowledges support
by the National Natural Science Foundation of China under grant
No. 12122301, 1223001, by the ``111'' project of the Ministry of
Education under grant No. B20019, and the sponsorship from Yangyang
Development Fund.
L.BeS. acknowledges the support provided by the Heising Simons Foundation through the Barbara Pichardo Future Faculty Fellowship from grant \# 2022-3927, and of the NASA-ATP award 80NSSC20K0509 and U.S. National Science Foundation AAG grant AST-2009122.
The simulation was run at the DiRAC Shared Memory Processing system at
the University of Cambridge, operated by the COSMOS Project at the
Department of Applied Mathematics and Theoretical Physics on behalf of
the STFC DiRAC HPC Facility (www.dirac.ac.uk). This equipment was
funded by BIS National E-infrastructure capital grant ST/J005673/1,
STFC capital grant ST/H008586/1 and STFC DiRAC Operations grant
vST/K00333X/1. DiRAC is part of the National E-Infrastructure. 
The analysis used the {\sc python} library {\sc pynbody}
\citep{pynbody}. The pattern speed analysis was performed on
Stardynamics, which was funded through Newton Advanced Fellowship
NA150272 awarded by the Royal Society and the Newton Fund.

\bigskip 
\noindent

\bigskip
\noindent
\section*{Data Availability}

\noindent
The simulation snapshot in this paper can be shared on reasonable request.

\label{lastpage}

\bibliographystyle{aj}
\bibliography{allrefs}

\appendix

\section{Other timesteps}
\label{app:othertimes}

\begin{figure*}
\centerline{
  \includegraphics[angle=0.,width=0.4\hsize]{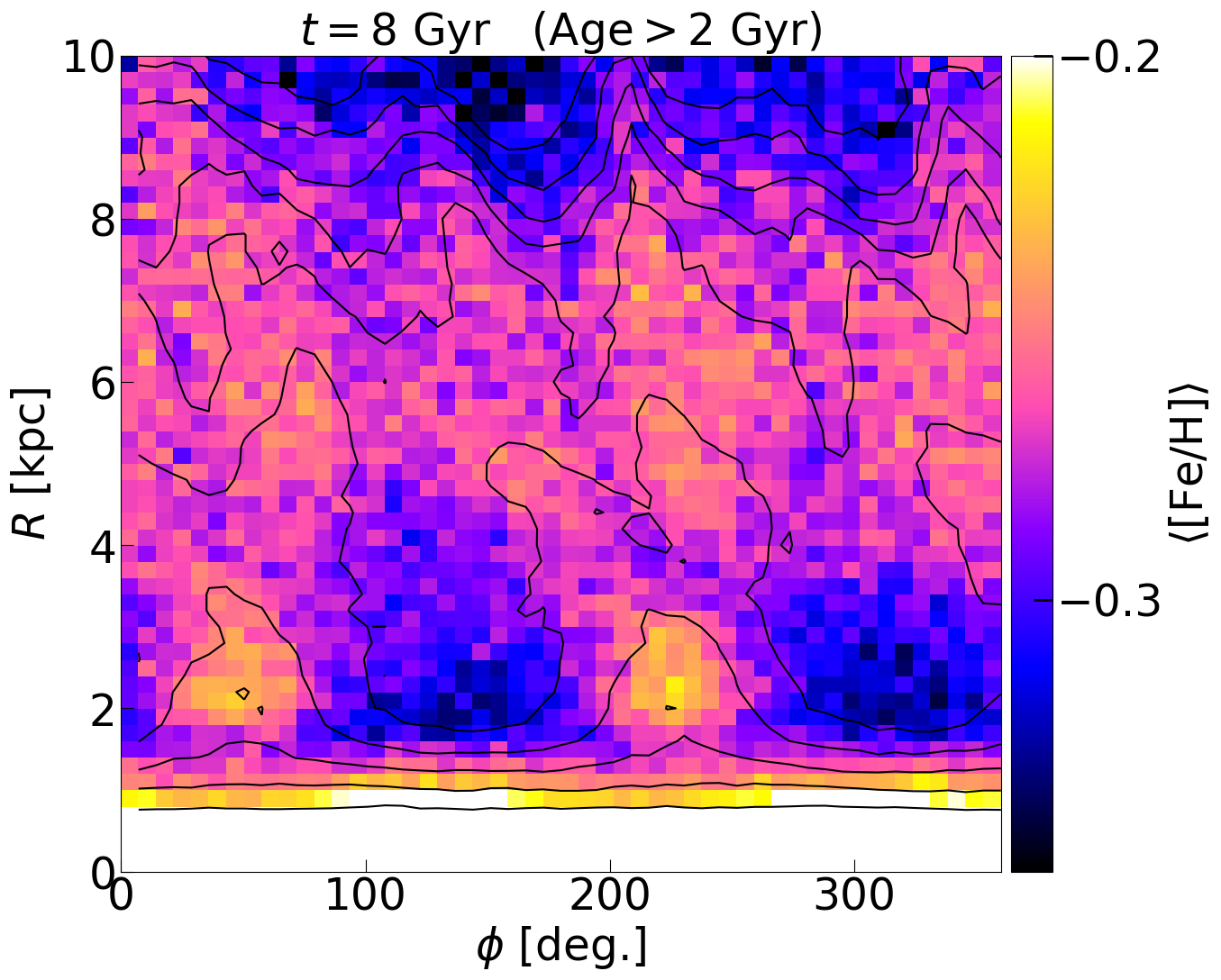}
  \includegraphics[angle=0.,width=0.4\hsize]{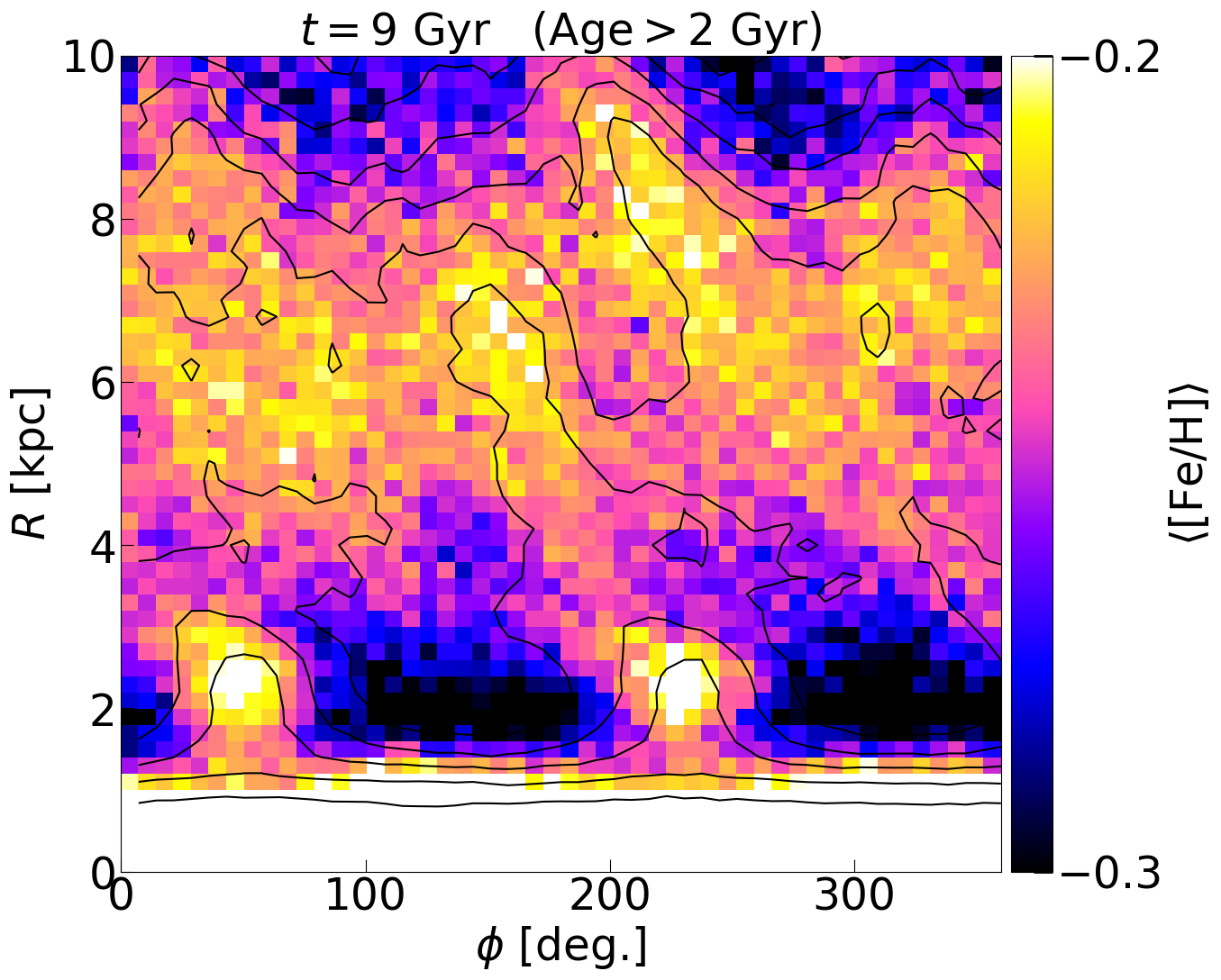} \
} 
\centerline{
  \includegraphics[angle=0.,width=0.4\hsize]{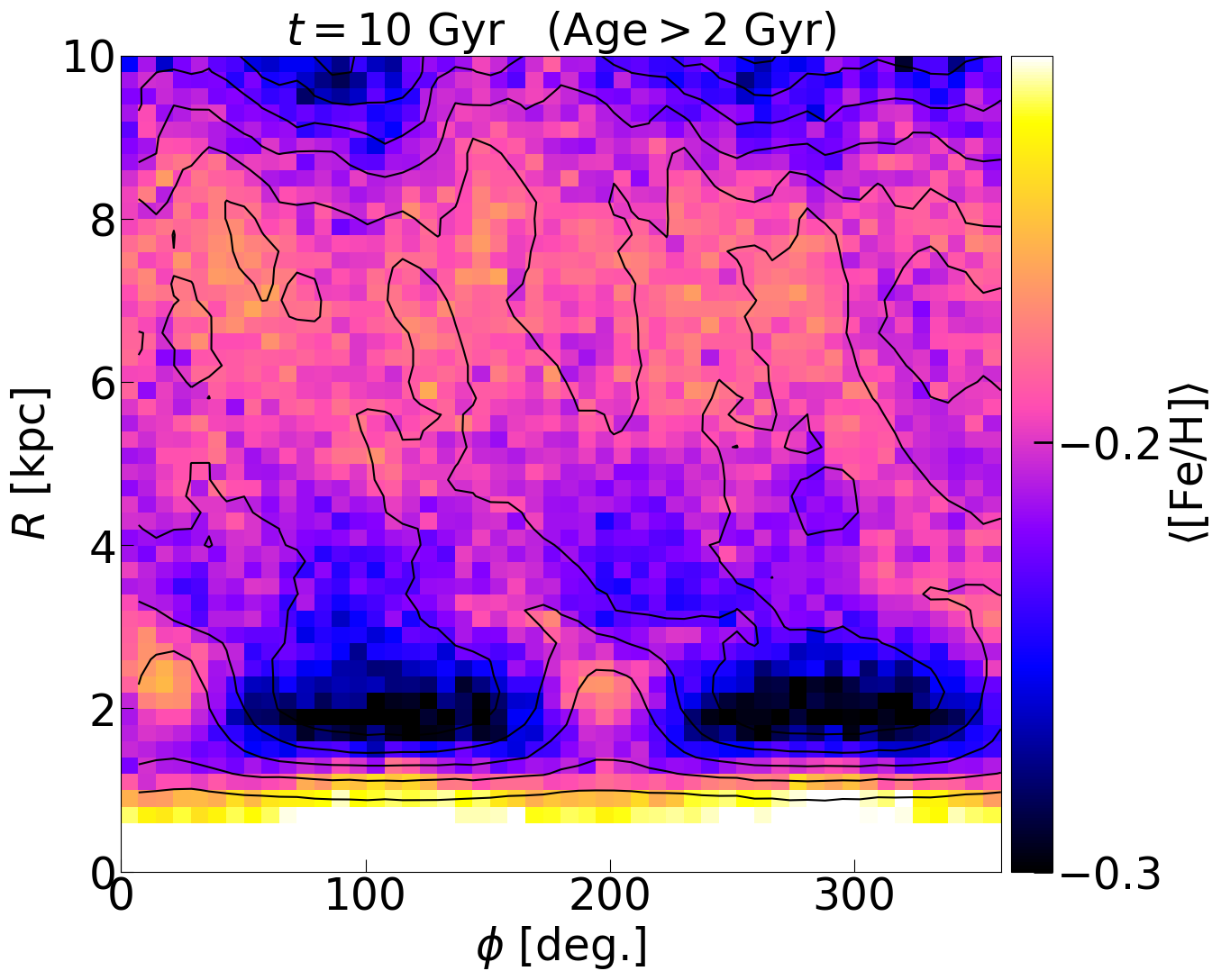}
  \includegraphics[angle=0.,width=0.4\hsize]{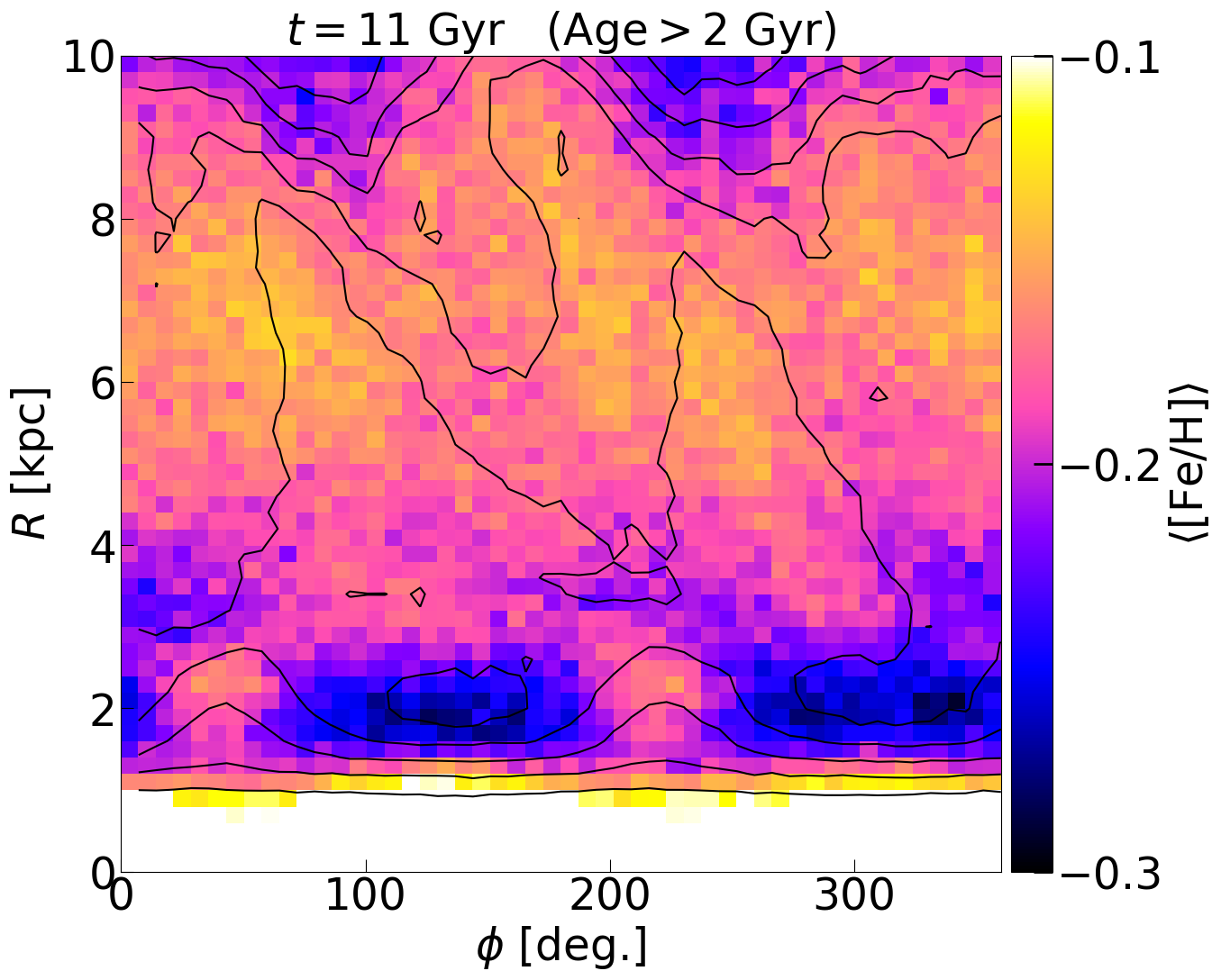} \
} 
\centerline{
  \includegraphics[angle=0.,width=0.4\hsize]{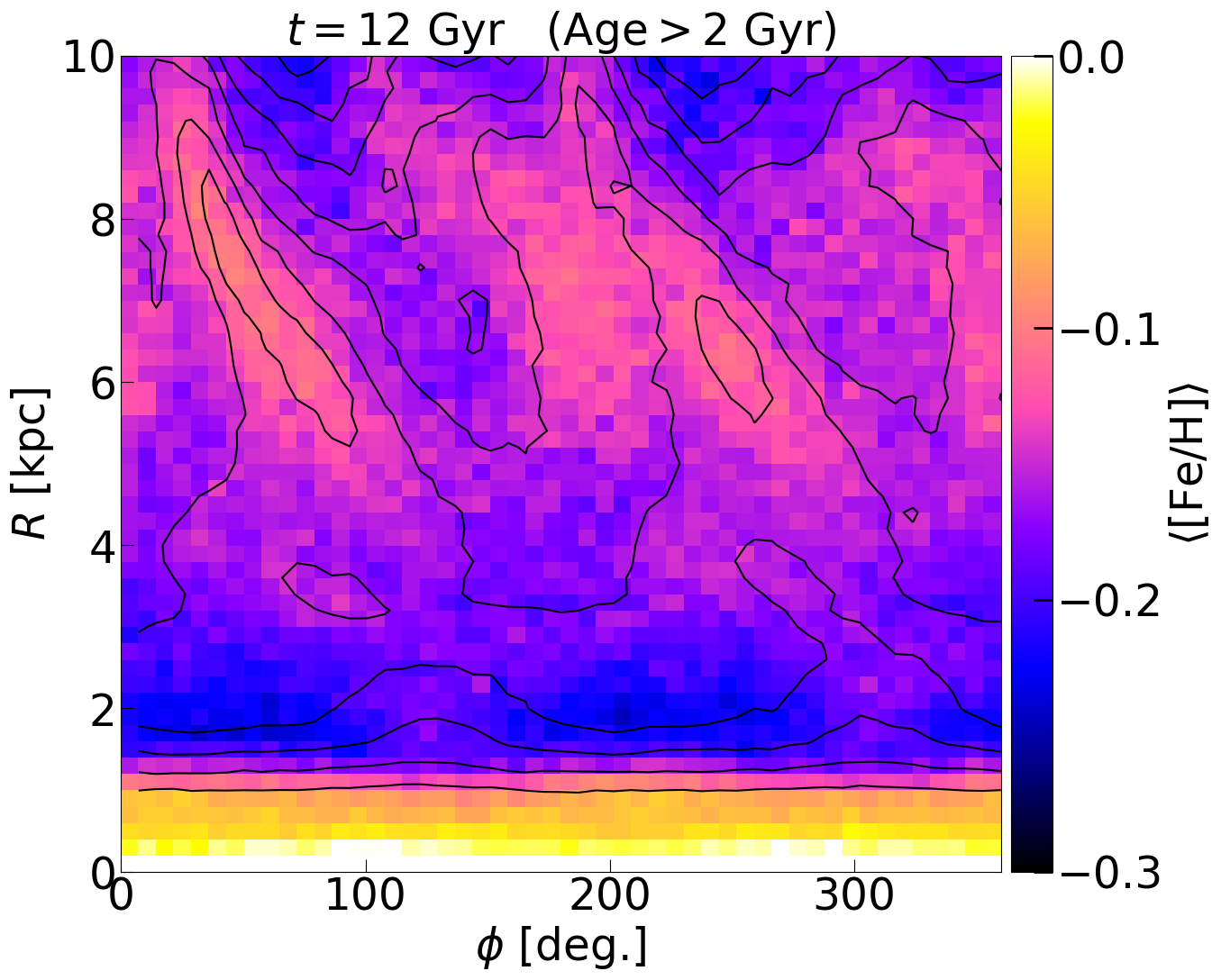}
  \includegraphics[angle=0.,width=0.4\hsize]{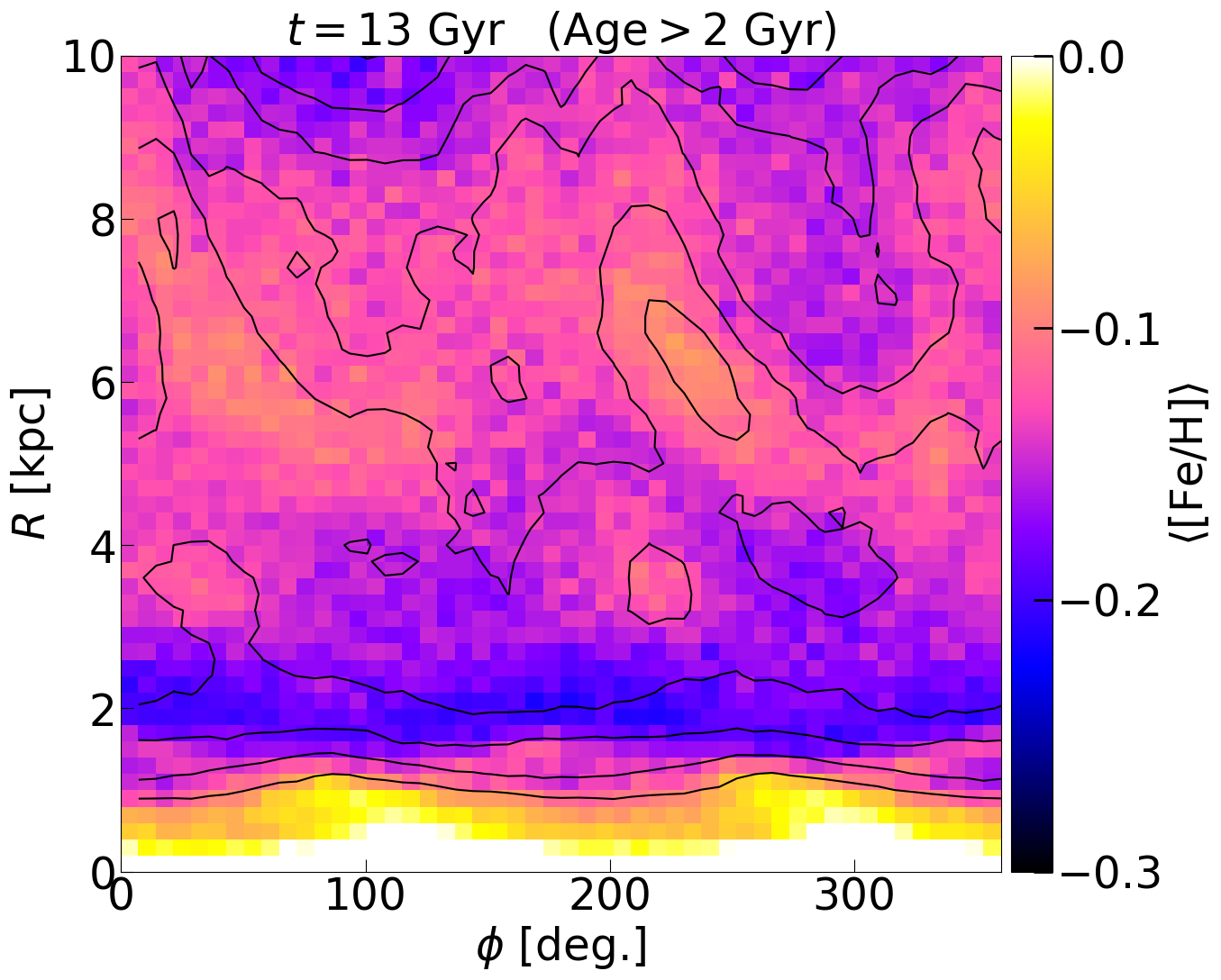} \
}
\caption{\avg{\feh} in cylindrical coordinates at $t=8-13\Gyr$, as labelled. Only stars older than $2\Gyr$ are considered (compare Fig.~\ref{f:azimuthal}). The contours correspond to the full surface density. Note the changing colour scale.
\label{f:fehmapswtime}}
\end{figure*}

\begin{figure*}
\centerline{
  \includegraphics[angle=0.,width=0.4\hsize]{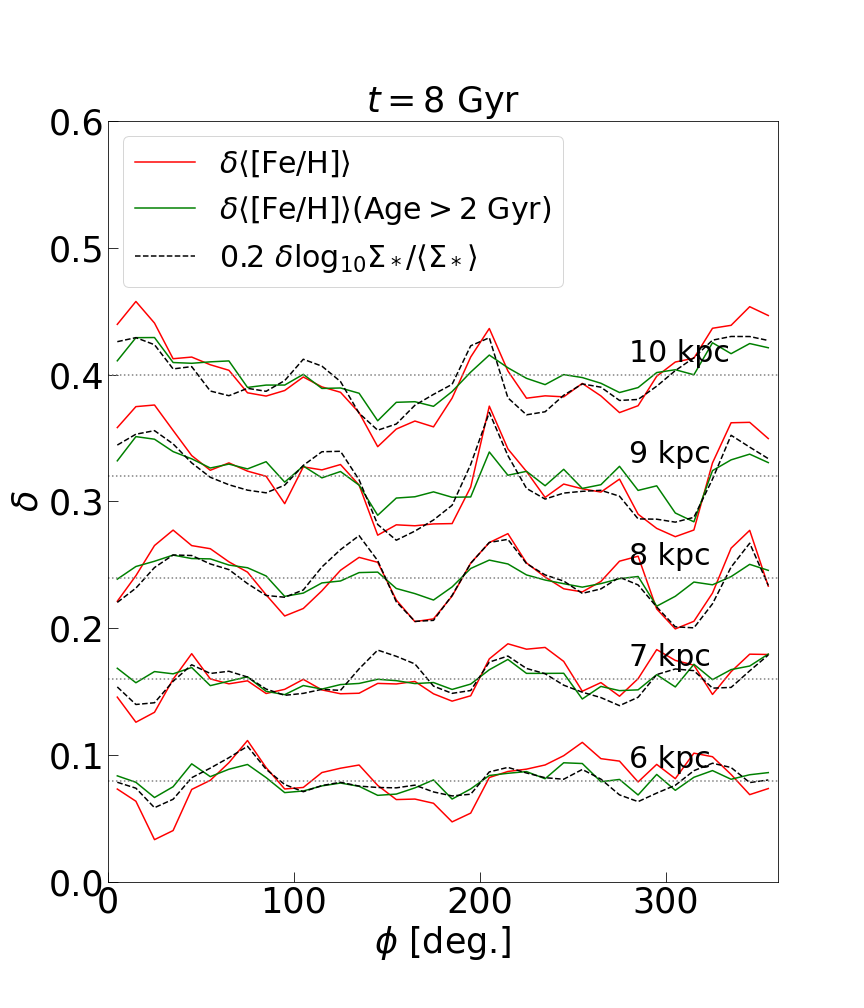}
  \includegraphics[angle=0.,width=0.4\hsize]{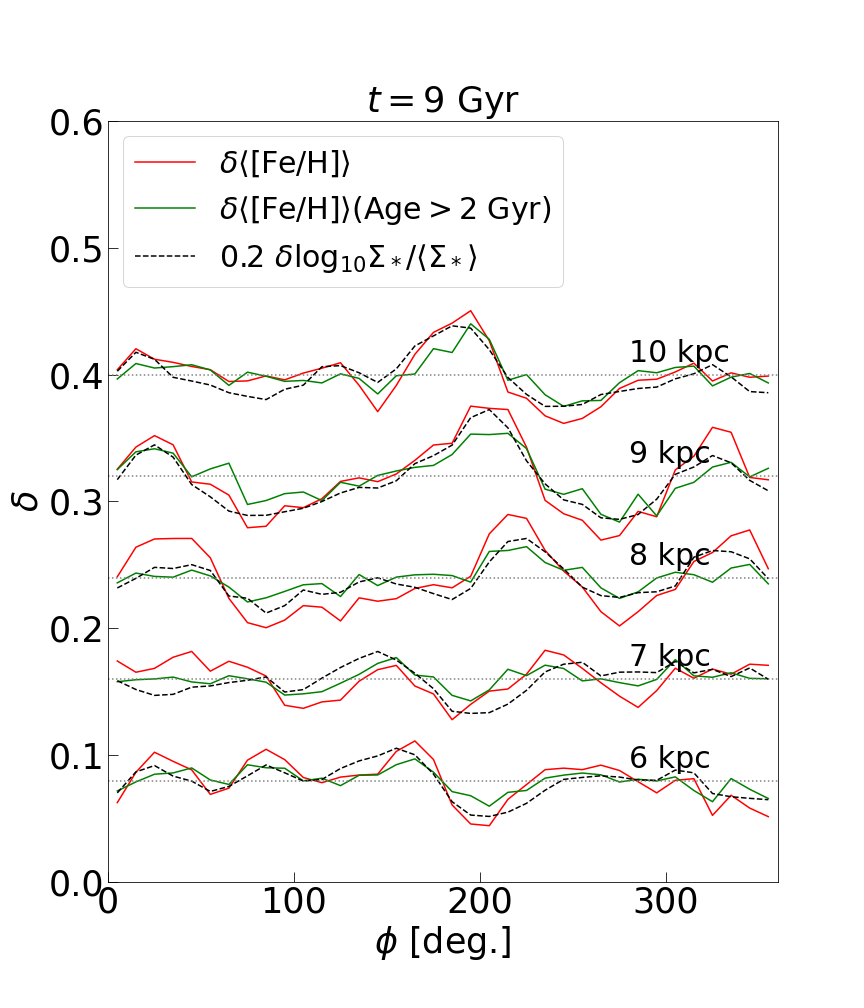} \
} 
\centerline{
  \includegraphics[angle=0.,width=0.4\hsize]{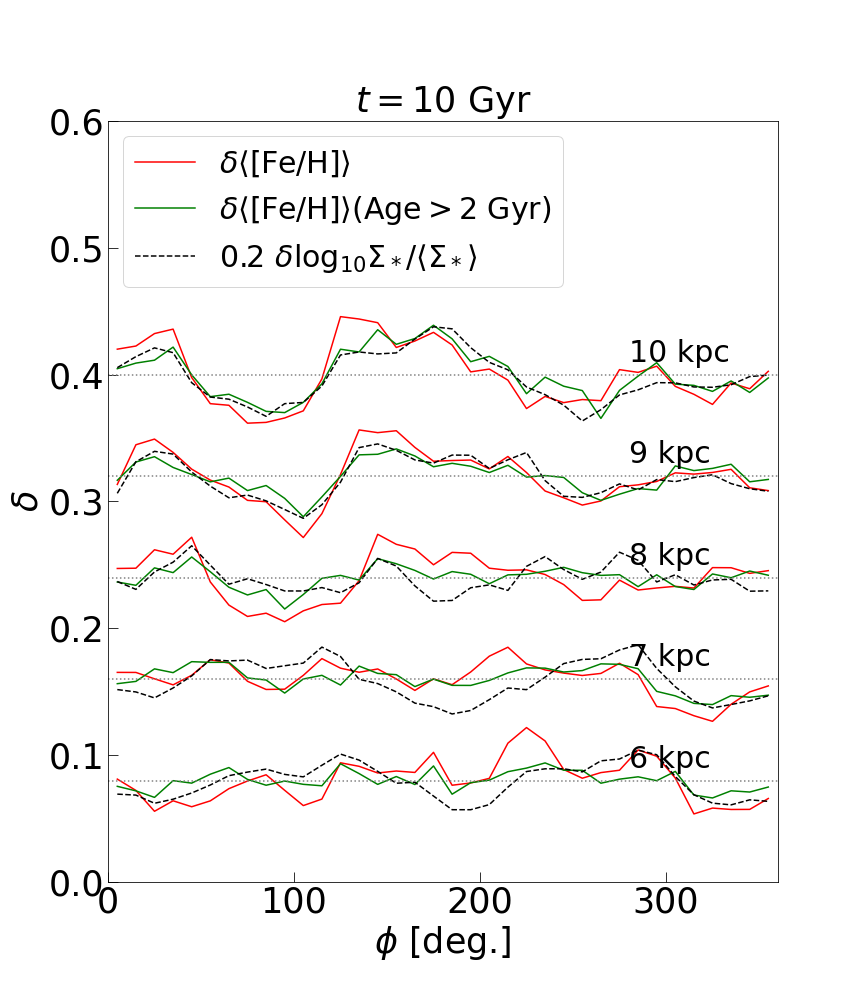}
  \includegraphics[angle=0.,width=0.4\hsize]{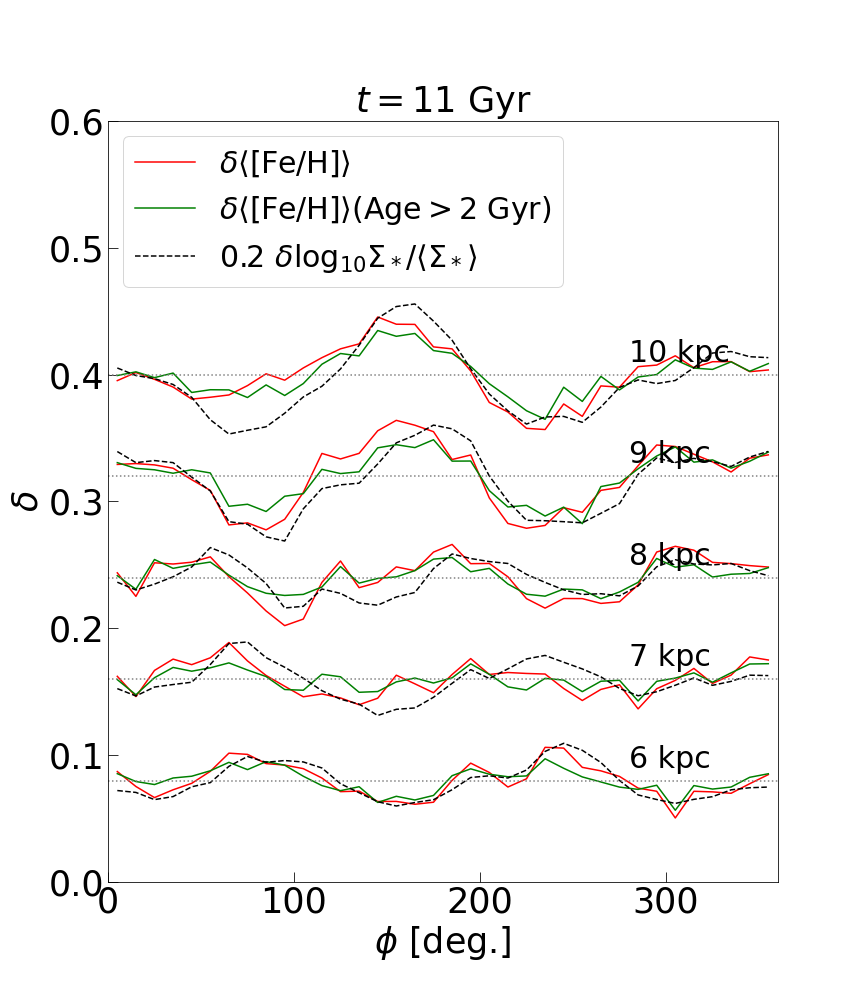} \
} 
\centerline{
  \includegraphics[angle=0.,width=0.4\hsize]{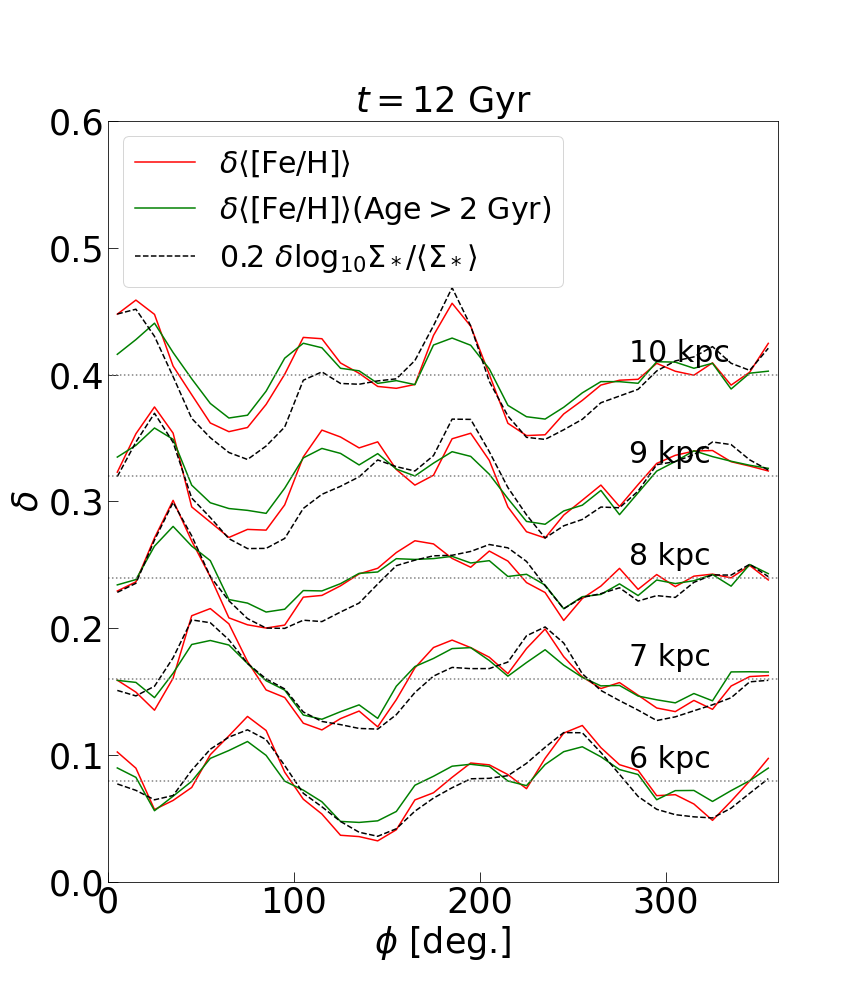}
  \includegraphics[angle=0.,width=0.4\hsize]{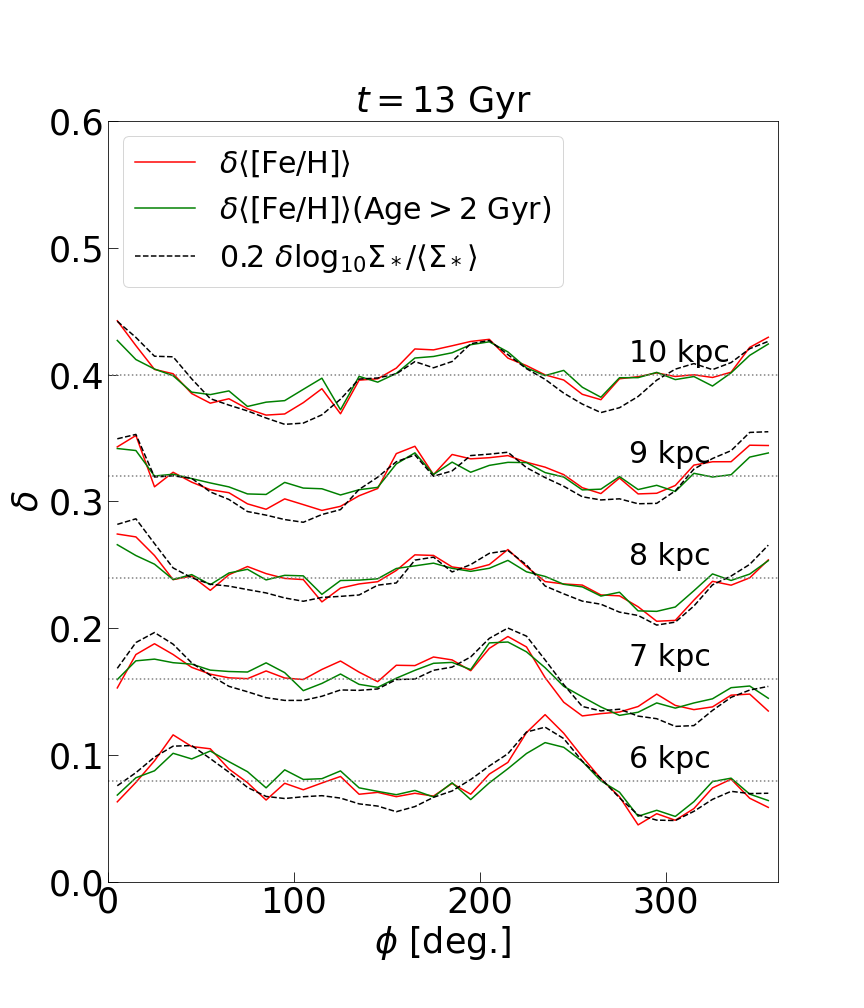} \
}
\caption{Same as the left panel of Fig.~\ref{f:razimuthal} at $t=8-13\Gyr$, as labelled.
\label{f:fehazprofswtime}}
\end{figure*}

We present \avfe\ maps (Fig.~\ref{f:fehmapswtime}) and azimuthal
profiles of $\Sigma_*$ and \avfe\ (Fig.~\ref{f:fehazprofswtime})
of the model at $t=8\Gyr$ to $13\Gyr$. The general behaviour described
in the main text holds at earlier times.

\section{Median \feh\ and radial action}
\label{app:medians}

\begin{figure}
\centerline{
\includegraphics[angle=0.,width=\hsize]{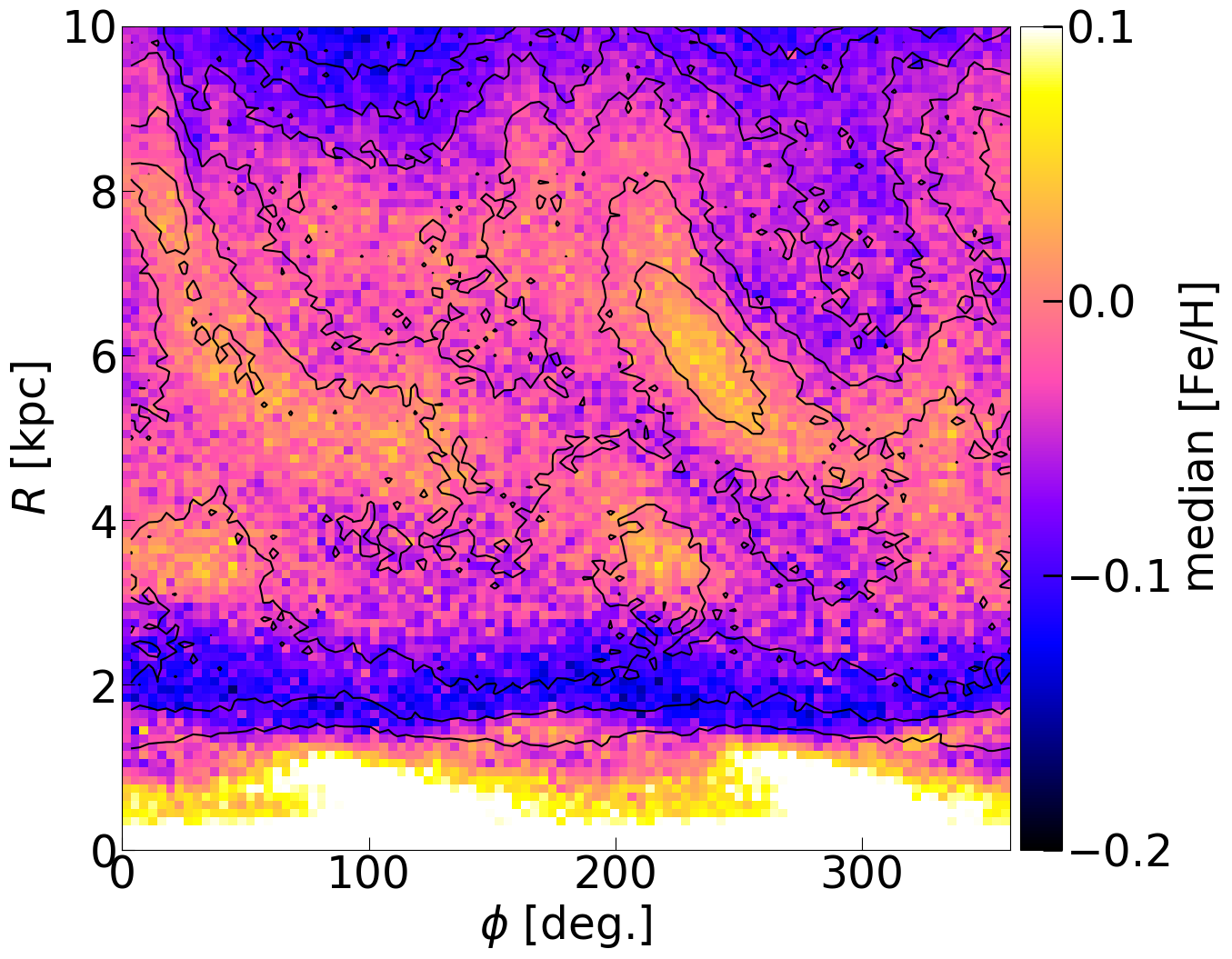}\
}
\centerline{
\includegraphics[angle=0.,width=\hsize]{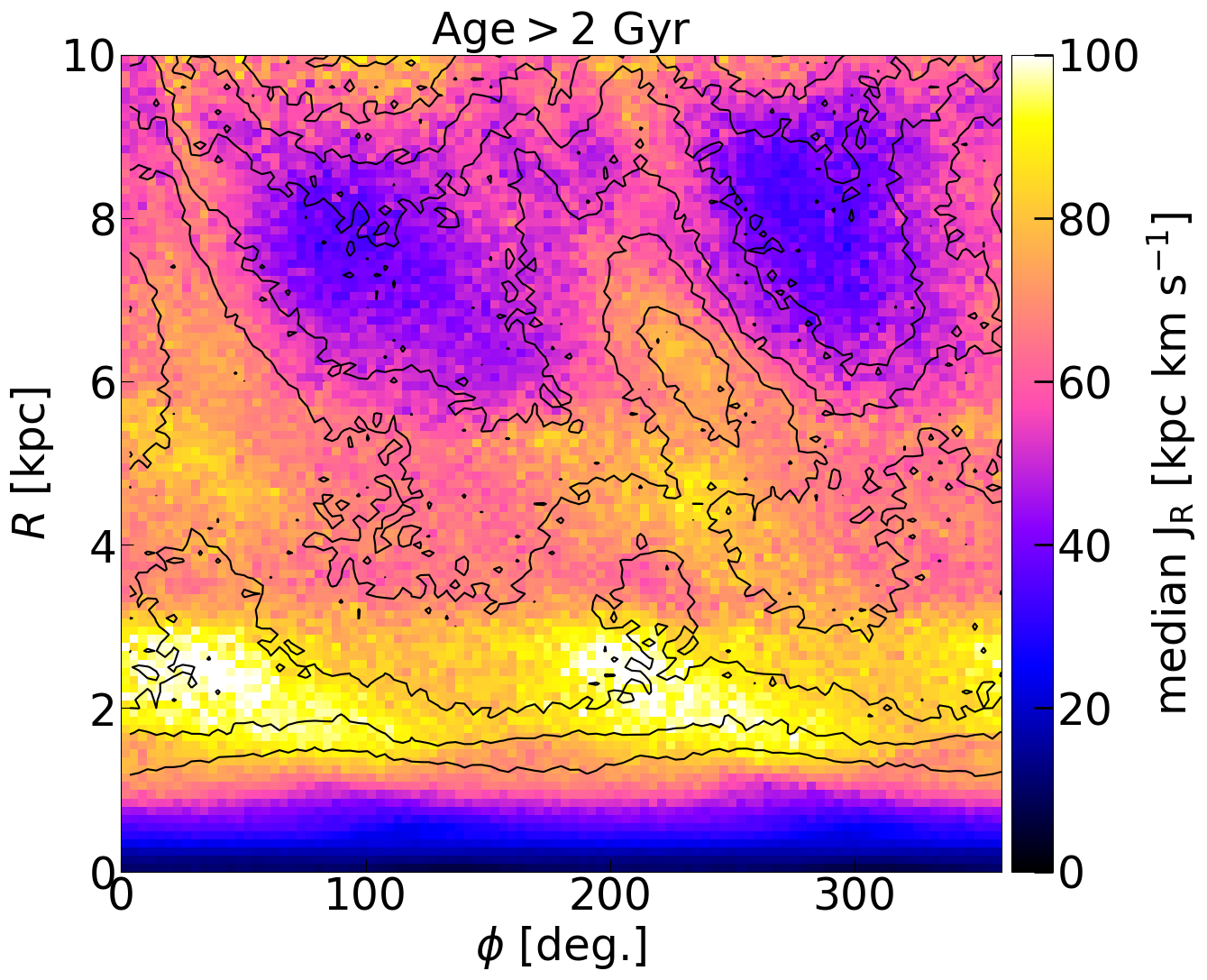}\
}
\centerline{
\includegraphics[angle=0.,width=\hsize]{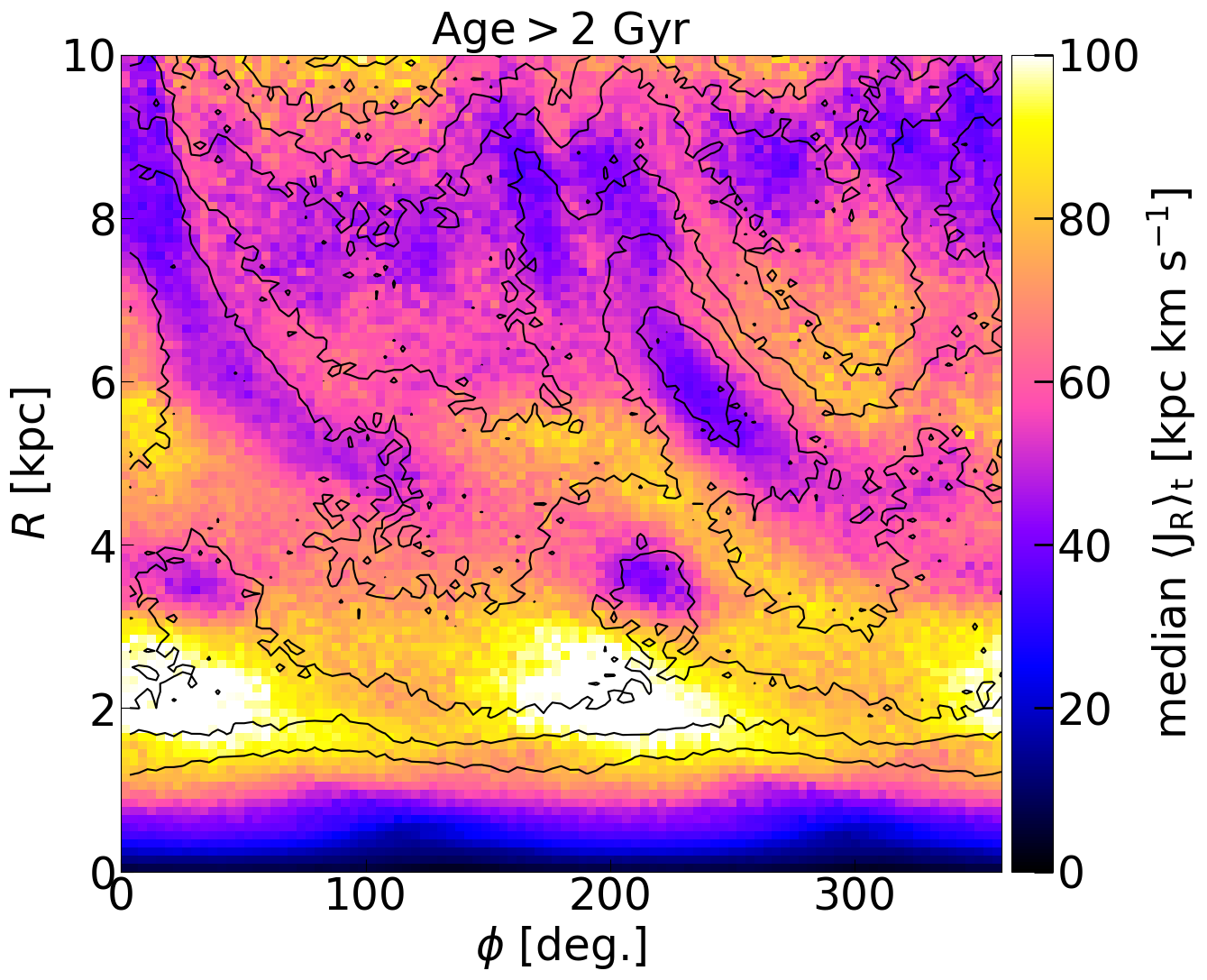}
}
\caption{Top: median \feh. Compare with the middle panel of Fig.~\ref{f:azimuthal}.
Middle and bottom: Median \act{R}\ and
\avgt{\act{R}}. Compare with Fig.~\ref{f:meanjrajrmaps}.
In all panels the contours correspond to the surface density. The disc
is rotating in a counter-clockwise sense, \ie\ rotation is in the
direction of increasing $\phi$. The model is shown at $13~\Gyr$.
\label{f:medianmaps}}
\end{figure}

\begin{figure}
\centerline{
\includegraphics[angle=0.,width=\hsize]{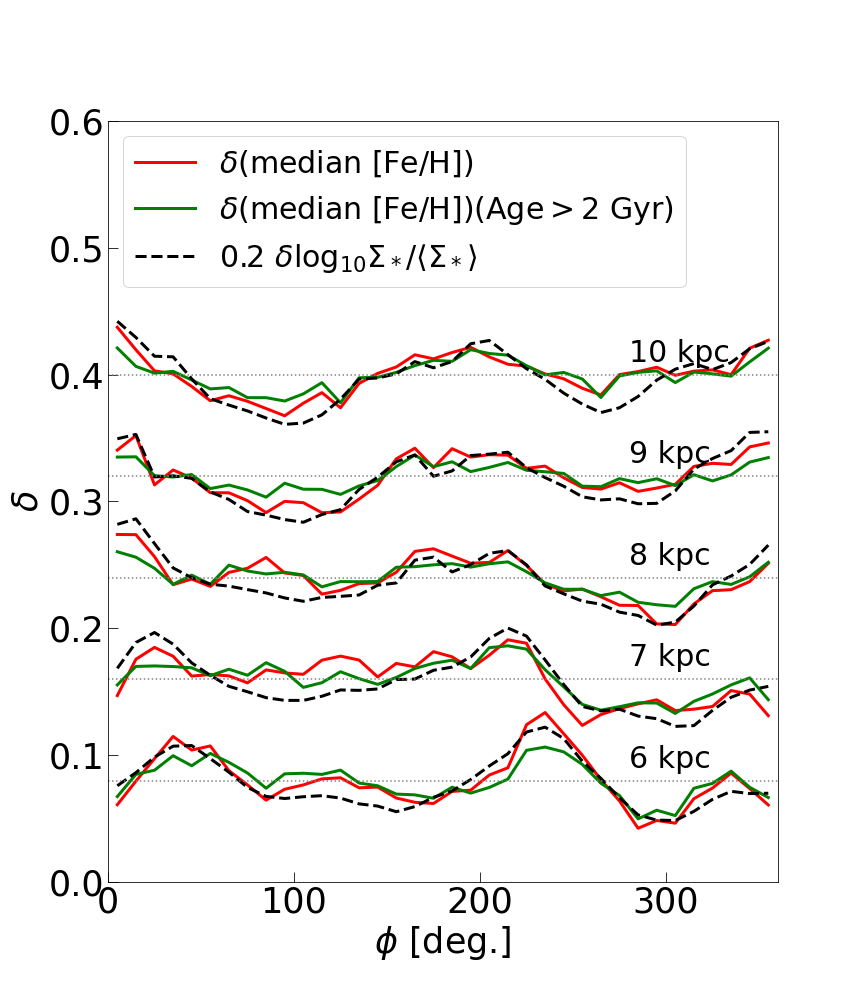}\
}
\centerline{
\includegraphics[angle=0.,width=\hsize]{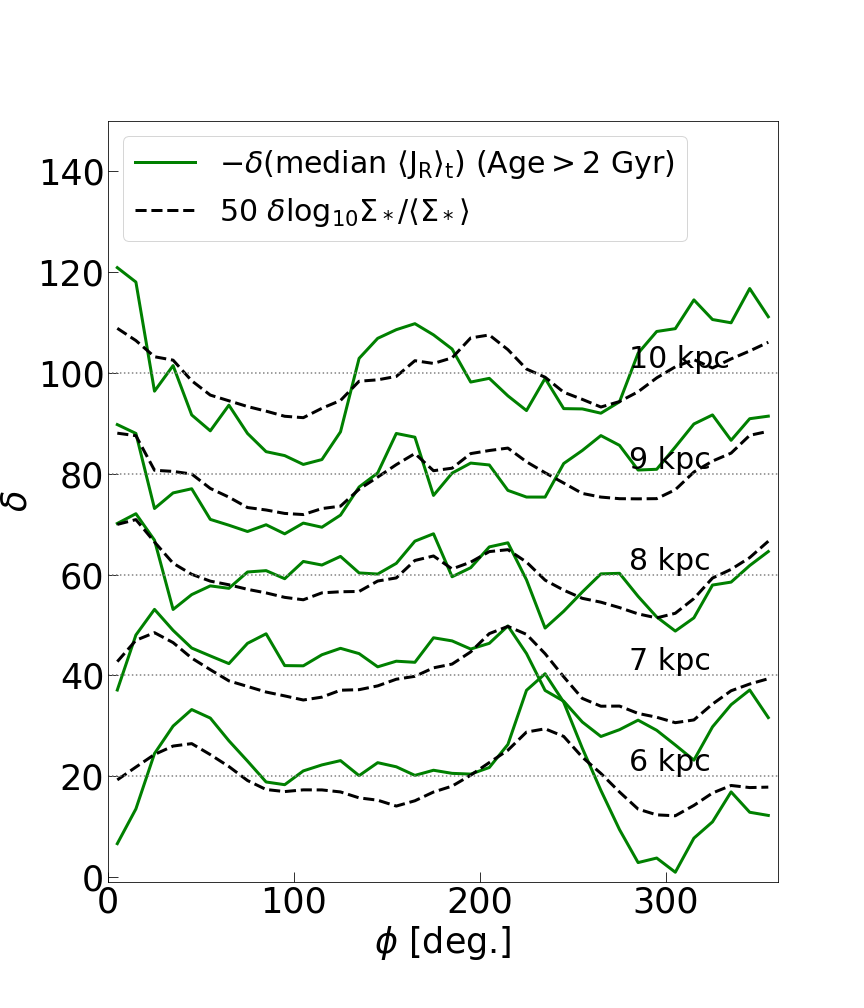}
}
\caption{The variation of median \feh\ (top) and
  $-\rm{median}~{\avgt{\act{R}}}$\ (bottom) compared with the density
  variation at a series of $500~\pc$-wide annuli centred at $6\kpc$ to
  $10\kpc$, as indicated. The average value at each radius is
  subtracted from each azimuthal profile and the profiles are then
  vertically offset by a fixed amount to show the variation. The
  azimuthal profiles of the density have been scaled for ease of
  comparison. The dotted horizontal lines indicate the zero for each
  radius. In the top panel, the red lines show profiles of median
  \feh\ for all stars, while the green lines show stars older than
  $2~\Gyr$ only (both panels). Compare these with
  Fig.~\ref{f:razimuthal} and the right panel of
  Fig.\ref{f:jrazimuthal}.
\label{f:medianazimuthal}}
\end{figure}

Since distributions of \feh\ and \act{R}\ are generally skewed, we
demonstrate that the results in the main text would still hold using
the medians instead of the means. Fig.~\ref{f:medianmaps} show maps of
medians, while Fig.~\ref{f:medianazimuthal} shows azimuthal profiles
of medians. Despite the $\sim 0.1$ dex difference in \feh\ between the
mean and the median, the trends are comparable. A similar result holds
for \act{R}.

\section{Other particles in action space}
\label{app:otherparticles}

\begin{figure*}
\centerline{
\includegraphics[angle=0.,width=0.9\hsize]{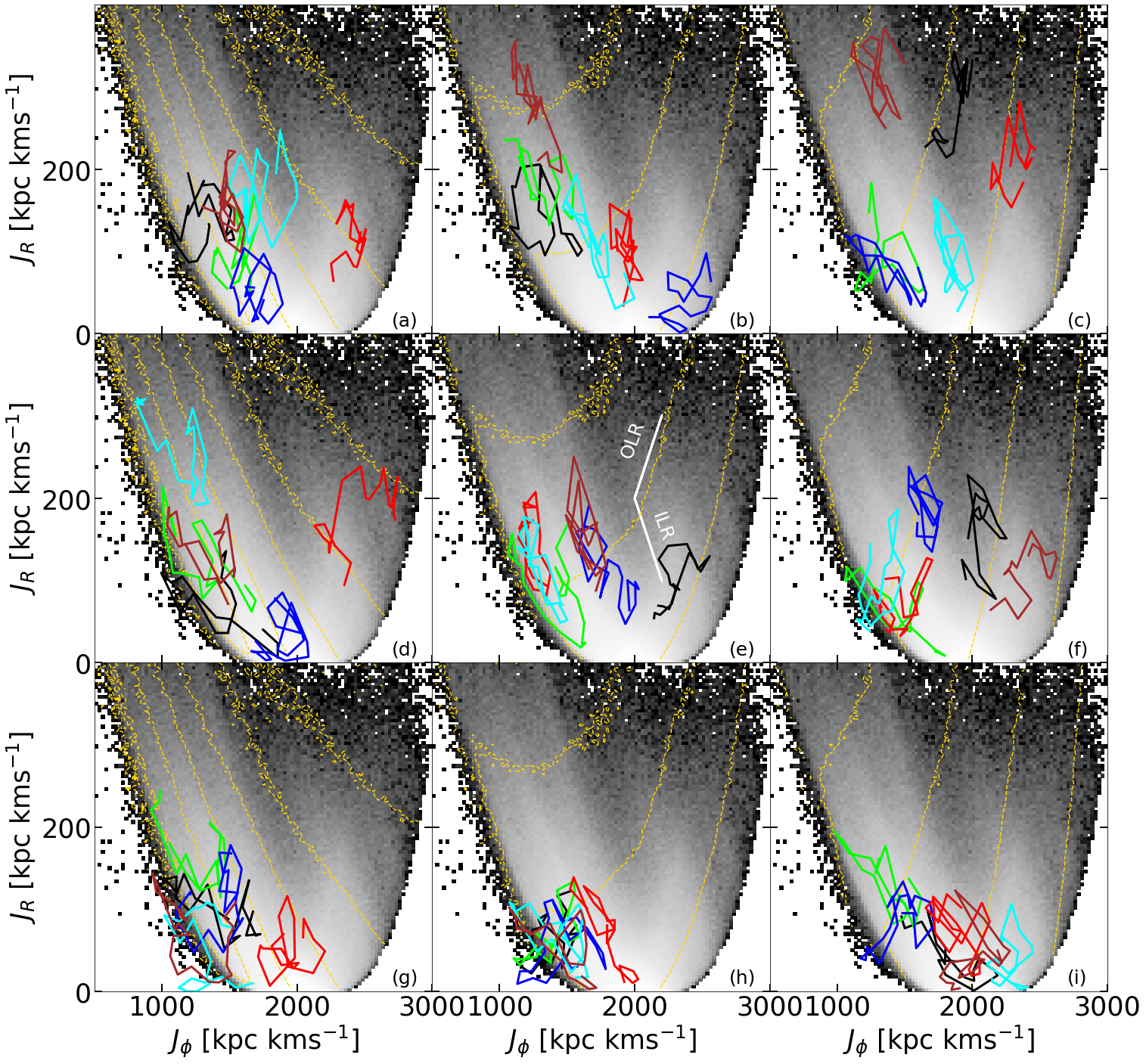}\
}
\caption{The evolution of a random set of 54 particles in the action
  space (\act{\phi},\act{R}). These particles have been selected as in
  Fig.~\ref{f:jrjpevol2} with the extra condition that the spread in
  \act{R}\ of each particle is large. In each panel we present 6 stars
  with different colours over a $1\Gyr$ time interval, plotting every
  $50\Myr$. The plotted particles are ordered by age, with the oldest
  in the top row ($8.9\leq \age/\Gyr \leq 12.7$), the intermediate age
  ones in the middle row ($5.8\leq \age/\Gyr \leq 8.9$) and the
  youngest in the bottom row ($2.4\leq \age/\Gyr \leq 5.7$). The
  dashed yellow lines show contours of constant $E_J$ for $\omp =
  15.4~\kmsk$ (left), $40.0~\kmsk$ (centre) and $61.5~\kmsk$
  (right). The white lines in panel (e) indicate the libration vectors
  at the OLR (upper line) and the ILR (lower line).  The background
  shows a log scale representation of the number density of stars.
\label{f:jrjpevol2}}
\end{figure*}

We consider a different selection of particles to track in the
$(\act{\phi},\act{R})$-space. We now add the additional selection that
the variation of \act{R}\ is large by requiring that its dispersion
for values computed over $1\Gyr$ is large: $25 < \sigma_{JR}/\kkms <
50$. The results are shown in Fig.~\ref{f:jrjpevol2}. The new
selection favours stars trapped at the Lindblad resonances and we see
that many stars appear to be trapped at the ILR of the slowest
spiral. Nevertheless even for these particles many of the tracks show
considerable jitter, as was the case in Fig.~\ref{f:jrjpevol},
reinforcing the conclusion that measuring actions assuming axisymmetry
introduces significant errors.

\end{document}